\documentclass[article]{elsarticle}

\usepackage{lineno,hyperref}
\modulolinenumbers[5]

\usepackage{bm}
\usepackage{amsmath}
\usepackage{subfigure}
\usepackage{booktabs}
\usepackage{makecell}
\usepackage{color}
\journal{Elsevier}

\begin{document}

\begin{frontmatter}

\title{Unified gas-kinetic scheme with simplified multi-scale numerical flux 
	for thermodynamic non-equilibrium flow in all flow regimes}

\author[a]{Rui Zhang}
\ead[Rui Zhang]{zhangruinwpu@mail.nwpu.edu.cn}
\author[a,b,c]{Sha Liu\corref{mycorrespondingauthor}}
\cortext[mycorrespondingauthor]{Corresponding author}
\ead[Sha Liu]{shaliu@nwpu.edu.cn}
\author[a,b]{Chengwen Zhong}
\ead[Chengwen Zhong]{zhongcw@nwpu.edu.cn}
\author[a,b]{Congshan Zhuo}
\ead[Congshan Zhuo]{zhuocs@nwpu.edu.cn}

\address[a]{School of Aeronautics, Northwestern Polytechnical University, Xi’an, Shaanxi 710072, China}
\address[b]{National Key Laboratory of Science and Technology on Aerodynamic Design and Research, 
	Northwestern Polytechnical University, Xi’an, Shaanxi 710072, China}
\address[c]{Institute of Extreme Mechanics, Northwestern Polytechnical University, Xi’an, Shaanxi 710072, China}

\begin{abstract}
In this paper, a unified gas-kinetic scheme (UGKS) with simplified multi-scale numerical flux is proposed for 
the thermodynamic non-equilibrium flow simulation involving the excitation of molecular vibrational degrees 
of freedom in all flow regimes. The present UGKS keep the basic conservation laws of the macroscopic flow 
variables and the microscopic gas distribution function in a discretized space. In order to improve the 
efficiency of the UGKS, a simplify multi-scale numerical flux is directly constructed from the characteristic 
difference solution of the kinetic model equation. In addition, a new BGK-type kinetic model for diatomic gases 
is proposed to describe the high-temperature thermodynamic non-equilibrium effect, which is a phenomenological 
relaxation model with the continuous distribution modes of rotational and vibrational energies. In present model, 
the equilibrium distribution functions is constructed by using a multi-dimensional Hermitian expansion around 
the Maxwellian distribution to achieve the correct Prandtl number and proper relaxation rate of heat fluxes. 
Furthermore, the application of the unstructured discrete velocity space (DVS) and a simple integration error 
correction reduce the number of velocity mesh significantly and make the present method be a efficient tool for 
simulations of flows in all flow regimes. 
The new scheme are examined in a series of cases, such as Sod’s shock tube, high non-equilibrium shock structure, 
hypersonic flow around a circular cylinder with Knudsen (Kn) number $Kn = 0.01$, and the rarefied hypersonic flow 
over a flat plate with a sharp leading edge. The present UGKS results agree well with the benchmark data of DSMC 
and the other validated methods. 
\end{abstract}

\begin{keyword}
\texttt Unified scheme\sep 
Diatomic molecules\sep 
Vibrational relaxation\sep 
Non-equilibrium flow
\end{keyword}

\end{frontmatter}

\linenumbers

\section{Introduction}\label{Introduction}
\par
Hypersonic rarefied gas flows are fundamental and crucial problems for the research and design of 
the hypersonic vehicles and atmospheric reentry spacecraft~\citep{JD_Schmisseur_2015,MarcSchouler_2020}. 
During atmospheric reentry, a space vehicle experiences complex atmospheric environments from free-molecular flow, 
transitional flow, slip flow to continuum flow regimes. Another situation is that the continuum and rarefied flow 
regimes can be encountered in different parts of the hypersonic vehicles even in the same atmospheric environment, 
due to the strong compression and expansion caused by its hypersonic speed. As a result, the complicated local 
rarefied and non-equilibrium regions often appear in the flow field~\citep{MS_Ivanov_1998,Harshal_Gijare_2019}. 
For example, at free-stream Mach (Ma) number $\rm{Ma}_{\infty}=6$ and Knudsen (Kn) 
number $\rm{Kn}_{\infty}=1.26\times10^{-4}$, the numerical results of X-38 vehicle show that the local Knudsen number 
around the vehicle can cover a wide range of values difference with four orders of magnitude~\citep{DingwuJiang_2016}, 
where the local Knudsen number is defined as the ratio of local molecular mean free path to the characteristic length 
scale of flow variable variation~\citep{GA_Bird_1994}.
\par
In such multi-scale hypersonic flows, both the rarefied gas effect and high-temperature non-equilibrium effect should be 
taken into account~\citep{GA_Bird_1994,JohnD_Anderson_Jr_2006}. In hypersonic rarefied gas flows, because of the fact that 
at high Mach number the flow temperature increases rapidly behind the shock, the molecular vibrational degrees of freedom 
will be excited, and even complex phenomena such as chemical reaction, dissociation and ionization may 
exist~\citep{JohnD_Anderson_Jr_2006}. 
As a result, the thermodynamic and thermochemical properties of the gas will change. On the other hand, both the 
inter-molecular and the gas-surface collisions play an important role in rarefied gas flows, such that the aerodynamic 
characteristics of the rarefied gas flows will change considerably as compared to their continuum behavior. Especially, 
the velocity slip and temperature jump phenomena can be prominent and significant in determining aerodynamic forces 
and heat fluxes on the body surface~\citep{ZhihuiLi_2009,R_Prakash_2019}. 
\par
The direct simulation Monte Carlo (DSMC) method, employing the Larsen–Borgnakke (LB) model~\citep{Claus_Borgnakke_1975} 
for the translational–internal energy exchange, is one of the most popular methods for the simulation of hypersonic 
flows~\citep{GA_Bird_1994}. However, the DSMC method becomes computationally expensive for simulating the transitional and 
near-continuum flows due to that the correspondingly cell size and time step have to be less than the molecular mean free path 
and collision time, respectively~\citep{Thomas_E_Schwartzentruber_2015,KunXu_A_2021}. On the other hand, the Navier-Stokes 
(N-S) equations with the Newtonian law of viscosity and the Fourier law of heat conduction accurately model the flows and can 
be solved efficiently in continuum flow regime, but the N-S solvers start to lose their validity when rarefied gas effect 
becomes serious.
\par  
The Boltzmann equation is a fundamental equation for kinetic theory of gases, where the flow physics from the continuum flow to 
the free-molecular flow regimes can be described on the kinetic scale~\citep{S_Chapman_1970}. 
However, it is extremely difficult to exactly solve the Boltzmann equation using a deterministic numerical approach for 
practical applications. In addition, the Boltzmann equation 
is valid for monoatomic gases while a large number of gases are diatomic and polyatomic in nature. Obviously, the problem 
will become more serious when a polyatomic gas including the translational and internal degrees of freedom is considered 
in the framework of the Wang Chang-Uhlenbeck (WCU) equation~\citep{CS_WangChang_1951}. Therefore, some simplified and 
tractable kinetic models~\citep{PL_Bhatnagar_1954,H_Holway_1966,EM_Shakhov_1968,VA_Rykov_1975,LeiWu_2015} 
have been proposed and widely used to approximate the solutions of complex Boltzmann and WCU equations.
The Bhatnagar–Gross–Krook (BGK) model~\citep{PL_Bhatnagar_1954} is the simplest and most widely employed simplification 
of the Boltzmann collision operator for monoatomic gases. However, the Prandtl (Pr) number of the BGK model is an unchangeable 
unit value ($\rm{Pr}=1$), while the exact value for a monoatomic gas is 2/3. In order to obtain the correct Prandtl number, 
a number of modified models, such as ellipsoidal statistical (ES) model~\citep{H_Holway_1966} and Shakhov-BGK 
model~\citep{EM_Shakhov_1968}, were introduced based on different physical considerations. In the polyatomic molecules, 
kinetic models with a correct Prandtl number were introduced by the Holway~\citep{H_Holway_1966} and 
Rykov~\citep{VA_Rykov_1975}, which are the extensions of the monoatomic ES model and Shakhov-BGK model, respectively. On the 
basis of Rykov model and ES model, a number of models considering the excitation of molecular vibrational degrees of freedom 
have been developed and applied to the simulation of hypersonic flows~\citep{ZhaoWang_2017,VA_Titarev_2018,BN_Todorova_2020,
Y_Dauvois_2021}. 
\par 
Several approaches have been used to solve the kinetic model equations for hypersonic flow problems. The discrete 
velocity method (DVM), also known as the discrete ordinate method (DOM), is widely used in the area of hypersonic rarefied 
flow simulation with translational–internal energy exchange~\citep{Peter_Clarke_2018,Florian_Bernard_2019,C_Baranger_2020}. 
The conventional DVM can obtain high efficiency and good accuracy in the simulation of high Knudsen number flows. However, 
it is very time-consuming in the simulation of low Kn number flows, and the extra numerical viscosity will harm the 
numerical accuracy. Thus, the single-scale conventional DVM can be hardly used to do simulate the multi-regime problems 
accurately and efficiently. However, many hypersonic flow problems involving the continuum and rarefied flow simultaneously 
in a single flow field and the multi-scale numerical method is required. Some modified discrete velocity method for gas 
flow in all flow regimes, such as improved DVM~\citep{LM_Yang_An_2018,LM_Yang_Improved_2018}, multi-scale 
DVM~\citep{Ruifeng_Yuan_Novel_2021}, gas kinetic unified algorithm (GKUA)~\citep{ZhiHui_Li_2004,JunLinWu_2021} and 
general synthetic iterative scheme (GSIS)~\citep{WeiSu_2020,WeiSu_2021} have been proposed.
\par
In recent years, the unified gas kinetic scheme (UGKS) proposed by Xu et al.~\citep{KunXu_A_2010, KunXu_Direct_2015} has 
been developed for all flow regimes. Differing from the typical DVM, a local time-dependent evolving solution of a kinetic 
model equation is designed to provide multi-scale numerical fluxes in the framework of UGKS, and both macroscopic flow variables 
and microscopic gas distribution function will be updated based on this evolution solution in a finite control volume. 
This local time-dependent solution couples the molecule transport and collision effects in a local time step. As a result, 
the cell size of UGKS can be determined by the requirements of numerical solution of practical applications, which is not 
passively limited by the molecular mean free path. On the other hand, the local time step can be chosen according to the 
Courant-Friedrichs-Lewy (CFL) condition. Consequently, the UGKS can precisely and efficiently capture the flow behaviors 
in the whole flow regimes from free-molecular flow to continuum flow~\citep{KunXu_2021}. At the present stage, the UGKS 
based on the Rykov model~\citep{VA_Rykov_1975} and vibrational model for diatomic gases has been used to describe and 
capture the hypersonic non-equilibrium flow~\citep{ZhaoWang_2017,ShaLiu_2014}. With similar physical process 
of UGKS, the discrete unified gas-kinetic scheme (DUGKS) proposed by Guo et al.~\citep{ZhaoliGuo_2013,ZhaoliGuo_2015} 
is another multi-scale scheme for all flow regimes, in which a simpler discrete characteristic difference solution of the 
kinetic model equation in space and time is employed to reconstruct the multi-scale numerical fluxes. However, only 
the microscopic gas distribution function is updated in the DUGKS, and macroscopic flow variables are calculated by 
a numerical quadrature of the discrete distribution functions. Recently, a simplified DUGKS (SDUGKS)~\citep{Mingliang_Zhong_2020} 
and conserved DUGKS (CDUGKS)~\citep{Hongtao_Liu_2018,JianfengChen_2019} have been developed because of the simplicity of 
constructing multi-scale fluxes in DUGKS. With those nice properties, the UGKS and DUGKS has been successfully 
applied to a variety of multi-scale transport problems in different flow regimes~\citep{KunXu_2021,ZhaoliGuo_P_2021}. 
\par
The use of the coupled macroscopic and microscopic equations make a better performance on the conservation in the UGKS compared 
with the DUGKS, especially when the unstructured discrete velocity space (DVS)~\citep{VA_Titarev_N_2017,RuifengYuan_2020} is 
adopted. However, when considering the kinetic model equation of the diatomic or polyatomic gases, the construction of local 
time-dependent evolving solution will become complicated and unwieldy~\citep{ZhaoWang_2017}. In order to develop an accurate 
and efficient numerical method for hypersonic thermodynamic non-equilibrium flow simulation in all flow regimes, a simplified 
multi-scale numerical flux based on the strategy of DUGKS is proposed and combined into the framework of UGKS in present work. 
The present algorithm based on a new phenomenological kinetic model equation for diatomic gases including the vibrational degrees 
of freedom, which is an extension of the Rykov model equation~\citep{VA_Rykov_1975}. The present BGK-type collision operator 
consist of the elastic collision term describing the relaxation of translational energy and inelastic collision term describing 
the relaxation of rotational and vibrational energies. Furthermore, by using the unstructured DVS and the integral error 
compensation~\citep{RuifengYuan_2020}, the computational efficiency of the present UGKS is significantly increased. 
\par
The rest of this paper is organized as follows. The kinetic model equation for diatomic gases involving rotational and 
vibrational degrees of freedom is introduced in Sec.~\ref{Kinetic_Model_Diatomic}. In Sec.~\ref{UGKS_Diatomic}, the basic 
algorithm of UGKS with simplified multi-scale numerical flux for diatomic molecules is described in detail. A series of 
numerical test cases are performed and discussed to validate the proposed method in Sec.~\ref{Numerical_Results}. 
Finally, the concluding remarks are given in Sec.~\ref{Conclusions}.
\par

\section{Kinetic model for diatomic gas involving internal molecular energy}\label{Kinetic_Model_Diatomic}
\subsection{Distribution function and moments}\label{Distribution_Function}
\par
In the present work, we consider the kinetic description of the diatomic gas involving the molecular 
rotational and vibrational energies which are treated classically. 
In this case, the system state can be described by the molecular number density distribution function 
$f({\bm{x}},{\bm{u}},{\bm{\xi}},\varepsilon_{rot},\varepsilon_{vib},t)$, where ${\bm{x}}$ and ${\bm{u}}$ 
are the $D$-dimensional physical space and particle velocity space, respectively, ${\bm{\xi}}$ is 
the velocity vector with the degrees of freedom $L=3-D$, which consists of the rest components of the 
particle velocity ${\bm{u}}$ in the three-dimensional space, the continuous variables $\varepsilon_{rot}$ 
($\varepsilon_{rot}>0$) and $\varepsilon_{vib}$ (${\varepsilon_{vib}>0}$) are the molecular rotational 
and vibrational energies, respectively, and $t$ is the time.
\par
The macroscopic conserved variables, such as the density $\rho$, the momentum density $\rho{\bm{U}}$ and 
the energy density $\rho E$ are defined as the moments of distribution function in the phase space 
$d{\bf{\Xi }} = d{\bm{u}}d{\bm{\xi}}d{\varepsilon_{rot}}d{\varepsilon_{vib}}$ as follows:
\begin{equation}\label{f_solve_rho}
	\rho = \int{mf({\bm{x}},{\bm{u}},{\bm{\xi}},\varepsilon_{rot},\varepsilon_{vib},t)d{\bf{\Xi }}},
\end{equation}
\begin{equation}\label{f_solve_rhoU}
	\rho {\bm{U}} = \int {{\bm{u}}mf({\bm{x}},{\bm{u}},{\bm{\xi}},\varepsilon_{rot},\varepsilon_{vib},t)d{\bf{\Xi }}},
\end{equation}
\begin{equation}\label{f_solve_Energy}
	\begin{aligned}
		\rho E &=\frac{1}{2}{\rho{|{\bm{U}}|^2}}+\frac{{{K_{tr}}+{K_{rot}}+{K_{vib}}\left(T \right)}}{2}\rho RT\\
		&=\int {\left[{\frac{1}{2}{m}({{|{\bm{u}}|}^2}+{{|{\bm{\xi}}|}^2})+\varepsilon_{rot}+\varepsilon_{vib}}
			\right] f({\bm{x}},{\bm{u}},{\bm{\xi 
			}},\varepsilon_{rot},\varepsilon_{vib},t)d{\bf{\Xi }}}.
	\end{aligned}
\end{equation}
Here $R$ is the specific gas constant, $m$ is the molecular mass and $T$ is the equilibrium temperature 
which correspond to equilibrium between the translational, rotational, and vibrational energy exchanges. 
The translational degrees of freedom is ${K_{tr}}=3$, and the rotational degrees of freedom ${K_{rot}}$ is 
equal to 2 for diatomic molecules. According to the harmonic oscillator model, the vibrational degrees of 
freedom ${K_{vib}}$ at temperature $T$ can be determined by the following formula~\citep{GA_Bird_1994}:
\begin{equation}\label{T_solve_Kvib}
	{K_{vib}}(T) = \frac{2{{\Theta_{vib}} \mathord{\left/{\vphantom {a b}} \right.\kern-\nulldelimiterspace} 
			{T}}}{e^{{\Theta_{vib}} \mathord{\left/{\vphantom {a b}} \right.\kern-\nulldelimiterspace} {T}}-1},
\end{equation}
where $\Theta_{vib}$ is the vibrational characteristic temperature (3371K for nitrogen, while 2256K for 
oxygen~\citep{GA_Bird_1994}).
\par
The energy density $\rho E$ is the sum of translational, rotational and vibrational energies, 
which are defined as follows:
\begin{equation}\label{f_solve_EnergyTra}
	\begin{aligned}
		\rho E_{tr} &= \frac{1}{2}{\rho {|{\bm{U}}|^2}} + \frac{{K_{tr}}}{2}\rho RT_{tr}\\
		&= \int {{\frac{1}{2}{m}({{|{\bm{u}}|}^2} + {{|{\bm{\xi}}|}^2})}f({\bm{x}},{\bm{u}},{\bm{\xi 
			}},\varepsilon_{rot},\varepsilon_{vib},t)d{\bf{\Xi }}},
	\end{aligned}
\end{equation}
\begin{equation}\label{f_solve_EnergyRot}
	\begin{aligned}
		\rho E_{rot} = \frac{K_{rot}}{2}\rho RT_{rot}
		= \int {{\varepsilon_{rot}}f({\bm{x}},{\bm{u}},{\bm{\xi }},\varepsilon_{rot},\varepsilon_{vib},t)
			d{\bf{\Xi }}},
	\end{aligned}
\end{equation}
\begin{equation}\label{f_solve_EnergyVib}
	\begin{aligned}
		\rho E_{vib} = \frac{K_{vib}(T_{vib})}{2}\rho RT_{vib}
		= \int {{\varepsilon_{vib}}f({\bm{x}},{\bm{u}},{\bm{\xi }},\varepsilon_{rot},\varepsilon_{vib},t)
			d{\bf{\Xi }}}.
	\end{aligned}
\end{equation}
Here $T_{tr}$, $T_{rot}$ and $T_{vib}$ are the translational, rotational and vibrational temperatures, 
respectively. 
According to Eq.~\eqref{f_solve_Energy} and Eqs.~\eqref{f_solve_EnergyTra}$\sim$\eqref{f_solve_EnergyVib}, 
the equilibrium temperature $T$ can be expressed in terms of the translational, rotational and vibrational 
temperatures as
\begin{equation}\label{T_TtraTrotTvib}
	T = \frac{{{K_{tr}}{T_{tr}} + {K_{rot}}{T_{rot}} + {K_{vib}}\left( {{T_{vib}}} \right){T_{vib}}}}{{{K_{tr}} 
			+ {K_{rot}} + {K_{vib}}\left( T \right)}}.
\end{equation}
Similarly, the joint translational-rotational temperature $T_{2}$ is defined as:
\begin{equation}\label{T2_TtraTrot}
	{T_2} = \frac{{{K_{tr}}{T_{tr}} + {K_{rot}}{T_{rot}}}}{{{K_{tr}} + {K_{rot}}}}.
\end{equation}
The corresponding equilibrium pressures $p$ and the pressure of translational motion $p_{tr}$ can 
be defined as:
\begin{equation}\label{p_T}
	{p} = {\rho RT},{\qquad}{p_{tr}} = \rho R{T_{tr}}.
\end{equation}
The heat flux $\bm{q}$ is the sum of the translational heat flux $\bm{q}_{tr}$, the rotational heat flux $\bm{q}_{rot}$
and the vibrational heat flux $\bm{q}_{vib}$, which are defined as:
\begin{equation}\label{f_solve_qtra}
	{{\bm{q}}_{tr}} = \frac{1}{2}\int {{\bm{c}}\left( {{{\left| {\bm{c}} \right|}^2} 
			+ {{\left| {\bm{\xi }} \right|}^2}} 
		\right)mf({\bm{x}},{\bm{u}},{\bm{\xi }},\varepsilon_{rot},\varepsilon_{vib},t)
		d{\bf{\Xi }}},
\end{equation}
\begin{equation}\label{f_solve_qrot}
	{{\bm{q}}_{rot}} = \int {{\bm{c}}{\varepsilon_{rot}}f({\bm{x}},{\bm{u}},{\bm{\xi }},
		\varepsilon_{rot},\varepsilon_{vib},t)
		d{\bf{\Xi }}},
\end{equation}
\begin{equation}\label{f_solve_qvib}
	{{\bm{q}}_{vib}} = \int {{\bm{c}}{\varepsilon_{vib}}f({\bm{x}},{\bm{u}},{\bm{\xi}},
		\varepsilon_{rot},\varepsilon_{vib},t)d{\bf{\Xi }}},
\end{equation}
where $\bm{c}=\bm{u}-\bm{U}$ is the peculiar velocity. The stress tension ${{\bf P}}$ is defined as:
\begin{equation}\label{f_solve_stress}
	{{\bf P}} = \int {{\bm c}{\bm c}mf({\bm{x}},{\bm{u}},{\bm{\xi }},\varepsilon_{rot},\varepsilon_{vib},t)
		d{\bf{\Xi }}}.
\end{equation}
\par

\subsection{Gas-kinetic model}\label{Kinetic_Model}
\par
In the absence of an external force, the kinetic model equation in $D$-dimensional space can be expressed as:
\begin{equation}\label{model_equ_f}
	\frac{{\partial f}}{{\partial t}} + {\bm{u}} \cdot \frac{{\partial f}}{{\partial {\bm{x}}}} 
	= \frac{{{f^{tr}} - f}}{\tau } 
	+ \frac{{{f^{rot}} - {f^{tr}}}}{{{Z_{rot}}\tau }} + \frac{{{f^{vib}} - {f^{tr}}}}{{{Z_{vib}}\tau }}
	= \frac{{{f^*} - f}}{\tau } : = \Omega \left( {{f^*},f} \right),
\end{equation}
where the BGK-type collision operator $ \Omega ({f^*},f)$ on the right of Eq.~\eqref{model_equ_f} describe the 
elastic collision (translational-translational relaxation) and inelastic collision (translational-rotational 
relaxation and translational-rotational-vibrational relaxation). The equilibrium distribution function $f^*$ 
is defined as:
\begin{equation}\label{define_f*}
	{f^*} = \left( {1 - \frac{1}{{{Z_{rot}}}} - \frac{1}{{{Z_{vib}}}}} \right){f^{tr}} 
	+ \frac{1}{{{Z_{rot}}}}{f^{rot}} + \frac{1}{{{Z_{vib}}}}{f^{vib}}.
\end{equation}
The equilibrium distribution functions ${f^{tr}}$, ${f^{rot}}$, and ${f^{vib}}$ are constructed by using a 
multi-dimensional Hermitian expansion around the Maxwellian equilibrium state. 
The coefficients of the Hermitian series are chosen by making the collision operator fulfill the mass, 
momentum, and energy conservation laws, meanwhile aiming to obtain the right relaxation rate of heat flux. 
The distribution functions ${f^{tr}}$, ${f^{rot}}$, and ${f^{vib}}$ are expressed as:
\begin{equation}\label{ftra}
	\begin{aligned}
		{f^{tr}} &= n{\left( {\frac{1}{{2\pi R{T_{tr}}}}} \right)^{\frac{{D + L}}{2}}}\exp 
		\left({-\frac{{{{\left| {\bm{c}}\right|}^2}+{{\left|{\bm{\xi}} \right|}^2}}}{{2R{T_{tr}}}}}\right)
		\frac{1}{{mR{T_{rot}}}}\exp \left( { - \frac{{{\varepsilon _{rot}}}}{{mR{T_{rot}}}}} \right)\Re 
		\left( {{T_{vib}}} \right) \\
		&\times \left\{ {1 + \frac{{{\bm{c}} \cdot {{\bm{q}}_{tr}}}}{{15R{T_{tr}}{p_{tr}}}}
			\left[ {\frac{{\left( {{{\left| {\bm{c}} \right|}^2} + {{\left| {\bm{\xi }} \right|}^2}} 
						\right)}}{{R{T_{tr}}}} - 5} \right] + \left( {1 - \delta } \right)\frac{{{\bm{c}} \cdot 
					{{\bm{q}}_{rot}}}}{{R{T_{tr}}{p_{rot}}}}\left( {\frac{{{\varepsilon _{rot}}}}{{mR{T_{rot}}}} - 1} 
			\right)} \right\},
	\end{aligned}
\end{equation}
\begin{equation}\label{frot}
	\begin{aligned}
		{f^{rot}} &= n{\left( {\frac{1}{{2\pi R{T_2}}}} \right)^{\frac{{D + L}}{2}}}\exp \left( { - \frac{{{{\left| 
			{\bm{c}} \right|}^2} + {{\left| {\bm{\xi }} \right|}^2}}}{{2R{T_2}}}} \right)\frac{1}{{mR{T_2}}}\exp 
		\left( { - \frac{{{\varepsilon _{rot}}}}{{mR{T_2}}}} \right)\Re \left( {{T_{vib}}} \right)\\
		&\times \left\{ {1 + {\omega _0}\frac{{{\bm{c}} \cdot {{\bm{q}}_{tr}}}}{{15R{T_2}{p_2}}}
			\left[ {\frac{{\left( {{{\left| {\bm{c}} \right|}^2}+{{\left|{\bm{\xi}}\right|}^2}}\right)}}{{R{T_2}}}
				- 5} \right] + {\omega _1}\left( {1 - \delta } \right)\frac{{{\bm{c}} \cdot 
			{{\bm{q}}_{rot}}}}{{R{T_2}{p_2}}}\left( {\frac{{{\varepsilon _{rot}}}}{{mR{T_2}}} - 1} \right)} \right\},
	\end{aligned}
\end{equation}
\begin{equation}\label{fvib}
	\begin{aligned}
		{f^{vib}} &= n{\left( {\frac{1}{{2\pi RT}}} \right)^{\frac{{D + L}}{2}}}\exp \left( 
		{ - \frac{{{{\left| {\bm{c}} \right|}^2} + {{\left| {\bm{\xi }} \right|}^2}}}{{2RT}}} \right)
		\frac{1}{{mRT}}\exp \left( { - \frac{{{\varepsilon _{rot}}}}{{mRT}}} \right)\Re \left( T \right)\\
		&\times \left\{ {1 + {\omega _2}\frac{{{\bm{c}} \cdot {{\bm{q}}_{tr}}}}{{15RTp}}
			\left[ {\frac{{\left( {{{\left| {\bm{c}} \right|}^2} + {{\left| {\bm{\xi }} \right|}^2}} \right)}}{{RT}}
				- 5} \right] + {\omega _3}\left( {1 - \delta } \right)\frac{{{\bm{c}} \cdot
					{{\bm{q}}_{rot}}}}{{RTp}}\left( {\frac{{{\varepsilon _{rot}}}}{{mRT}} - 1} \right)} \right\},
	\end{aligned}
\end{equation}
with
\begin{equation}\label{fvib_part_vib}
	\Re \left( T \right) = \frac{{{{\varepsilon_{vib}} ^{{{{K_{vib}}\left( T \right)} \mathord{\left/
				{\vphantom {{{K_{vib}}\left( T \right)} 2}} \right.
				\kern-\nulldelimiterspace} 2} - 1}}{{\left( {mRT} \right)}^{ - {{{K_{vib}}\left( T \right)} 
			\mathord{\left/{\vphantom {{{K_{vib}}\left( T \right)} 2}} \right.
				\kern-\nulldelimiterspace} 2}}}}}{{\Gamma \left( {{{{K_{vib}}\left( T \right)} \mathord{\left/
			{\vphantom {{{K_{vib}}\left( T \right)} 2}} \right.
			\kern-\nulldelimiterspace} 2}} \right)}}\exp \left\{ { - \frac{\varepsilon _{vib}}{{mRT}}} \right\},
\end{equation}
where $\Gamma$ is the gamma function, and $n$ is the molecular number density.
\par
In Eq.~\eqref{model_equ_f}, $\tau$ is the characteristic relaxation time determined by the dynamic 
viscosity $\mu$ and translational pressure $p_{tr}$ with $\tau = {\mu \mathord{\left/{\vphantom {\mu  
{{p_{tr}}}}} \right.\kern-\nulldelimiterspace} {{p_{tr}}}}$. The dynamic viscosity $\mu$ is related to the 
inter-molecular interactions. For variable hard-sphere (VHS) molecules, the dynamic viscosity is
\begin{equation}\label{T_solve_mu}
	\mu  = {\mu _{ref}}{\left( {\frac{{{T_{tr}}}}{{{T_{ref}}}}} \right)^\omega },
\end{equation}
where $\omega$ is the viscosity index, which is 0.74 for nitrogen and 0.77 for oxygen~\citep{GA_Bird_1994}. 
$\mu_{ref}$ is the reference viscosity at the reference temperature $T_{ref}$. The relationship between the 
mean free path $\lambda$ and the dynamic viscosity $\mu$ for VHS molecules is
\begin{equation}\label{mfp_mu_T}
	\lambda = \frac{{2\mu \left( {7 - 2\omega } \right)\left( {5 - 2\omega } \right)}}
	{{15\rho {{\left( {2\pi RT} \right)}^{{1 
						\mathord{\left/{\vphantom {1 2}} \right.\kern-\nulldelimiterspace} 2}}}}}.
\end{equation}
By using dimensionless parameters, such as Knudsen number, Mach number and Reynolds (Re) number, we can 
obtain:
\begin{equation}\label{Kn_Ma_Re}
	{\mathop{\rm Kn}\nolimits}  = \sqrt {\frac{{2\gamma }}{\pi }} \frac{{\left( {7 - 2\omega } \right)
			\left( {5 - 2\omega } \right)}}{{15}}\frac{{{\mathop{\rm Ma}\nolimits} }}{{{\mathop{\rm Re}\nolimits} }},
\end{equation}
where the definition of Knudsen number, Mach number and Reynolds number are ${\rm Kn}=\lambda / L_{c}$, 
${\rm Ma}={|{\bm{U}}|}/{\sqrt {\gamma RT} }$ and ${\rm Re}={\rho |{\bm{U}}|L_{c}}/{\mu}$, respectively, 
and ${L_c}$ is the characteristic scale of the flow.
\par
The $Z_{rot}$ and $Z_{vib}$ are rotational and vibrational collision numbers, respectively. 
The rotational collision number $Z_{rot}$ can be introduced from the 
DSMC~\citep{Christos_Tantos_2016} as follows:
\begin{equation}\label{Solve_Zrot}
	{Z_{rot}} = \frac{{\Pr }}{{30}}\left( {7 - 2\omega } \right)\left( {5 - 2\omega } \right)Z_{rot}^{D},
\end{equation}
where $Z_{rot}^D$ is~\citep{JunLinWu_2021}
\begin{equation}\label{Zrot_DSMC}
	Z_{rot}^D = \frac{{{K_{tr}}}}{{{K_{tr}} + {K_{rot}}}}{Z_R},
\end{equation}
The value of ${Z_{R}}$ can be determined by approximating the theoretical formulas and comparing with the 
experimental data~\citep{JG_Parker_1959,ID_Boyd_1990}. 
In the present work, the Parker~\citep{JG_Parker_1959} formula Eq.~\eqref{ZR_Parker} with ${Z_{R}^\infty}=15.7$ 
and ${T^*}=80.0K$ is adopted.
\begin{equation}\label{ZR_Parker}
	{Z_R} = \frac{{Z_R^\infty }}{{1 + \left( {{{{\pi ^{{3 \mathord{\left/
	{\vphantom {3 2}} \right.\kern-\nulldelimiterspace} 2}}}} \mathord{\left/
			{\vphantom {{{\pi ^{{3 \mathord{\left/{\vphantom {3 2}} \right.\kern-\nulldelimiterspace} 2}}}} 2}} \right.
	\kern-\nulldelimiterspace} 2}} \right)\sqrt {{{{T^*}} \mathord{\left/
		{\vphantom {{{T^*}} {{T_{tr}}}}} \right.\kern-\nulldelimiterspace} {{T_{tr}}}}}  + \left( {{{{\pi ^2}} 
		\mathord{\left/{\vphantom {{{\pi ^2}} 4}} \right.\kern-\nulldelimiterspace} 4} + \pi } \right)\left( {{{{T^*}} 
		\mathord{\left/{\vphantom {{{T^*}} {{T_{tr}}}}} \right.\kern-\nulldelimiterspace} {{T_{tr}}}}} \right)}},
\end{equation}
\par
In Eqs.~\eqref{ftra}$\sim$\eqref{fvib}, the parameter $\delta$ depends on the inter-molecular potential, 
and $\delta={1/1.55}$ when the viscosity index 
is close to unity. The values of the parameters $\omega _0$, $\omega _1$, $\omega _2$ and $\omega _3$
are chosen to achieve proper relaxation of the translational and rotational heat fluxes, which are set to
$\omega _0 = \omega _1 = \omega _3 =0.2354$ and $\omega _2 = 0.3049$ in the present work.
In the spatial homogeneous case, the relaxation rate of the heat fluxes can be determined according to 
Eq.~\eqref{model_equ_f}. Multiplying the Eq.~\eqref{model_equ_f} (ignore the convection term) by vector 
$(\frac{1}{2}{\bm{c}}({{|{\bm{c}}|}^2}+{{|{\bm{\xi}}|}^2}),{\bm{c}}\varepsilon_{rot},
{\bm{c}}\varepsilon_{vib})^T$ and integrating the resulting equations with respect to 
$d{\bf{\Xi }}$, one can obtain:
\begin{equation}\label{heat_flux_tra_relaxation}
	\frac{{\partial {{\bm{q}}_{tr}}}}{{\partial t}} =  - \left[ {\frac{2}{3}  + \frac{1}{{3{Z_{rot}}}}\left( 
		{1 - {\omega _0}} \right)+ \frac{1}{{3{Z_{vib}}}}\left( {1 - {\omega _2}} \right)} \right]
	\frac{1}{{{\tau}}}{{\bm{q}}_{tr}},
\end{equation}
\begin{equation}\label{heat_flux_rot_relaxation}
	\frac{{\partial {{\bm{q}}_{rot}}}}{{\partial t}} =  - \left[ {\delta  + \frac{1}{{{Z_{rot}}}}
		\left( {1 - {\omega _1}} \right)\left( {1 - \delta } \right) + \frac{1}{{{Z_{vib}}}}
		\left( {1 - {\omega _3}} \right)\left( {1 - \delta } \right)} 
	\right]\frac{1}{\tau }{{\bm{q}}_{rot}},
\end{equation}
\begin{equation}\label{heat_flux_vib_relaxation}
	\frac{{\partial {{\bm{q}}_{vib}}}}{{\partial t}} =  - \frac{1}{\tau }{{\bm{q}}_{vib}}.
\end{equation}
It is obvious that the values of the rotational and vibrational collision numbers $Z_{rot}$, $Z_{vib}$ and 
the parameters $\omega _0$, $\omega _1$, $\omega _2$ and $\omega _3$ affect
the relaxation rate of the translational and rotational heat fluxes.
\par
According to the moments of distribution function $f$ and 
the definition of the equilibrium distribution functions ${f^{tr}}$, ${f^{rot}}$, and ${f^{vib}}$, 
it is easy to verify that the collision operator satisfy the following formulas:
\begin{equation}\label{coll_f_int_mass}
	\int {m\Omega \left( {{f^*},f} \right)d{\bf{\Xi }}}  = 0,
\end{equation}
\begin{equation}\label{coll_f_int_momentum}
	\int {{\bm{u}}m\Omega \left({{f^*},f}\right)d{\bf{\Xi }}} 
	= {\bm{0}},
\end{equation}
\begin{equation}\label{coll_f_int_energy}
	\int {\left( {m\frac{{{{\left| {\bm{u}} \right|}^2} + {{\left| {\bm{\xi}} \right|}^2}}}{2} + 
			\varepsilon_{rot} + \varepsilon_{vib}} \right)\Omega \left( {{f^*},f} \right)
	d{\bf{\Xi }}}  = 0,
\end{equation}
\begin{equation}\label{coll_f_int_energy_rot}
	\int {{\varepsilon_{rot}}\Omega \left( {{f^*},f} \right)d{\bf{\Xi }}}  
	= \frac{1}{{{Z_{rot}}\tau }}\left( {\rho R{T_2}-{\rho E_{rot}}}\right) 
	+ \frac{1}{{{Z_{vib}}\tau }}\left( {\rho RT - {\rho E_{rot}}} \right),
\end{equation}
\begin{equation}\label{coll_f_int_energy_vib}
	\int {\varepsilon_{vib} \Omega \left( {{f^*},f} \right)d{\bf{\Xi }}}  = \frac{1}{{{Z_{vib}}\tau }}\left( 
	{\frac{{{K_{vib}}\left( T \right)}}{2}\rho RT - {\rho E_{vib}}} \right).
\end{equation}
It is shown from the above five equations that the conservative properties of mass, momentum and
energy are satisfied by the present model equation. However, it should be noticed that the 
rotational and vibrational energies are not conservative due to the energy conversion between the 
translational, rotational, and vibrational energies.
\par

\subsection{Reduced distribution function}\label{Reduced_Distribution_Function}
\par
The transport process of the distribution function depends only on the $D$-dimensional particle velocity 
$\bm {u}$ and is irrelevant to $\bm {\xi}$, $\varepsilon_{rot}$ and $\varepsilon_{vib}$. In order to save 
computational memory and cost, the reduced distribution functions $G({\bm{x}},{\bm {u}},t)$, 
$H({\bm{x}},{\bm {u}},t)$, $R({\bm{x}},{\bm {u}},t)$, $B({\bm{x}},{\bm {u}},t)$
are introduced~\citep{CK_Chu_1965} in the numerical computations,
\begin{equation}\label{define_phi}
	\left( {\begin{array}{*{20}{c}}
			G \\
			H \\
			R \\
			B
	\end{array}} \right) 
	= \int {\bm {\vartheta} f({\bm{x}},{\bm{u}},{\bm{\xi}},
		\varepsilon_{rot},\varepsilon_{vib},t)d{\bm{\xi}}d\varepsilon_{rot}d\varepsilon_{vib}},
\end{equation}
where the vector $\bm {\vartheta} ={\left({m,m{{\left|{\bm{\xi}}\right|}^2},\varepsilon_{rot},
		\varepsilon_{vib}}\right)^T}$. 
The macroscopic flow variable $\bm {Q}=(\rho,{\rho {\bm{U}}},\rho E,\rho E_{rot},\rho E_{vib})^T$
can be solved by the moments of the reduced distribution functions as follows:
\begin{equation}\label{phi_solve_mac}
	\bm {Q} = \int {{\bf {\Phi}}\left( {\bm{u}} \right) \cdot \left( {\begin{array}{*{20}{c}}
				G \\
				H \\
				R \\
				B
		\end{array}} \right) d{\bm{u}}},
\end{equation}
with
\begin{equation}\label{PHI_matrix}
	{\bf {\Phi}}\left( {\bm{u}} \right) = \left[ {\begin{array}{*{20}{c}}
			1&0&0&0\\
			{\bm{u}}&0&0&0\\
			{\frac{1}{2}{{\left| {\bm{u}} \right|}^2}}&{\frac{1}{2}}&1&1\\
			0&0&1&0\\
			0&0&0&1
	\end{array}} \right].
\end{equation}
The translational, rotational, and vibrational heat fluxes ${{\bm{q}}_{tr}}$, ${{\bm{q}}_{rot}}$, 
${{\bm{q}}_{vib}}$ and the 
stress tension ${\bf P}$ are calculated by:
\begin{equation}\label{phi_solve_qtra}
	{{\bm{q}}_{tr}} = \int {\frac{1}{2}{\bm{c}}\left( {{{\left| {\bm{c}} \right|}^2}G + H} \right)d{\bm{u}}},
\end{equation}
\begin{equation}\label{phi_solve_qrot}
	{{\bm{q}}_{rot}} = \int {{\bm{c}}Rd{\bm{u}}},
\end{equation}
\begin{equation}\label{phi_solve_qvib}
	{{\bm{q}}_{vib}} = \int {{\bm{c}}Bd{\bm{u}}},
\end{equation}
\begin{equation}\label{phi_solve_stress}
	{{\bf P}} = \int {\bm {c} \bm {c}Gd{\bm{u}}}.
\end{equation}
Multiplying Eq.~\eqref{model_equ_f} by vector $\bm {\vartheta}$ and integrating the resulting equations 
same as Eq.~\eqref{define_phi}, the model equation~\eqref{model_equ_f} can be transformed into the 
following evolution equation for the reduced distribution functions,
\begin{equation}\label{model_reduced_equ_phi}
	\frac{{\partial \phi }}{{\partial t}} + {\bm{u}} \cdot \frac{{\partial \phi }}{{\partial {\bm{x}}}} 
	= \frac{{{\phi ^*} - \phi }}{\tau }: = \Omega \left( {{\phi ^*},\phi } \right),	
\end{equation}
where ${\phi} = G, H, R$ or $B$, and the reduced equilibrium distribution functions 
${\phi}^* = {G^*}, {H^*}, {R^*}$ or ${B^*}$ are given by:
\begin{equation}\label{eqdf_G*}
	{G^*} = \left( {1 - \frac{1}{{{Z_{rot}}}} - \frac{1}{{{Z_{vib}}}}} \right){G^{tr}} + 
	\frac{1}{{{Z_{rot}}}}{G^{rot}} + \frac{1}{{{Z_{vib}}}}{G^{vib}},
\end{equation}
\begin{equation}\label{eqdf_H*}
	{H^*} = \left( {1 - \frac{1}{{{Z_{rot}}}} - \frac{1}{{{Z_{vib}}}}} \right){H^{tr}} + 
	\frac{1}{{{Z_{rot}}}}{H^{rot}} + \frac{1}{{{Z_{vib}}}}{H^{vib}},
\end{equation}
\begin{equation}\label{eqdf_R*}
	{R^*} = \left( {1 - \frac{1}{{{Z_{rot}}}} - \frac{1}{{{Z_{vib}}}}} \right){R^{tr}} + 
	\frac{1}{{{Z_{rot}}}}{R^{rot}} + \frac{1}{{{Z_{vib}}}}{R^{vib}},
\end{equation}
\begin{equation}\label{eqdf_B*}
	{B^*} = \left( {1 - \frac{1}{{{Z_{rot}}}} - \frac{1}{{{Z_{vib}}}}} \right){B^{tr}} + 
	\frac{1}{{{Z_{rot}}}}{B^{rot}} + \frac{1}{{{Z_{vib}}}}{B^{vib}},
\end{equation}
with
\begin{equation}\label{eqdf_Gtra}
	{G^{tr}}({\bm{x}},{\bm{u}},t) = {g^{eq}}\left( {{T_{tr}}} \right)\left[ {1 + \frac{{{\bm{c}} \cdot 
				{{\bm{q}}_{tr}}}}{{15R{T_{tr}}{p_{tr}}}}\left( {\frac{{{{\left| {\bm{c}} \right|}^2}}}{{R{T_{tr}}}}
			- D - 2} \right)} \right],
\end{equation}
\begin{equation}\label{eqdf_Grot}
	{G^{rot}}({\bm{x}},{\bm{u}},t) = {g^{eq}}\left( {{T_2}} \right)\left[ {1 + {\omega _0}\frac{{{\bm{c}} \cdot 
				{{\bm{q}}_{tr}}}}{{15R{T_2}{p_2}}}\left( {\frac{{{{\left| {\bm{c}} \right|}^2}}}{{R{T_2}}} - D - 2}
		\right)} \right],
\end{equation}
\begin{equation}\label{eqdf_Gvib}
	{G^{vib}}({\bm{x}},{\bm{u}},t) = {g^{eq}}\left( T \right)\left[ {1 + {\omega _2}\frac{{{\bm{c}} \cdot 
		{{\bm{q}}_{tr}}}}{{15RTp}}\left( {\frac{{{{\left| {\bm{c}} \right|}^2}}}{{RT}} - D - 2} \right)} \right],
\end{equation}
\begin{equation}\label{eqdf_Htra}
	{H^{tr}}({\bm{x}},{\bm{u}},t) = {g^{eq}}\left( {{T_{tr}}} \right)R{T_{tr}}\left( {3 - D} \right)\left[ {1 + 
		\frac{{{\bm{c}} \cdot {{\bm{q}}_{tr}}}}{{15R{T_{tr}}{p_{tr}}}}\left( {\frac{{{{\left| {\bm{c}} 
							\right|}^2}}}{{R{T_{tr}}}} - D} \right)} \right],
\end{equation}
\begin{equation}\label{eqdf_Hrot}
	{H^{rot}}({\bm{x}},{\bm{u}},t) = {g^{eq}}\left( {{T_2}} \right)R{T_2}\left( {3 - D} \right)\left[ {1 + {\omega 
		_0}\frac{{{\bm{c}} \cdot {{\bm{q}}_{tr}}}}{{15R{T_2}{p_2}}}\left( {\frac{{{{\left| {\bm{c}} \right|}^2}}}
			{{R{T_2}}} - D} \right)} \right],
\end{equation}
\begin{equation}\label{eqdf_Hvib}
	{H^{vib}}({\bm{x}},{\bm{u}},t) = {g^{eq}}\left( T \right)RT\left( {3 - D} \right)\left[ {1 + {\omega _2}
		\frac{{{\bm{c}} \cdot {{\bm{q}}_{tr}}}}{{15RTp}}\left( {\frac{{{{\left| {\bm{c}} \right|}^2}}}{{RT}} - D}
		\right)} \right],
\end{equation}
\begin{equation}\label{eqdf_Rtra}
	{R^{tr}}({\bm{x}},{\bm{u}},t) = R{T_{rot}}\left[ {{G^{tr}} + \left( {1 - \delta } \right)\frac{{{\bm{c}} 
				\cdot {{\bm{q}}_{rot}}}}{{R{T_{tr}}{p_{rot}}}}{g^{eq}}\left( {{T_{tr}}} \right)} \right],
\end{equation}
\begin{equation}\label{eqdf_Rrot}
	{R^{rot}}({\bm{x}},{\bm{u}},t) = R{T_2}\left[ {{G^{rot}} + {\omega _1}\left( {1 - \delta } \right)
		\frac{{{\bm{c}}\cdot {{\bm{q}}_{rot}}}}{{R{T_2}{p_2}}}{g^{eq}}\left( {{T_2}} \right)} \right],
\end{equation}
\begin{equation}\label{eqdf_Rvib}
	{R^{vib}}({\bm{x}},{\bm{u}},t) = RT\left[ {{G^{vib}} + {\omega _3}\left( {1 - \delta } \right)\frac{{{\bm{c}} 
				\cdot {{\bm{q}}_{rot}}}}{{RTp}}{g^{eq}}\left( T \right)} \right],
\end{equation}
\begin{equation}\label{eqdf_Btra}
	{B^{tr}}({\bm{x}},{\bm{u}},t) = \frac{{{K_{vib}}\left( {{T_{vib}}} \right)}}{2}R{T_{vib}}{G^{tr}},
\end{equation}
\begin{equation}\label{eqdf_Brot}
	{B^{rot}}({\bm{x}},{\bm{u}},t) = \frac{{{K_{vib}}\left( {{T_{vib}}} \right)}}{2}R{T_{vib}}{G^{rot}},
\end{equation}
\begin{equation}\label{eqdf_Bvib}
	{B^{vib}}({\bm{x}},{\bm{u}},t) = \frac{{{K_{vib}}\left( T \right)}}{2}RT{G^{vib}}.
\end{equation}
In the above equations, the $g^{eq}$ is the Maxwellian equilibrium distribution function,
\begin{equation}\label{eqdf_Maxwellian}
	{g^{eq}}\left(\bm {u}; {\rho ,{\bm{U}},T} \right) = \rho {\left( {\frac{1}{{2\pi RT}}} \right)^{{D 
				\mathord{\left/
		{\vphantom {D 2}} \right.\kern-\nulldelimiterspace} 2}}}\exp \left( { - \frac{{{{\left| {{\bm{u}} - 
		{\bm{U}}} \right|}^2}}}{{2RT}}} \right).
\end{equation}
\par
According to the above proof of conservation property and the definition of the reduced  
distribution functions, it is easy to verify that the reduced collision operators satisfy the 
following formulas:
\begin{equation}\label{co_int_mass}
	\int {\Omega \left( {{G^*},G} \right)} d{\bm{u}} = 0,
\end{equation}
\begin{equation}\label{co_int_momentum}
	\int {{\bm{u}}\Omega \left( {{G^*},G} \right)} d{\bm{u}} = {\bm{0}},
\end{equation}
\begin{equation}\label{co_int_energy}
	\int {\left\{ {\frac{1}{2}\left[ {{{\left| {\bm{u}} \right|}^2}\Omega \left( {{G^*},G} \right) + \Omega 
				\left( {{H^*},H} \right)} \right] + \Omega \left( {{R^*},R} \right) + 
			\Omega \left( {{B^*},B} \right)} \right\}d{\bm{u}}}  = 0.
\end{equation}
\begin{equation}\label{co_int_energy_rot}
	\int {\Omega \left( {{R^*},R} \right)d{\bm{u}}}  = \frac{1}{{{Z_{rot}}\tau }}\left( {\rho R{T_2} - 
		{\rho E_{rot}}} \right) + \frac{1}{{{Z_{vib}}\tau }}\left( {\rho RT - {\rho E_{rot}}} \right),
\end{equation}
\begin{equation}\label{co_int_energy_vib}
	\int {\Omega \left( {{B^*},B} \right)d{\bm{u}}}  = \frac{1}{{{Z_{vib}}\tau }}\left( {\frac{{{K_{vib}}
				\left( T \right)}}{2}\rho RT - {\rho E_{vib}}} \right).
\end{equation}
It should be noted that the moments of the collision operators $\Omega ({{R^*},R})$ related to rotational 
energy and $\Omega ({{B^*},B})$ related to vibrational energy are not equal to zero.
\par

\section{Unified gas-kinetic scheme with simplified multi-scale numerical flux}\label{UGKS_Diatomic}
\subsection{Dimensionless analysis}\label{Dimensionless}
\par
In the calculations, the dimensionless quantities normalized by the reference length, density, temperature, 
and velocity are introduced as follows:
\begin{equation}\label{Ref_quantities}
	{L_{ref}} = {L_c},{\qquad}{\rho _{ref}} = {\rho _\infty },{\qquad}{T_{ref}} = {T_\infty },{\qquad}{U_{ref}} 
	= \sqrt {2R{T_{ref}}},
\end{equation}
where $L_{c}$ is the characteristic length scale of the flow, ${\rho_\infty}$, ${T_\infty}$ are the density and 
temperature of the free-stream, respectively. Then the following dimensionless quantities can be obtained:
\begin{equation}\label{Dimensionless_quantities_basic}
	\hat L = \frac{L}{{{L_{ref}}}},{\qquad}\hat \rho  = \frac{\rho }{{{\rho _{ref}}}},{\qquad}\hat T = 
	\frac{T}{{{T_{ref}}}},{\qquad}\hat U = \frac{U}{{{U_{ref}}}},{\qquad}\hat t = 
	\frac{t}{{{L_{ref}}U_{ref}^{ - 1}}},
\end{equation}
\begin{equation}\label{Dimensionless_quantities_other}
	\begin{aligned}
		&	\hat n = \frac{n}{{L_{ref}^{ - 3}}},
		{\qquad}\hat m = \frac{m}{{{\rho _{ref}}L_{ref}^3}},
		{\qquad}\hat \mu  = \frac{\mu }{{{\rho _{ref}}{U_{ref}}{L_{ref}}}},
	    {\qquad}\hat E = \frac{E}{{U_{ref}^2}},\\
		&   \hat p = \frac{p}{{{\rho _{ref}}U_{ref}^2}},
		{\qquad}{\bm{\hat q}} = \frac{{\bm{q}}}{{{\rho _{ref}}U_{ref}^3}},
		{\qquad}\hat \tau  = \frac{\tau }{{{L_{ref}}U_{ref}^{ - 1}}},
		{\qquad}\hat R = \frac{R}{{U_{ref}^2T_{ref}^{ - 1}}}.
	\end{aligned}	
\end{equation}
\par
Finally, we can obtain a complete dimensionless system. In the following, all variables without the “hat” 
are nondimensionalized for simplicity unless stated otherwise.
\par

\subsection{Discretization of the particle velocity space}\label{DVM}
\par
The macroscopic flow variables can be obtained by integrating the distribution functions in the continuous
velocity space. However, in order to capture the non-equilibrium distributions, the particle velocity space is 
discrete in the UGKS. With the particle DVS, the moments of the distribution functions can be obtained 
by numerical quadrature over the DVS,
\begin{equation}\label{phi_solve_mac_dvm}
	\bm {Q} = \left( {\begin{array}{*{20}{c}}
			\rho \\
			{\rho {\bm{U}}}\\
			\rho E\\
			{{\rho E_{rot}}}\\
			{{\rho E_{vib}}}
	\end{array}} \right) = \sum\limits_k {W_k{\bf {\Phi}}\left( {{{\bm{u}}_k}} \right) }
	\cdot {\left( {\begin{array}{*{20}{c}}
				G_k \\
				H_k \\
				R_k \\
				B_k
		\end{array}} \right)},
\end{equation}
where $G_k, H_k, R_k, B_k$ are the discrete distribution functions in DVS, ${\bf {\Phi}}({{\bm{u}}_k})$ 
is the discrete form of ${\bf {\Phi}}({\bm{u}})$ in particle velocity space. 
$W_k$ is the associated quadrature weight at the 
discrete velocity point ${\bm u}_k$, $k$ is the index of discrete velocity points.
The discrete reduced kinetic equation~\eqref{model_reduced_equ_phi} in velocity space is as follows:
\begin{equation}\label{model_equ_rd_phi}
	\frac{{\partial \phi_{k} }}{{\partial t}} + {\bm{u}_{k}} \cdot \frac{{\partial \phi_{k} }}
	{{\partial {\bm{x}}}} = 
	\frac{{{\phi ^*_{k}} - \phi_{k} }}{\tau }: = \Omega \left( {{\phi ^*_{k}},\phi_{k} } \right).
\end{equation}
\par 
The uniform particle DVS with Newton-Cotes numerical quadrature is frequently employed 
to capture the non-equilibrium distributions, while it results in a great demand for computation and storage, 
especially for three-dimensional flows. To reduce the amount of discrete velocity points and improve the 
computational efficiency, many technologies have been proposed to ease this problem, such as the adaptive 
velocity space technology~\citep{SongzeChen_2012} and the unstructured DVS~\citep{RuifengYuan_2020}. 
In the present study, the unstructured DVS with midpoint 
integration formula is adopted ($W_{k}$ is chosen as the volume in the unstructured DVS). 
Compared to the Cartesian DVS, the unstructured DVS is more flexible which can refine and coarsen 
the grid points according to the specific flows~\citep{JianfengChen_2019,YajunZhu_2020}. Besides, 
the unstructured DVS is easier to employ the velocity space decomposition approaches than the adaptive 
velocity space in parallel computing. Theoretically, 
the accuracy of this type of integration is slightly lower than that of the Newton-Cotes one on structured 
Cartesian mesh. Therefore, the integral error compensation~\citep{RuifengYuan_2020} has been 
proposed to reduce the integration error and allows more flexible discretization for particle velocity space. 
As a result, the total computational cost on unstructured DVS is often less than ten to thirty percent of the 
structured DVS with high order numerical quadrature in two dimensional flows, and less than three to ten 
percent in three dimensional flows~\citep{JianfengChen_2019}. 
\par

\subsection{General framework of unified gas-kinetic scheme}\label{UGKS}
\par
The construction of UGKS for diatomic gases is based on the reduced kinetic equation~\eqref{model_equ_rd_phi} 
in the classical finite volume framework. The physical space is divided into a set of control volumes $V_{i}$.
The temporal discretization is denoted by $t_{n}$ for the n-th time step. Integrating Eq.~\eqref{model_equ_rd_phi} 
on a control volume $V_{i}$ from time $t_{n}$ to $t_{n}+\Delta t$, we can obtain:
\begin{equation}\label{cdugks_mic_dis_equ_ugks}
	\phi _{i,k}^{n + 1} - \phi _{i,k}^n + \frac{1}{{\left| {{V_i}} \right|}}\int_{{t_n}}^{{t_n} + \Delta t} 
	{F\left( {{\phi _{ij,k}}\left( t \right)} \right)dt}  = \frac{{\Delta t}}{2}\left( {\frac{{\phi _{i,k}^{*,n}
	 - \phi _{i,k}^n}}{{\tau _i^n}} + \frac{{\phi _{i,k}^{*,n + 1} - \phi _{i,k}^{n + 1}}}{{\tau _i^{n + 1}}}} \right),
\end{equation}
where the trapezoidal rule is used for the time integration of collision term. The $|V_{i}|$ is the cell volume 
and $\phi_{i,k}^{n}$ is the cell-averaged value of the distribution function, e.g.,
\begin{equation}\label{cdugks_cell_averaged_phi}
	\phi _{i,k}^n = \frac{1}{{\left| {{V_i}} \right|}}\int_{{V_i}} {\phi \left( {{{\bm{x}}_i},{{\bm{u}}_k},
			{t_n}} \right)} dV,
\end{equation}
and the micro-flux ${F\left( {{\phi _{ij,k}}\left( t \right)} \right)}$ across the cell interface is defined as:
\begin{equation}\label{cdugks_mic_flux_ugks}
	{F\left( {{\phi _{ij,k}}\left( t \right)} \right)} = \sum\limits_{j \in N\left( i \right)} {\left( {{{\bm{u}}_k} \cdot 
			{{\bm{n}}_{ij}}} \right){A_{ij}}\phi _{ij,k}{(t)}}.
\end{equation}
The sign $j$ denotes the neighboring cells of cell $i$ and $N(i)$ is the set of all of the neighbors 
of cell $i$. $ij$ denotes the variable at the interface between cell $i$ and $j$. 
$A_{ij}$ is the interface area, ${\bm {n}}_{ij}$ is the outward unit 
vector normal to the interface $ij$ from cell $i$ to cell $j$. 
${\Delta t}$ is the time step which can be determined by the CFL number less than one.
\par
In the evolution process of original UGKS, a time-dependent distribution function $\phi_{ij,k}(t)$ at the cell 
interface constructed from the analytic solution of kinetic model equation is used to 
calculate the time step-averaged micro-flux $F(\phi_{ij,k}(t))$~\cite{KunXu_A_2010}. In order to simplify the 
calculation of multi-scale numerical flux, the distribution function $\phi_{ij,k}({t_n}+s)$ at the discrete time 
step ${t_n}+s$ ($0 < s \leq {\Delta t}$) is used to evaluate the micro-flux in the present work. 
\begin{equation}\label{cdugks_mic_dis_equ}
	\phi _{i,k}^{n + 1} - \phi _{i,k}^n + \frac{{\Delta t}}{{\left| {{V_i}} \right|}}F(\phi_{ij,k}({t_n}+s))
	=\frac{{\Delta t}}{2}\left({\frac{{\phi _{i,k}^{*,n} 
		- \phi _{i,k}^n}}{{\tau _i^n}}+\frac{{\phi _{i,k}^{*,n+1} - \phi _{i,k}^{n+1}}}{{\tau _i^{n+1}}}} \right),
\end{equation}
and the micro-flux $F(\phi_{ij,k}({t_n}+s))$ across the cell interface is
\begin{equation}\label{cdugks_mic_flux}
	F(\phi_{ij,k}({t_n}+s)) = \sum\limits_{j \in N\left( i \right)} {\left( {{{\bm{u}}_k} \cdot 
			{{\bm{n}}_{ij}}} \right){A_{ij}}{\phi_{ij,k}({t_n}+s)}}.
\end{equation}
\par
Note that the update rule of $\phi _{i,k}^{n+1}$ given by Eq.~\eqref{cdugks_mic_dis_equ} is implicit, 
due to the unknown macroscopic flow variables at time $t_{n}+{\Delta t}$
are required for evaluation of ${\phi _{i,k}^{*,n + 1}}$ and $\tau _i ^{n+1}$. 
In order to remove this implicit requirement, macroscopic flow variables are 
also updated in UGKS~\citep{KunXu_A_2010,KunXu_Direct_2015}. 
Once the macroscopic flow variables at time $t_{n}+{\Delta t}$ are obtained, then the implicit evolution
equation~\eqref{cdugks_mic_dis_equ} can be transformed into the following explicit one:
\begin{equation}\label{cdugks_mic_update_equ}
	{\rm{ }}\phi _{i,k}^{n + 1} = {\left[ {1 + \frac{{\Delta t}}{{2\tau _i^{n + 1}}}} \right]^{ - 1}}\left[ 
	{\frac{{\Delta t}}{2}\left( {\frac{{\phi _{i,k}^{*,n} - \phi _{i,k}^n}}{{\tau _i^n}} + 
			\frac{{\phi _{i,k}^{*,n + 
						1}}}{{\tau _i^{n + 1}}}} \right) + \phi _{i,k}^n - \frac{{\Delta t}}{{\left| {{V_i}} 
				\right|}}F(\phi_{ij,k}(t_n+{s})}) \right].
\end{equation}
\par
Multiplying Eq.~\eqref{cdugks_mic_dis_equ} by ${\bf{\Phi}} \left( {\bm{u}}_k \right)$ and integrating 
the resulting equations in the particle velocity space, the macroscopic flow variables $\bm {Q}_i^{n+1}$
can be be updated as follows:
\begin{equation}\label{cdugks_mac_update_equ}
	\bm {Q}_i^{n + 1} = \bm {Q}_i^n - \frac{{\Delta t}}{{\left| {{V_i}} \right|}}\sum\limits_k {W_k} 
	{\bf{\Phi}} \left( {\bm{u}}_k \right) \cdot 	
	\left( 
	{\begin{array}{*{20}{c}}
			F(G_{ij,k}(t_n+{s})) \\[1mm]
			F(H_{ij,k}(t_n+{s})) \\[1mm]
			F(R_{ij,k}(t_n+{s})) \\[1mm]
			F(B_{ij,k}(t_n+{s}))
	\end{array}} \right)
	+ {\frac{{\Delta t}}{2}}
	\left( {\bm {S}_i^{n+1}+\bm {S}_i^n} \right),
\end{equation}
where the source term
\begin{equation}\label{cdugks_mac_define_S}
	\bm {S} = \left( 
	{\begin{array}{*{20}{c}}
			0\\
			{\bm{0}}\\
			0\\
			{\frac{1}{{{Z_{rot}}\tau }}\left( {\rho R{T_2} - {\rho E_{rot}}} \right) + 
				\frac{1}{{{Z_{vib}}\tau }}\left( {\rho RT - {\rho E_{rot}}} \right)}\\[1mm]
			{\frac{1}{{{Z_{vib}}\tau }}\left({\frac{{{K_{vib}}\left( T \right)}}{2}\rho RT
					- {\rho E_{vib}}} \right)}
	\end{array}} 
	\right).
\end{equation} 
Note that for the conserved variables $\rho$, $\rho \bm{U}$ and $\rho E$, the source terms $\bm{S}$ are zero, 
thus can be directly updated as follows:
\begin{equation}\label{cdugks_mac_update_equ_rho_U_E}
	\begin{aligned}
	&\left( {\begin{array}{*{20}{c}}
			{\rho _i^{n + 1}}\\[1mm]
			{\left( {\rho {\bm{U}}} \right)_i^{n + 1}}\\[1mm]
			{({\rho E})_i^{n + 1}}
	\end{array}} \right) 
	= \left( {\begin{array}{*{20}{c}}
			{\rho _i^n}\\[1mm]
			{\left( {\rho {\bm{U}}} \right)_i^n}\\[1mm]
			{({\rho E})_i^n}
	\end{array}} \right)\\
	&- \frac{{\Delta t}}{{\left| {{V_i}} \right|}}\sum\limits_k {{W_k}\left\{ {\sum\limits_{j \in N\left( i 
				\right)} {\left( {{{\bm{u}}_k} \cdot {{\bm{n}}_{ij}}} \right){A_{ij}}\left[ {\begin{array}{*{20}{c}}
						{G_{ij,k}(t_n+{s})} \\[1mm]
						{{{\bm{u}}_k}G_{ij,k}(t_n+{s})}\\[1mm]
		{\frac{1}{2}\left( {{{\left| {{{\bm{u}}_k}} \right|}^2}G_{ij,k}(t_n+{s}) + H_{ij,k}(t_n+{s})} \right)
				+ R_{ij,k}(t_n+{s}) + B_{ij,k}(t_n+{s})}
				\end{array}} \right]} } \right\}}.
	\end{aligned}
\end{equation} 
Given Equations~\eqref{cdugks_mic_flux},~\eqref{cdugks_mac_update_equ} and 
~\eqref{cdugks_mac_define_S}, $\rho E_{rot}$ and $\rho E_{vib}$
can be updated using the following formulas,  
\begin{equation}\label{cdugks_mac_update_equ_Erot}
	\begin{aligned}
		\left( {{\rho E_{rot}}} \right)_i^{n + 1} 
		&= {\left\{ {1 + \frac{{\Delta t}}{2}\left[ {\frac{1}{{\left( {{Z_{rot}}\tau } \right)_i^{n + 1}}} + 
			\frac{1}{{\left( {{Z_{vib}}\tau } \right)_i^{n + 1}}}} \right]} \right\}^{ - 1}}\\
		&\times \left\{ \begin{aligned}
	&\frac{{\Delta t}}{2}\left[ {\frac{{\left( {\rho R{T_2}} \right)_i^n - \left( {{\rho E_{rot}}} 
						\right)_i^n}}
		{{\left( {{Z_{rot}}\tau } \right)_i^n}} + \frac{{\left( {\rho RT} \right)_i^n - \left( {{\rho E_{rot}}} 
		\right)_i^n}}{{\left( {{Z_{vib}}\tau } \right)_i^n}} + \frac{{\left( {\rho R{T_2}} \right)_i^{n + 
		1}}}{{\left( {{Z_{rot}}\tau } \right)_i^{n + 1}}} + \frac{{\left( {\rho RT} \right)_i^{n + 1}}}{{\left( 
				{{Z_{vib}}\tau } \right)_i^{n + 1}}}} \right]\\
			&+ \left( {{\rho E_{rot}}} \right)_i^n - \frac{{\Delta t}}{{\left| {{V_i}} \right|}}\sum\limits_k 
			{{W_k}\left( 
				{\sum\limits_{j \in N\left( i \right)} {\left( {{{\bm{u}}_k} \cdot {{\bm{n}}_{ij}}} 
						\right){A_{ij}}R_{ij,k}(t_n+{s})} } \right)} 
		\end{aligned} \right\}
	\end{aligned},
\end{equation}
\begin{equation}\label{cdugks_mac_update_equ_Evib}
	\begin{aligned}
		({\rho E_{vib}})_i^{n + 1} 
		&= {\left[ {1 + \frac{{\Delta t}}{{2\left( {{Z_{vib}}\tau } \right)_i^{n + 
							1}}}} \right]^{ - 1}}\\
		&\times \left\{ \begin{aligned}
			&\frac{{\Delta t}}{2}\left[ {\frac{{\left( {0.5{K_{vib}}\left( T \right)\rho RT} \right)_i^n 
						- \left( {{\rho E_{vib}}} \right)_i^n}}{{\left( {{Z_{vib}}\tau } \right)_i^n}} + 
				\frac{{\left( {0.5{K_{vib}}\left( T \right)\rho RT} 
						\right)_i^{n + 1}}}{{\left( {{Z_{vib}}\tau } \right)_i^{n + 1}}}} \right]\\
			&+ \left( {{\rho E_{vib}}} \right)_i^n - \frac{{\Delta t}}{{\left| {{V_i}} \right|}}\sum\limits_k
			{{W_k}\left( 
				{\sum\limits_{j \in N\left( i \right)} {\left( {{{\bm{u}}_k} \cdot {{\bm{n}}_{ij}}} \right)
						{A_{ij}}B_{ij,k}(t_n+{s})} } \right)} 
		\end{aligned} \right\}
	\end{aligned}.
\end{equation}
\par
Here, the calculation procedure and details of updating macroscopic flow variables 
(Eqs.~\eqref{cdugks_mac_update_equ_rho_U_E},~\eqref{cdugks_mac_update_equ_Erot} 
and~\eqref{cdugks_mac_update_equ_Evib}) are summarized as follows:\\
\textbf{(a)} Solve Eq.~\eqref{cdugks_mac_update_equ_rho_U_E} to update the conserved variables 
$\rho _i^{n+1}$, $({\rho \bm{U}}) _i^{n+1}$ and $({\rho E})_i^{n+1}$.\\
\textbf{(b)} In order to update the rotational energy $(\rho E_{rot})_i^{n+1}$ 
(Eq.~\eqref{cdugks_mac_update_equ_Erot})
and the vibrational energy $(\rho E_{vib})_i^{n+1}$ (Eq.~\eqref{cdugks_mac_update_equ_Evib}), 
the temperature $T_i^{n+1}$ and $(T_{2})_i^{n+1}$ must be solved firstly. 
Given Eq.~\eqref{f_solve_Energy}, one can obtain:
\begin{equation}\label{cdugks_mac_update_equ_T}
	T_{i}^{n+1} = \frac{{\left( {2({\rho E})_{i}^{n+1} - \rho _{i}^{n+1}{{\left| {{\bm{U}}_{i}^{n+i}} 
						\right|}^2}}
			\right)}}{{\rho _{i}^{n + 1}R\left( {5 + {K_{vib}}\left( {T_{i}^{n + 1}} \right)} \right)}},
\end{equation}
where the vibrational degrees of freedom ${K_{vib}}(T_i^{n+1})$ (Eq.~\eqref{T_solve_Kvib}) is dependent on 
the temperature $T_i^{n+1}$ in equation ~\eqref{cdugks_mac_update_equ_T}. 
Thus, the temperature $T_i^{n+1}$ is solved by the iteration method as:
\begin{equation}\label{cdugks_mac_update_equ_solve_T}
	\begin{aligned}
		T_i^{n + 1,0} &= \frac{{\left( {2({\rho E})_i^{n + 1} - \rho _i^{n + 1}{{\left| {{\bm{U}}_i^{n + i}} 
		\right|}^2}} \right)}}{{\rho _i^{n + 1}R\left( {5 + {K_{vib}}\left( {T_i^n} \right)} \right)}},\\
		T_i^{n + 1,m} &= \frac{{\left( {2({\rho E})_i^{n + 1} - \rho _i^{n + 1}{{\left| {{\bm{U}}_i^{n + i}} 
				\right|}^2}} \right)}}{{\rho _i^{n + 1}R\left( {5 + \frac{{{{2{\Theta _{vib}}} \mathord{\left/
			{\vphantom {{2{\Theta _{vib}}} {T_i^{n + 1,m - 1}}}} \right.
		\kern-\nulldelimiterspace} {T_i^{n + 1,m - 1}}}}}{{\exp \left( {{{{\Theta _{vib}}} \mathord{\left/
		{\vphantom {{{\Theta _{vib}}} {T_i^{n + 1,m - 1}}}} \right.
		\kern-\nulldelimiterspace} {T_i^{n + 1,m - 1}}}} \right) - 1}}} \right)}},
	\end{aligned}
\end{equation}
where $m$ is the iterative steps and 10 times’ iterations for solving $T_i^{n+1}$ is enough.
Once the temperature $T_i^{n+1}$ is obtained, the vibrational degrees of freedom ${K_{vib}}(T_i^{n+1})$ 
at the equilibrium state can be solved using Eq.~\eqref{T_solve_Kvib}.\\ 
\textbf{(c)} Solve Eq.~\eqref{cdugks_mac_update_equ_Evib} to update the vibrational energy $(\rho E_{vib})_i^{n+1}$.\\
\textbf{(d)} Calculate the vibrational degrees of freedom ${K_{vib}}((T_{vib})_i^{n+1})$ and 
the vibrational temperature $(T_{vib})_i^{n+1}$ as:\\
\begin{equation}\label{cdugks_mac_update_equ_Kvib_Tvib}
	\begin{aligned}
		&{K_{vib}}\left( {\left( {{T_{vib}}} \right)_i^{n + 1}} \right) = \frac{{2\left( {{\rho E_{vib}}} 
		\right)_i^{n + 1}\ln \left( 
		{{{\left( {\rho _i^{n + 1}R{\Theta _{vib}}} \right)} \mathord{\left/
		{\vphantom {{\left( {\rho _i^{n + 1}R{\Theta _{vib}}} \right)} {\left( {{E_{vib}}} \right)_i^{n + 1}}}}
		\right. \kern-\nulldelimiterspace} {\left( {{E_{vib}}} \right)_i^{n + 1}}} + 1} \right)}}
		{{\rho _i^{n + 1}R{\Theta _{vib}}}},\\
		&\left( {{T_{vib}}} \right)_i^{n + 1} = \frac{{2\left( {{\rho E_{vib}}} \right)_i^{n + 1}}}
		{{\rho _i^{n + 1}R{K_{vib}}\left( {\left( {{T_{vib}}} \right)_i^{n + 1}} \right)}}.
	\end{aligned}
\end{equation}
\textbf{(e)} Calculate the translational-rotational equilibrium temperature $(T_{2})_i^{n+1}$ as:\\
\begin{equation}\label{cdugks_mac_update_equ_T2}
	\left( {{T_2}} \right)_i^{n + 1} = \frac{{\left[ {5 + {K_{vib}}\left( {T_i^{n + 1}} \right)} 
			\right]T_i^{n + 1} - 
	{K_{vib}}\left( {\left( {{T_{vib}}} \right)_i^{n + 1}} \right)\left( {{T_{vib}}} \right)_i^{n + 1}}}{5}.
\end{equation}
\textbf{(f)}  Solve Eq.~\eqref{cdugks_mac_update_equ_Erot} to update the rotational energy $(\rho E_{rot})_i^{n+1}$.\\
\par
Eqs.~\eqref{cdugks_mic_update_equ} and~\eqref{cdugks_mac_update_equ} are the update rules for 
the microscopic distribution functions and the macroscopic flow variables, respectively. 
In this system, in order to update the gas distribution functions in Eq.~\eqref{cdugks_mic_update_equ},
${\phi _{i,k}^{*,n + 1}}$ and $\tau _i ^{n+1}$ depend on the macroscopic flow variables at (n+1)-th step, 
which can be provided by solving the Eq.~\eqref{cdugks_mac_update_equ}. Therefore, 
Eqs.~\eqref{cdugks_mic_update_equ} and~\eqref{cdugks_mac_update_equ} are uniquely determined once 
the micro-flux $F(\phi_{ij,k}(t_n+{s}))$ across the cell interface is obtained. 
\par

\subsection{Simplified multi-scale numerical flux}\label{FLUX}
\par
The construction of distribution function at the interface is very important and it is about whether the scheme is 
multi-scale and applicable to all flow regimes. In the evolution process of original UGKS~\cite{KunXu_A_2010}, a local 
time-dependent analytical solution of the model equation to describe the evolution of the interface distribution 
function during the time step is used to calculate the micro-flux and the macro-flux. However, the analytical solution 
of the kinetic model equation will become extremely intricate when the non-equilibrium diatomic gases including the 
rotational~\cite{ShaLiu_2014} and vibrational~\cite{ZhaoWang_2017} degrees of freedom are considered. In original 
DUGKS~\cite{ZhaoliGuo_2013}, a discrete temporal difference scheme along the characteristic line of the model equation 
is used at the interface to get the distribution function $\phi_{ij,k}^{n+{1/2}}$. In present method, the construct of 
DUGKS is adopted and a simplified multi-scale numerical flux will be proposed. we evolve the initial distribution 
function inside the cell to the interface taking into account the collision process with a time step $s$ 
($0 < s \leq {\Delta t}$) through a temporal difference scheme of the model equation~\eqref{model_equ_rd_phi}.
\par
The kinetic model equation~\eqref{model_equ_rd_phi} is integrated within a time step  
$s$ along the characteristic line ${\bm x}+{\bm u}_k{}t$ whose end 
point ${\bm x}_{ij}$ is the middle point of cell interface $ij$, the obtained characteristic line solution 
is as follows:
\begin{equation}\label{cdugks_flux_dis_equ}
	\begin{aligned}
	\phi \left( {{{\bm{x}}_{ij}},{{\bm{u}}_k},{t_n}+s} \right) &- \phi \left( {{{\bm{x}}_{ij}} -
	{{\bm{u}}_k}s,{{\bm{u}}_k},{t_n}} \right) = s\Omega \left( {{{\bm{x}}_{ij}},{{\bm{u}}_k},{t_n} + s} \right)\\
	&= s\frac{{{\phi ^*}\left( {{{\bm{x}}_{ij}},{{\bm{u}}_k},{t_n} + s} \right) - \phi \left( {{{\bm{x}}_{ij}},
	{{\bm{u}}_k},{t_n} + s} \right)}}{{\tau _{ij}({t_n}+s)}}.
	\end{aligned}
\end{equation} 
Finally the distribution function $\phi \left( {{{\bm{x}}_{ij}},{{\bm{u}}_k},{t_n} + s} \right)$ at the cell 
interface is calculated as
\begin{equation}\label{cdugks_solve_phi}
	\phi \left( {{{\bm{x}}_{ij}},{{\bm{u}}_k},{t_n} + s} \right) = \frac{{\tau _{ij}({t_n}+s)}}
{{\tau _{ij}({t_n}+s) + s}}\phi \left( {{{\bm{x}}_{ij}} - {{\bm{u}}_k}s,{{\bm{u}}_k},{t_n}} \right) 
	+ \frac{s}{{\tau _{ij}({t_n}+s) + s}}{\phi ^*}\left( {{{\bm{x}}_{ij}},{{\bm{u}}_k},{t_n} + s} \right).
\end{equation} 
In steady flow calculation, the local physical time step $s={\bf{min}}({\Delta t}_{i}, {\Delta t}_{j})$ is used 
to speed up convergence (fully-implicit treated for the the convection term in Eq.~\eqref{cdugks_mic_dis_equ}), 
and the ${\Delta t}_{i}$ and ${\Delta t}_{j}$ are chosen according to the CFL condition. 
While in unsteady flow calculation, the global time step $s={\bf{min}}({\Delta t}_{1}/2, ..., {\Delta t}_{N}/2)$ 
is adopted to keep the second-order time accuracy (the midpoint rule for the time integration of the convection
term), and the $N$ is total number of cells.
\par
\begin{figure}[!t]
	\centering
	\includegraphics[scale=1.0]{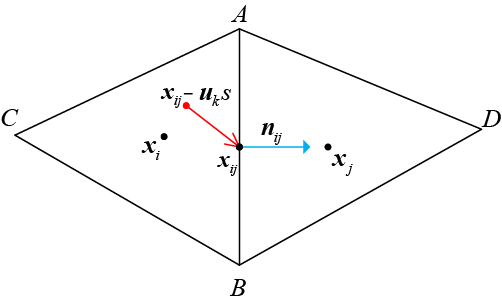}
	\caption{Sketch of two neighboring cells and the particle
		trajectories on a general unstructured mesh.}
	\label{fig_surface_reconstruction}
\end{figure}

According to Eq.~\eqref{cdugks_solve_phi}, once the distribution function 
${\phi}\left({{{\bm{x}}_{ij}}-{{\bm{u}}_k}s,{{\bm{u}}_k},{t_n}}\right)$ and the 
equilibrium distribution function $\phi^*({{{\bm{x}}_{ij}},{{\bm{u}}_k},{t_n} + s})$
at interface center are obtained, then the distribution functions 
$\phi \left( {{{\bm{x}}_{ij}},{{\bm{u}}_k},{t_n} + s} \right)$ can be recovered. 
With the Taylor expansion around the cell center, the distribution function 
${\phi}\left({{{\bm{x}}_{ij}}-{{\bm{u}}_k}s,{{\bm{u}}_k},{t_n}}\right)$ is approximated as
\begin{equation}\label{cdugks_flux_solve_phi_bar_plus}
	{\phi}\left( {{{\bm{x}}_{ij}} - {{\bm{u}}_k}s,{{\bm{u}}_k},{t_n}} \right) = {\phi}\left( 
	{{{\bm{x}}_c},{{\bm{u}}_k},{t_n}} \right) + \left( {{{\bm{x}}_{ij}} - {{\bm{u}}_k}s -
		{{\bm{x}}_c}} \right) \cdot L\left( {\nabla {{\phi}},{{\bm{x}}_c}} \right)\nabla 
	{\phi}\left( {{{\bm{x}}_c},{{\bm{u}}_k},{t_n}} \right),
\end{equation}
where ${\bm x}_c$ represent the central coordinates of the cell which the particles migrate from. 
As shown in Fig.~\ref{fig_surface_reconstruction}, 
the ${\bm x}_c$ equals to ${\bm x}_i$ if ${{{\bm{u}}_k}\cdot{{\bm{n}}_{ij}}} > 0$, 
or ${\bm x}_j$ otherwise.
The gradient $\nabla {\phi}({{{\bm{x}}_c},{{\bm{u}}_k},{t_n}})$ at the cell center is calculated 
using the least square method. The function $L\left( {\nabla {{\phi}},{{\bm{x}}_c}} \right)$ 
in Eq.~\eqref{cdugks_flux_solve_phi_bar_plus} denotes the gradient limiter which is used to suppress numerical 
oscillations, and the Venkatakrishnan limiter~\citep{V_Venkatakrishnan_1995} for flow computations 
on unstructured mesh is adopted.
\par
The macroscopic flow variables $\bm {Q}_{ij}({t_n}+s)$ at 
time $t_{n}+s$ used to evaluate the equilibrium distribution functions 
$\phi^*({{{\bm{x}}_{ij}},{{\bm{u}}_k},{t_n} + s})$ 
are also calculated from ${\phi}\left({{{\bm{x}}_{ij}}-{{\bm{u}}_k}s,{{\bm{u}}_k},{t_n}}\right)$ as follows:
\begin{equation}\label{cdugks_halftime_solve_rho}
	\rho _{ij}({t_n}+s) = \sum\limits_k {{W_k}G\left( {{{\bm{x}}_{ij}}
	 - {{\bm{u}}_k}s,{{\bm{u}}_k},{t_n}} \right)},
\end{equation}
\begin{equation}\label{cdugks_halftime_solve_U}
	\left( {\rho {\bm{U}}} \right)_{ij}({t_n}+s) = \sum\limits_k {{W_k}{{\bm{u}}_k}G\left( {{{\bm{x}}_{ij}}
	 - {{\bm{u}}_k}s,{{\bm{u}}_k},{t_n}} \right)},
\end{equation}
\begin{equation}\label{cdugks_halftime_solve_E}
	\left( {\rho E} \right)_{ij}({t_n}+s) = \sum\limits_k {{W_k}\left\{ \begin{array}{l}
	\frac{1}{2}\left[ {{{\left| {{{\bm{u}}_k}} \right|}^2}G\left( {{{\bm{x}}_{ij}} - {{\bm{u}}_k}s,{{\bm{u}}_k},{t_n}}
	 \right) + H\left( {{{\bm{x}}_{ij}} - {{\bm{u}}_k}s,{{\bm{u}}_k},{t_n}} \right)} \right]\\
	+ R\left( {{{\bm{x}}_{ij}} - {{\bm{u}}_k}s,{{\bm{u}}_k},{t_n}} \right) + B\left( {{{\bm{x}}_{ij}} -
		 {{\bm{u}}_k}s,{{\bm{u}}_k},{t_n}} \right)
		\end{array} \right\}},
\end{equation}
\begin{equation}\label{cdugks_halftime_solve_Erot}
	\begin{aligned}
	&\left( {\rho {E_{rot}}} \right)_{ij}({t_n}+s) 
	= {\left[ {1 + \frac{s}{{\tau _{ij}({t_n}+s)}}\left( {\frac{1}{{\left( {{Z_{rot}}} \right)_{ij}({t_n}+s)}}
				 + \frac{1}{{\left( {{Z_{vib}}} \right)_{ij}({t_n}+s)}}} \right)} \right]^{ - 1}}\\
	&\times \left\{ {\sum\limits_k {{W_k}R\left( {{{\bm{x}}_{ij}} - {{\bm{u}}_k}s,{{\bm{u}}_k},{t_n}} \right)}  
		+ \frac{s}{{\tau _{ij}({t_n}+s)}}\left[ {\left( {\frac{{\rho R{T_2}}}{{{Z_{rot}}}}} \right)_{ij}({t_n}+s) + \left( {\frac{{\rho RT}}{{{Z_{vib}}}}} \right)_{ij}({t_n}+s)} \right]} \right\},
	\end{aligned}
\end{equation}
\begin{equation}\label{cdugks_halftime_solve_Evib}
	\begin{aligned}
	&\left( {\rho {E_{vib}}} \right)_{ij}({t_n}+s) 
	= {\left[ {1 + \frac{s}{{\left( {{Z_{vib}}\tau } \right)_{ij}({t_n}+s)}}} \right]^{ - 1}}\\
	&\times \left[ {\sum\limits_k {{W_k}B\left( {{{\bm{x}}_{ij}} - {{\bm{u}}_k}s,{{\bm{u}}_k},{t_n}} \right)}
		+ \frac{s}{{\left( {{Z_{vib}}\tau } \right)_{ij}({t_n}+s)}}\frac{{{K_{vib}}\left( T \right)}}{2}
		\left( {\rho RT} \right)_{ij}({t_n}+s)} \right].
	\end{aligned}
\end{equation}
\par
In addition to the above variables, according to Eqs.~\eqref{cdugks_flux_dis_equ}
and~\eqref{heat_flux_tra_relaxation}$\sim$\eqref{heat_flux_vib_relaxation}, the translational, rotational, 
and vibrational heat fluxes $(\bm{q}_{tr})_{ij}({t_n}+s)$, $(\bm{q}_{rot})_{ij}({t_n}+s)$, 
$(\bm{q}_{vib})_{ij}({t_n}+s)$ can also be obtained from the distribution function
${\phi}\left({{{\bm{x}}_{ij}}-{{\bm{u}}_k}s,{{\bm{u}}_k},{t_n}}\right)$ as:
\begin{equation}\label{cdugks_halftime_solve_qtra}
	\left( {{{\bm{q}}_{tr}}} \right)_{ij}({t_n}+s) = \frac{{\tau _{ij}({t_n}+s)
	\sum\limits_k {{W_k}\frac{1}{2}{{\bm{c}}_k}\left[ {{{\left| {{{\bm{c}}_k}} \right|}^2}
	G\left( {{{\bm{x}}_{ij}} - {{\bm{u}}_k}s,{{\bm{u}}_k},{t_n}} \right) + H\left( {{{\bm{x}}_{ij}} 
		- {{\bm{u}}_k}s,{{\bm{u}}_k},{t_n}} \right)} \right]} }}{{\tau _{ij}({t_n}+s) + s 
	- \frac{s}{3}\left[ {\left( {1 - \frac{1}{{\left( {{Z_{rot}}} \right)_{ij}({t_n}+s)}} 
	- \frac{1}{{\left( {{Z_{vib}}} \right)_{ij}({t_n}+s)}}} \right) + \frac{{{\omega _0}}}{{\left( {{Z_{rot}}} \right)_{ij}({t_n}+s)}} + \frac{{{\omega _2}}}{{\left( {{Z_{vib}}} \right)_{ij}({t_n}+s)}}} \right]}},
\end{equation}
\begin{equation}\label{cdugks_halftime_solve_qrot}
	\begin{aligned}
	&\left( {{{\bm{q}}_{_{rot}}}} \right)_{ij}({t_n}+s) \\
	&= \frac{{\tau _{ij}({t_n}+s)\sum\limits_k {{W_k}{{\bm{c}}_k}R\left( {{{\bm{x}}_{ij}} - {{\bm{u}}_k}s,
	{{\bm{u}}_k},{t_n}} \right)} }}{{\tau _{ij}({t_n}+s) + s + s\left( {\delta  - 1} \right)\left[ {\left( {1 - \frac{1}
	{{\left( {{Z_{rot}}} \right)_{ij}({t_n}+s)}} - \frac{1}{{\left( {{Z_{vib}}} \right)_{ij}({t_n}+s)}}} \right)
 + \frac{{{\omega _1}}}{{\left( {{Z_{rot}}} \right)_{ij}({t_n}+s)}} + \frac{{{\omega _3}}}{{\left( {{Z_{vib}}} \right)_{ij}({t_n}+s)}}} \right]}},
	\end{aligned}
\end{equation}
\begin{equation}\label{cdugks_halftime_solve_qvib}
	\left( {{{\bm{q}}_{vib}}} \right)_{ij}^{n + {1 \mathord{\left/
			{\vphantom {1 2}} \right.
	\kern-\nulldelimiterspace} 2}} = \frac{{\tau _{ij}({t_n}+s)}}{{\tau _{ij}({t_n}+s) + s}}\sum\limits_k {{W_k}{{\bm{c}}_k}B\left( {{{\bm{x}}_{ij}} 
	- {{\bm{u}}_k}s,{{\bm{u}}_k},{t_n}} \right)}.
\end{equation}
\par
Up to now, the equilibrium distribution function $\phi^*({{{\bm{x}}_{ij}},{{\bm{u}}_k},{t_n} + s})$ 
at the cell interface can be obtained from the macroscopic flow variables. Thus, the distribution function 
$\phi({{{\bm{x}}_{ij}},{{\bm{u}}_k},{t_n} + s})$ at the cell interface is solved by 
Eq.~\eqref{cdugks_solve_phi}. As a result, the micro-flux $F(\phi_{ij,k}({t_n}+s))$ 
can be obtained using Eq.~\eqref{cdugks_mic_flux}.
\par
In summary, the calculation procedure of the UGKS with simplified multi-scale numerical flux from time 
level $t_n$ to $t_{n+1}$ is summarized in the following steps:
\\
\textbf{Step 1.} Given the initial macroscopic flow variables ${\bm{Q}}_{i}^{0}$ and calculate the
equilibrium distribution functions $\phi_{i,k}^{*,0}$.
\\
\textbf{Step 2.} Compute the micro-flux $F(\phi_{ij,k}({t_n}+s))$ across the cell interface of control volumes.

\textbf{(a)} Calculate the distribution function ${\phi}\left({{{\bm{x}}_{ij}}-{{\bm{u}}_k}s,{{\bm{u}}_k},{t_n}}\right)$ 
according to Eq.~\eqref{cdugks_flux_solve_phi_bar_plus}.

\textbf{(b)} Calculate the macroscopic flow variables ${\bm{Q}}_{ij}({t_n}+s)$ using 
Eqs.~\eqref{cdugks_halftime_solve_rho}$\sim$\eqref{cdugks_halftime_solve_Evib} and the translational, 
rotational, and vibrational heat fluxes 
using Eqs.~\eqref{cdugks_halftime_solve_qtra}$\sim$\eqref{cdugks_halftime_solve_qvib} from 
the distribution function ${\phi}\left({{{\bm{x}}_{ij}}-{{\bm{u}}_k}s,{{\bm{u}}_k},{t_n}}\right)$.

\textbf{(c)} Calculate the equilibrium distribution functions $\phi^*({{{\bm{x}}_{ij}},{{\bm{u}}_k},{t_n} + s})$ 
from the macroscopic flow variables ${\bm{Q}}_{ij}({t_n}+s)$ and heat fluxes.

\textbf{(d)} Calculate the distribution functions $\phi({{{\bm{x}}_{ij}},{{\bm{u}}_k},{t_n} + s})$ 
at the cell interface using Eq.~\eqref{cdugks_solve_phi}.

\textbf{(e)} Handle the interface boundary conditions that will be discussed in the next subsection.

\textbf{(f)} Calculate the micro-flux $F(\phi_{ij,k}({t_n}+s))$ using Eq.~\eqref{cdugks_mic_flux}.
\\
\textbf{Step 3.} Update the macroscopic flow variables ${\bm{Q}}_{i}^{n+1}$ in each cell $i$ according
to Eqs.~\eqref{cdugks_mac_update_equ_rho_U_E}$\sim$\eqref{cdugks_mac_update_equ_Evib}.
\\
\textbf{Step 4.} Update the distribution function $\phi _{i,k}^{n+1}$ in each cell $i$ according
to Eq.~\eqref{cdugks_mic_update_equ}.
\par

\subsection{Boundary conditions}\label{Boundary_Conditions}
\par
The distribution functions on the boundary surface consist of two portions. A portion of
distribution functions from the inner fluid field to the boundary surface are solved similarly with the 
inner surface, while another portion reentering from the boundary to the inner fluid field need to be 
handled according to the different boundary conditions. 
Assuming that the ${\bm {n}}_{bi}$ is the outward unit vector normal to the boundary interface
from the inner fluid field to the boundary surface. In the present study, we first consider the isothermal 
wall boundary condition with the constant surface temperature $T_{w}$. 
The reflection law is assumed to be completely diffusive, and  
the reflected distribution functions (${{\bm{u}}_k} \cdot {{\bm{n}}_{bi}} < 0$) are Maxwellian as follows:
\begin{equation}\label{boundary_wall}
	\phi \left( {{{\bm{x}}_w},{{\bm{u}}_k} \cdot {{\bm{n}}_{bi}} < 0,{t_n} + s} \right) = 
	\left( {\begin{aligned}
			&{{g^{eq}}\left( {{\rho _w},{{\bm{U}}_w},{T_w}} \right)}\\
			&{\left( {3 - D} \right)R{T_w}{g^{eq}}\left( {{\rho _w},{{\bm{U}}_w},{T_w}} \right)}\\
			&{R{T_w}{g^{eq}}\left( {{\rho _w},{{\bm{U}}_w},{T_w}} \right)}\\
			&{\frac{{{K_{vib}}\left( {{T_w}} \right)}}{2}R{T_w}{g^{eq}}\left( {{\rho _w},{{\bm{U}}_w},
					{T_w}} \right)}
	\end{aligned}} \right),
\end{equation}
where ${\bm x}_{w}$ is the coordinates of wall boundary interface and
the ${g^{eq}}({{\rho _w},{{\bm{U}}_w},{T_w}})$ is the Maxwellian equilibrium distribution function.
The density $\rho _w$ is determined by the condition that no particles can go through the wall, i.e.,
\begin{equation}\label{boundary_wall_solve_density_equation}
	\sum\limits_{{{\bm{u}}_k} \cdot {{\bm{n}}_{bi}} < 0} {{W_k}\left( {{{\bm{u}}_k} \cdot {{\bm{n}}_{bi}}} 
		\right){g^{eq}}\left( 
		{{\rho _w},{{\bm{U}}_w},{T_w}} \right)}  + \sum\limits_{{{\bm{u}}_k} \cdot {{\bm{n}}_{bi}} > 0} 
	{{W_k}\left( 
		{{{\bm{u}}_k} \cdot {{\bm{n}}_{bi}}} \right)G\left( {{{\bm{x}}_w},{{\bm{u}}_k},{t_n}+s} \right)} = 0,
\end{equation}
which gives
\begin{equation}\label{boundary_wall_solve_density}
	{\rho _w} =  - \frac{{\sum\limits_{{{\bm{u}}_k} \cdot {{\bm{n}}_{bi}} > 0} {{W_k}\left( {{{\bm{u}}_k} \cdot 
					{{\bm{n}}_{bi}}} 
				\right)G\left( {{{\bm{x}}_w},{{\bm{u}}_k},{t_n} + s} \right)} }}{{\sum\limits_{{{\bm{u}}_k}
				\cdot {{\bm{n}}_{bi}} < 0} {{W_k}\left( {{{\bm{u}}_k} \cdot {{\bm{n}}_{bi}}} \right){g^{eq}}
				\left( {1,{{\bm{U}}_w},{T_w}} \right)} }}.	
\end{equation}
\par
The inlet and outlet boundary conditions for supersonic flow will also be considered in this work.
The distribution functions entering the flow field are Maxwellian 
${g^{eq}}({{\rho _\infty},{{\bm{U}}_\infty},{T_\infty}})$
determined by the density, velocity, and temperature of the free-stream conditions,
\begin{equation}\label{boundary_inlet}
	\phi \left( {{{\bm{x}}_{in}},{{\bm{u}}_k} \cdot {{\bm{n}}_{bi}} < 0,{t_n} + s} \right) = 
	\left( {\begin{aligned}
			&{{g^{eq}}\left( {{\rho _\infty },{{\bm{U}}_\infty },{T_\infty }} \right)}\\
			&{\left( {3 - D} \right)R{T_\infty }{g^{eq}}\left( {{\rho _\infty },{{\bm{U}}_\infty },{T_\infty }} 
				\right)}\\
			&{R{T_\infty }{g^{eq}}\left( {{\rho _\infty },{{\bm{U}}_\infty },{T_\infty }} \right)}\\
			&{\frac{{{K_{vib}}\left( {{T_\infty }} \right)}}{2}R{T_\infty }{g^{eq}}\left( {{\rho _\infty },
					{{\bm{U}}_\infty },{T_\infty }} \right)}
	\end{aligned}} \right),
\end{equation}
where the ${\bm x}_{in}$ is the coordinate of inlet boundary interface.
\par
The non-equilibrium extrapolation scheme~\citep{ChenWu_2016} is used for 
the supersonic outlet boundary. The distribution function of the particle reentering the flow field 
from the boundary surface are divided into the equilibrium part and the non-equilibrium part,
\begin{equation}\label{boundary_outlet_nne}
	\phi \left( {{{\bm{x}}_{out}},{{\bm{u}}_k} \cdot {{\bm{n}}_{bi}} < 0,{t_n} + s} \right) = {\phi ^*}\left( 
	{{{\bm{x}}_{out}},{{\bm{u}}_k} \cdot {{\bm{n}}_{bi}} < 0,{t_n} + s} \right) + {\phi ^{neq}}\left( 
	{{{\bm{x}}_{out}},{{\bm{u}}_k} \cdot {{\bm{n}}_{bi}} < 0,{t_n} + s} \right).
\end{equation}
The equilibrium distribution function 
${\phi ^*}\left( {{{\bm{x}}_{out}},{{\bm{u}}_k} \cdot {{\bm{n}}_{bi}} < 0,{t_n} + s} \right)$
can be calculated using the macroscopic flow variables which have been solved by 
${\bar \phi} \left( {{{\bm{x}}_{out}},{{\bm{u}}_k} ,{t_n} + s} \right) $.
The non-equilibrium part ${\phi ^{neq}}\left( {{{\bm{x}}_{out}},{{\bm{u}}_k}\cdot{{\bm{n}}_{bi}} < 0,{t_n}+s}\right)$ 
is approximated by non-equilibrium part of the inner fluid cell ${\bm x}_{i}$ near the boundary face,
\begin{equation}\label{boundary_outlet_nne_neq}
	{\phi ^{neq}}\left( {{{\bm{x}}_{out}},{{\bm{u}}_k} \cdot {{\bm{n}}_{bi}} < 0,{t_n} + s} \right) \approx 
	\phi \left( {{{\bm{x}}_i},{{\bm{u}}_k} \cdot {{\bm{n}}_{bi}} < 0,{t_n}} \right) - {\phi ^*}
	\left( {{{\bm{x}}_i},{{\bm{u}}_k} \cdot {{\bm{n}}_{bi}} < 0,{t_n}} \right).
\end{equation}
\par

\subsection{Integration error correction for stress and heat flux}
\par
Although the unstructured DVS adopted in the present work is more flexible and efficient, 
the discrete velocity points is relatively arbitrary compared with the Cartesian velocity space.
Besides, the integration accuracy of the midpoint integration on unstructured DVS 
is lower than that of the Newton-Cotes integration on Cartesian DVS. 
For those reasons, the macroscopic flow variables obtained by the numerical quadrature of distribution functions 
will inevitably introduce a large integration error.
\par 
In the present UGKS, the update of macroscopic flow variables $\bm{Q}$ are based on the fluxes 
(Eqs.~\eqref{cdugks_mac_update_equ_rho_U_E}$\sim$\eqref{cdugks_mac_update_equ_Evib})
rather than the direct integration of the distribution functions. 
Therefore, the integration error caused by the source term $\bm{S}$, which will result in the damage of 
conservation property, does not directly introduce in the macroscopic discrete equations. 
However, the stress tension and 
heat flux are obtained directly by numerical integration of the high order moments. 
In this case, the influence of integration error needs to be considered and 
corresponding integration error correction is adopted.
\par
The stress tension ${\bf {P}}$ is calculated as: 
\begin{equation}\label{stress_correct_equ1_G}
	{\bf {P}} = \int {\bm{c}\bm{c}Gd{\bm{u}}}  = \sum\limits_k {{W_k}\bm{c}\bm{c}{G_k}}  + {\bf{e}},
\end{equation}
where ${\bf{e}}$ is the integration error which can be expressed as:
\begin{equation}\label{stress_correct_define_delta_e}
	{\bf{e}} = \sum\limits_k {\left\{ {\int_{\partial {{\bm{u}}_k}} {{\bm{cc}}\sum\limits_{n = 1}^\infty  
				{\frac{1}{{n!}}{{\left[ {\left( {{\bm{u}} - {{\bm{u}}_k}} \right) \cdot \frac{\partial }{{\partial 
										{\bm{u}}}}} \right]}^n}G\left( {{{\bm{u}}_k}} \right)} d{\bm{u}}} } \right\}}.
\end{equation}
\par
Noticed that the moments of distribution function $G$ can be decomposed into the moments of equilibrium 
part $G^*$ which can be solved analytically and the moments of non-equilibrium part expressed as $G-G^*$.
Thus, the stress tension ${\bf P}$ can also be calculated as:
\begin{equation}\label{stress_correct_equ2_Geq_Gneq}
	{{\bf P}} = \int {\bm{c}\bm{c}{G^*}d{\bm{u}}}  + \int {\bm{c}\bm{c}\left({G-{G^*}}\right)d{\bm{u}}} = \int 
	{\bm{c}\bm{c}{G^*}d{\bm{u}}} + \sum\limits_k {{W_k}\bm{c}\bm{c}\left( {{G_k}-{G_k^*}}\right)} 
	+\bf{e^*},
\end{equation}
where $\bf{e^*}$ is the integration error which can be expressed as:
\begin{equation}\label{stress_correct_define_delta_e_Geq}
	{\bf{e^*}} = \sum\limits_k {\left\{ {\int_{\partial {{\bm{u}}_k}} {{\bm{cc}}\sum\limits_{n = 1}^\infty   
				{\frac{1}{{n!}}{{\left[ {\left( {{\bm{u}} - {{\bm{u}}_k}} \right) \cdot \frac{\partial }{{\partial 
			{\bm{u}}}}}  \right]}^n}\left[ {G\left( {{{\bm{u}}_k}} \right) - {G^*}\left( 
						{{{\bm{u}}_k}} \right)} \right]} d{\bm{u}}} }  \right\}}.
\end{equation}
\par
We noticed that the Chapman–Enskog expansion gives $G={G^*}+O(\tau)$, and the integration error $\bf{e^*}$
is approximate of order $\tau$, which result in ${\bf{e^*}} = O(\tau){\bf{e}}$. Therefore, the modified 
formula~\eqref{stress_correct_equ2_Geq_Gneq} can obtain more accurate numerical integration results 
compared with formula~\eqref{stress_correct_equ1_G}, especially in the continuum flow regime ($\tau \ll 1$).
In the present work, the stress tension ${\bf{P}}$ is calculated using 
Eq.~\eqref{stress_correct_equ2_Geq_Gneq} as follows:
\begin{equation}\label{stress_correct}
	{\bf{P}} = \int {\bm{c}\bm{c}{G^*}d{\bm{u}}}  + \sum\limits_k {{W_k}\bm{c}\bm{c}({{G_k}-{G_k^*}})} = 
	\rho R{T_a}{\bm{\delta}} + 
	\sum\limits_k {{W_k}\bm{c}\bm{c}({{G_k}-{G_k^*}})},
\end{equation}
where ${\bm{\delta}}$ is the usual Kronecker delta and $T_a$ is
\begin{equation}\label{stress_correct_define_Ta}
	T_a = \left( {1 - \frac{1}{{{Z_{rot}}}} - \frac{1}{{{Z_{vib}}}}} \right){T_{tr}} + \frac{1}{{{Z_{rot}}}}{T_2} + 
	\frac{1}{{{Z_{vib}}}}T.
\end{equation}
\par
Similarly, the translational, rotational and vibrational heat fluxes 
${{\bm q}_{tr}}$, ${{\bm q}_{rot}}$ and ${{\bm q}_{vib}}$ can be calculated as:
\begin{equation}\label{heatflux_correct}
	\begin{aligned}
		{{\bm{q}}_{tr}} &= {\left[ {\frac{2}{3}  + \frac{{\left( {1 - {\omega _0}} \right)}}{{3{Z_{rot}}}} + 
				\frac{{\left( {1 - {\omega _2}} \right)}}{{3{Z_{vib}}}}} \right]^{ - 1}}\sum\limits_k 
		{{W_k}\frac{1}{2}{{\bm{c}}_k}\left[ {{{\left| {{{\bm{c}}_k}} \right|}^2}\left( {{G_k} - G_k^*} \right) 
				+ \left( {{H_k} - H_k^*} \right)} \right]},\\
		{{\bm{q}}_{rot}} &= {\left[ {\delta  + \frac{{\left( {1 - {\omega _1}} \right)\left( {1 - \delta } 
	\right)}}{{{Z_{rot}}}} + \frac{{\left( {1 - {\omega _3}} \right)\left( {1 - \delta } 
		\right)}}{{{Z_{vib}}}}} \right]^{ - 1}}\sum\limits_k {{W_k}{{\bm{c}}_k}\left( {{R_k} - R_k^*} \right)},\\
		{{\bm{q}}_{vib}} &= \sum\limits_k {{W_k}{{\bm{c}}_k}\left( {{B_k} - B_k^*} \right)}.
	\end{aligned}
\end{equation}
\par

\section{Numerical results and discussions}\label{Numerical_Results}
\subsection{Shock tube}\label{Sod}
The Sod’s shock tube problem from continuum to free-molecular regimes is computed to validate the 
present method for unsteady flow in one-dimensional case. 
The computational domain is $\left[-0.5,0.5\right]$ and a uniform mesh with 100 cells is used.
For the velocity space, 101 discrete points are uniformly distributed in $\left[-10,10\right]$.
The initial conditions are given by
\begin{equation}\label{sod_initial_conditions}
	\left\{ {\begin{array}{*{20}{l}}
			{\left( {{\rho _1},{u_1},{T_{tr,1}},{T_{rot,1}},{T_{vib,1}}} \right) 
				= \left( {1,0,2,2,2} \right),}&{x \le 0,}\\
			{\left( {{\rho _2},{u_2},{T_{tr,2}},{T_{rot,2}},{T_{vib,2}}} \right) 
				= \left( {0.125,0,1.6,1.6,1.6} \right),}&{x > 0.}
	\end{array}} \right.
\end{equation}
The gas is modeled as hard-sphere molecules and the viscosity index $\omega$ is $0.5$. 
Thus, the mean free path $\lambda_1$ is then changed by adjusting the $\mu_1$ according to Eq~\eqref{mfp_mu_T}.
The rotational and vibrational collision numbers are $Z_{rot}=3$ and $Z_{vib}=30$, respectively.
\par
\begin{figure}[!t]
	\centering
	\subfigure[Density]{
		\begin{minipage}[!t]{0.45\textwidth}
			\centering
			\includegraphics[width=1.0\textwidth]{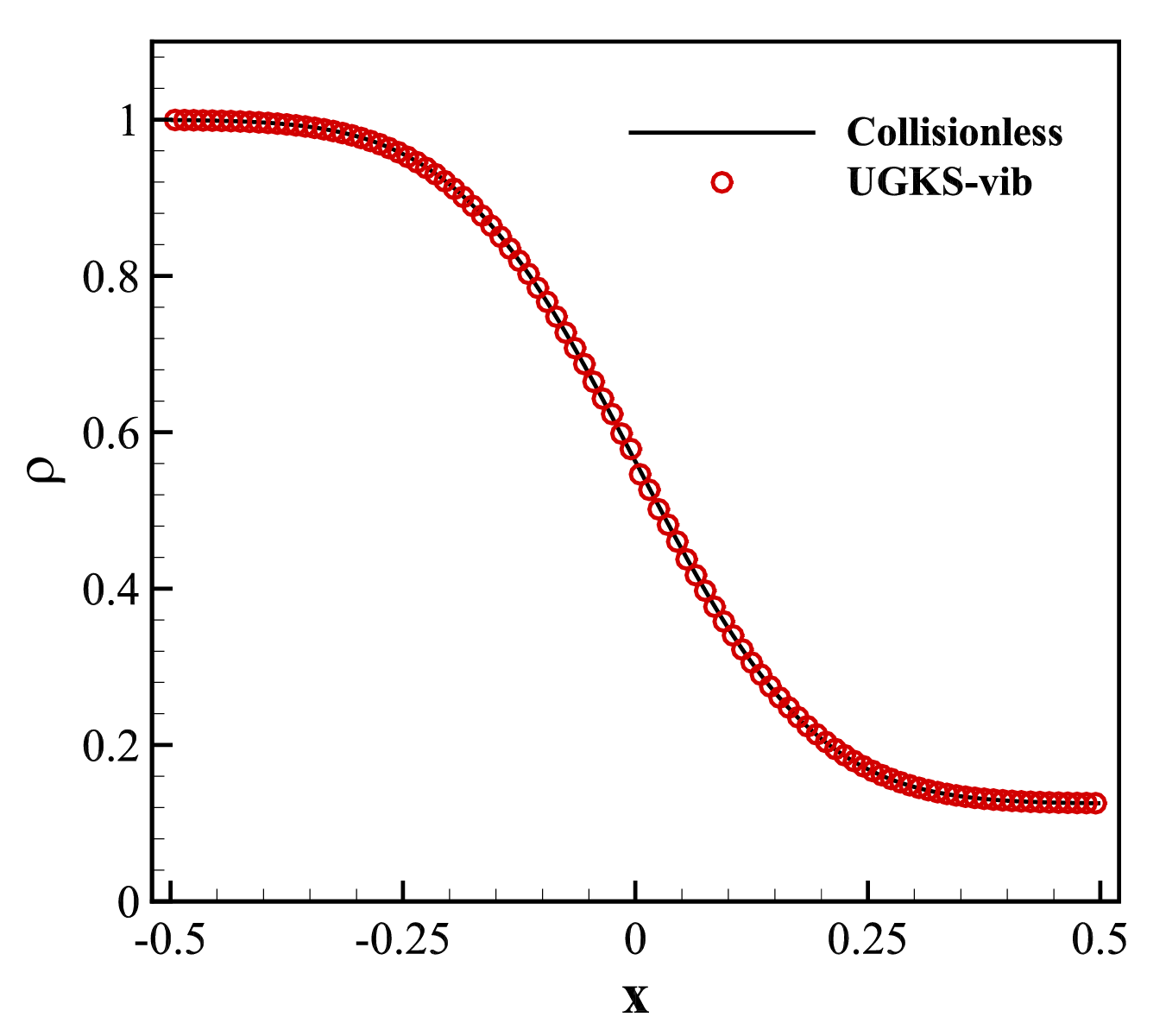}	    
		\end{minipage}
		\label{fig_sod_kn12_rho}
	}
	\subfigure[Pressure]{
		\begin{minipage}[!t]{0.45\textwidth}
			\centering
			\includegraphics[width=1.0\textwidth]{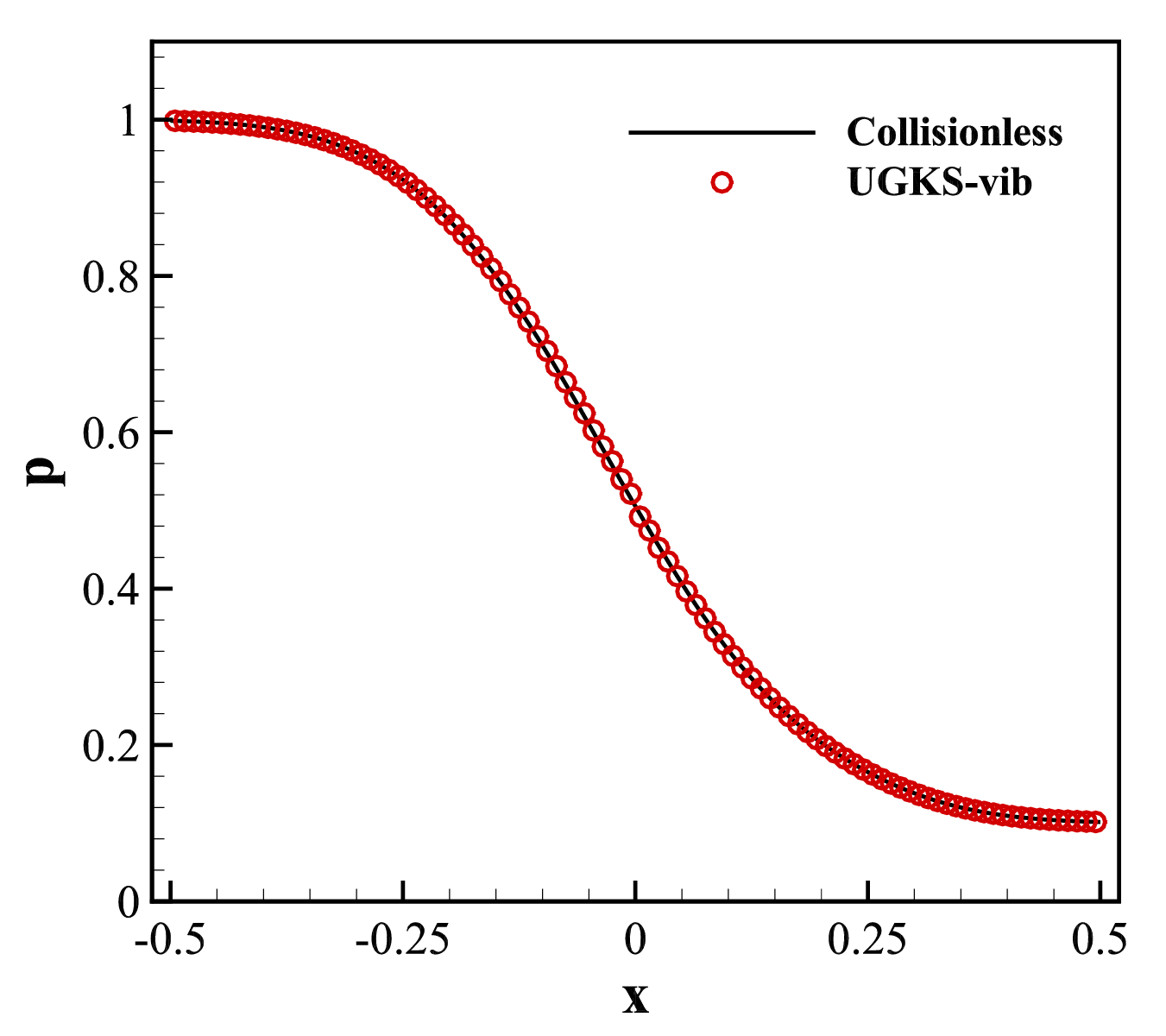}		
		\end{minipage}
		\label{fig_sod_kn12_pre}
	}
	\subfigure[Velocity]{
		\begin{minipage}[!t]{0.45\textwidth}
			\centering
			\includegraphics[width=1.0\textwidth]{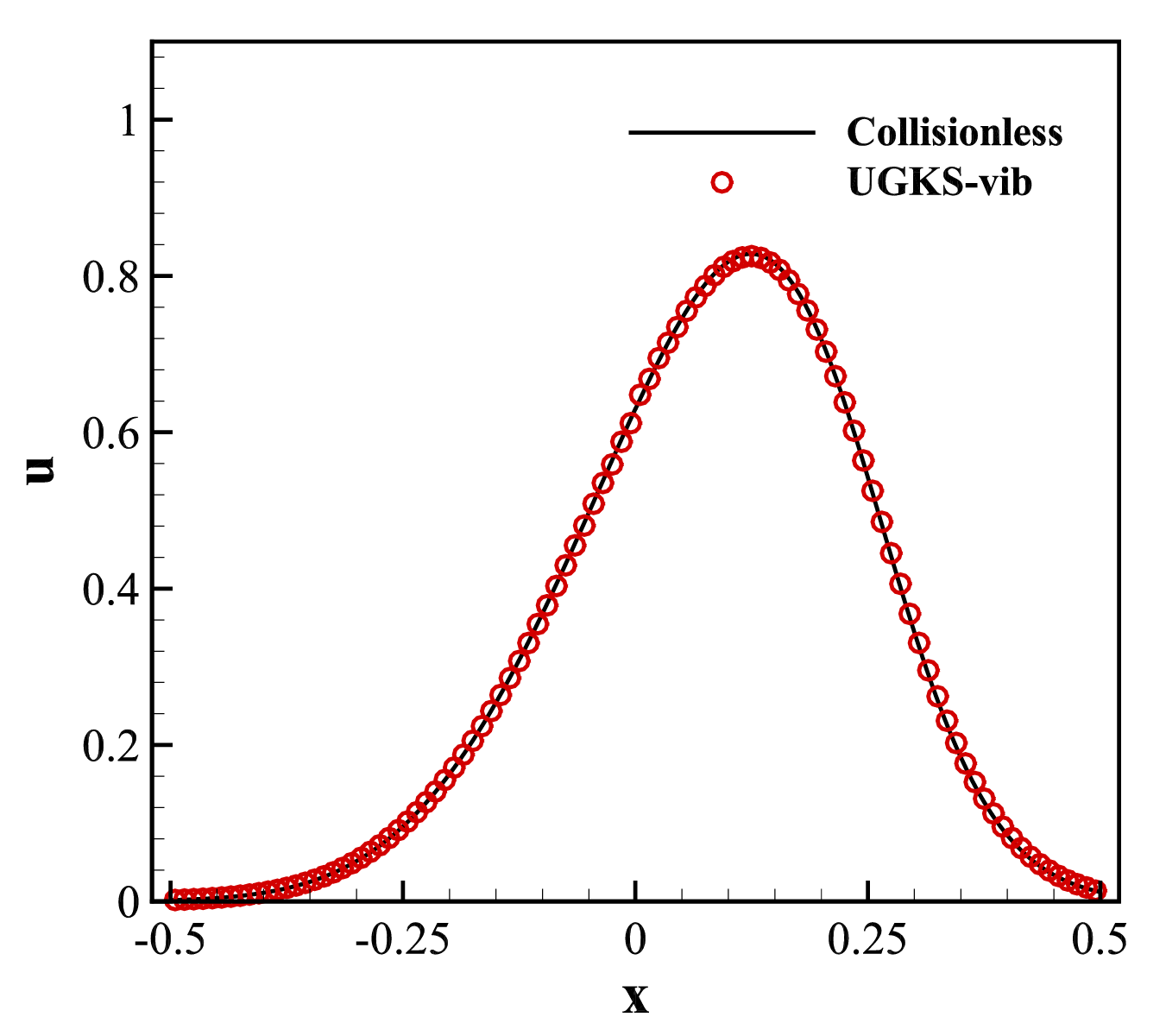}		
		\end{minipage}
		\label{fig_sod_kn12_vel}
	}
	\subfigure[Temperature]{
		\begin{minipage}[!t]{0.45\textwidth}
			\centering
			\includegraphics[width=1.0\textwidth]{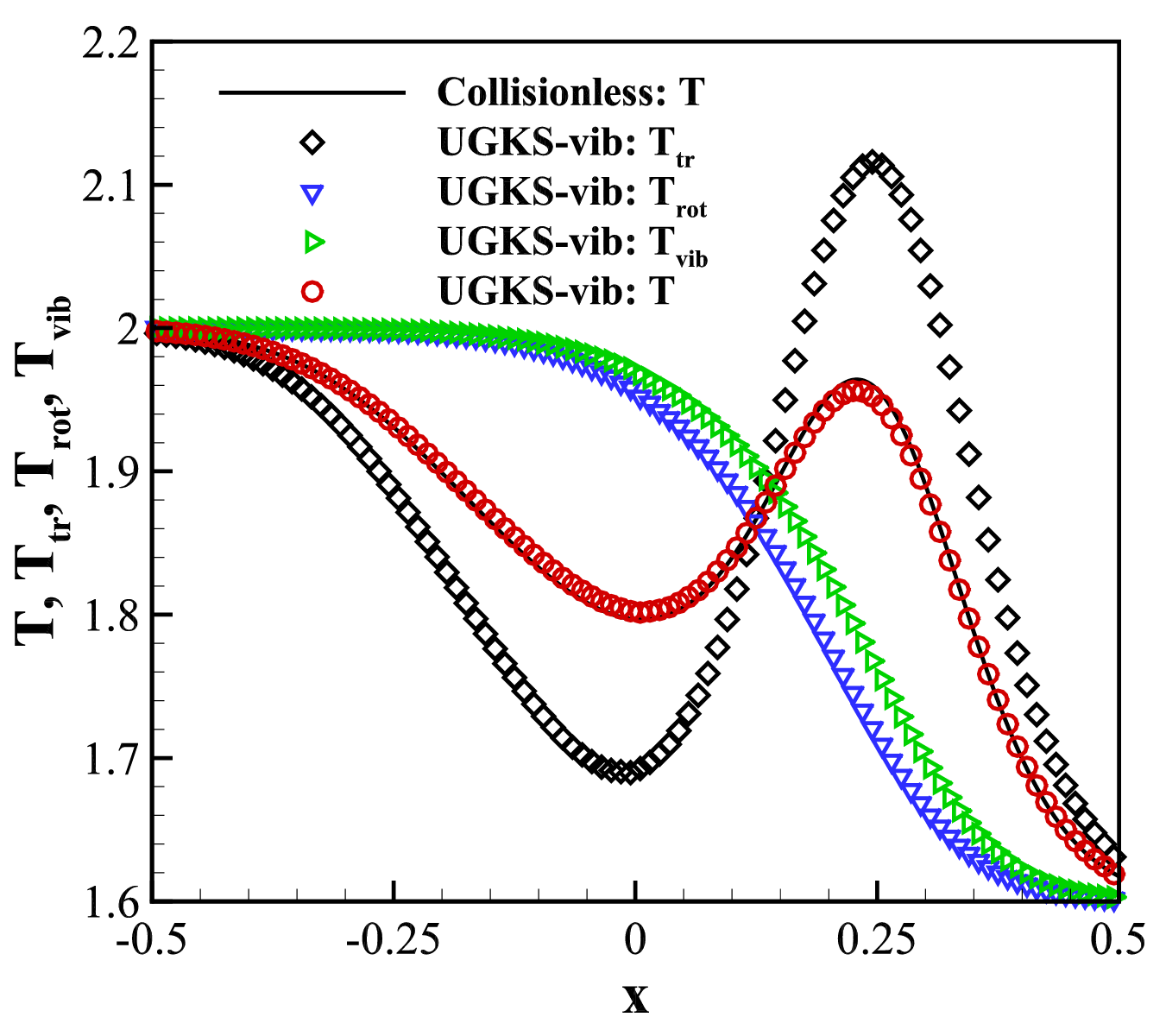}		
		\end{minipage}
		\label{fig_sod_kn12_tem}
	}
	\caption{The density, pressure, velocity and temperature profiles of the Sod’s shock tube at $\rm Kn=12.77$.}
	\label{fig_sod_kn12}
\end{figure}
\begin{figure}[!t]
	\centering
	\subfigure[Density]{
		\begin{minipage}[!t]{0.45\textwidth}
			\centering
			\includegraphics[width=1.0\textwidth]{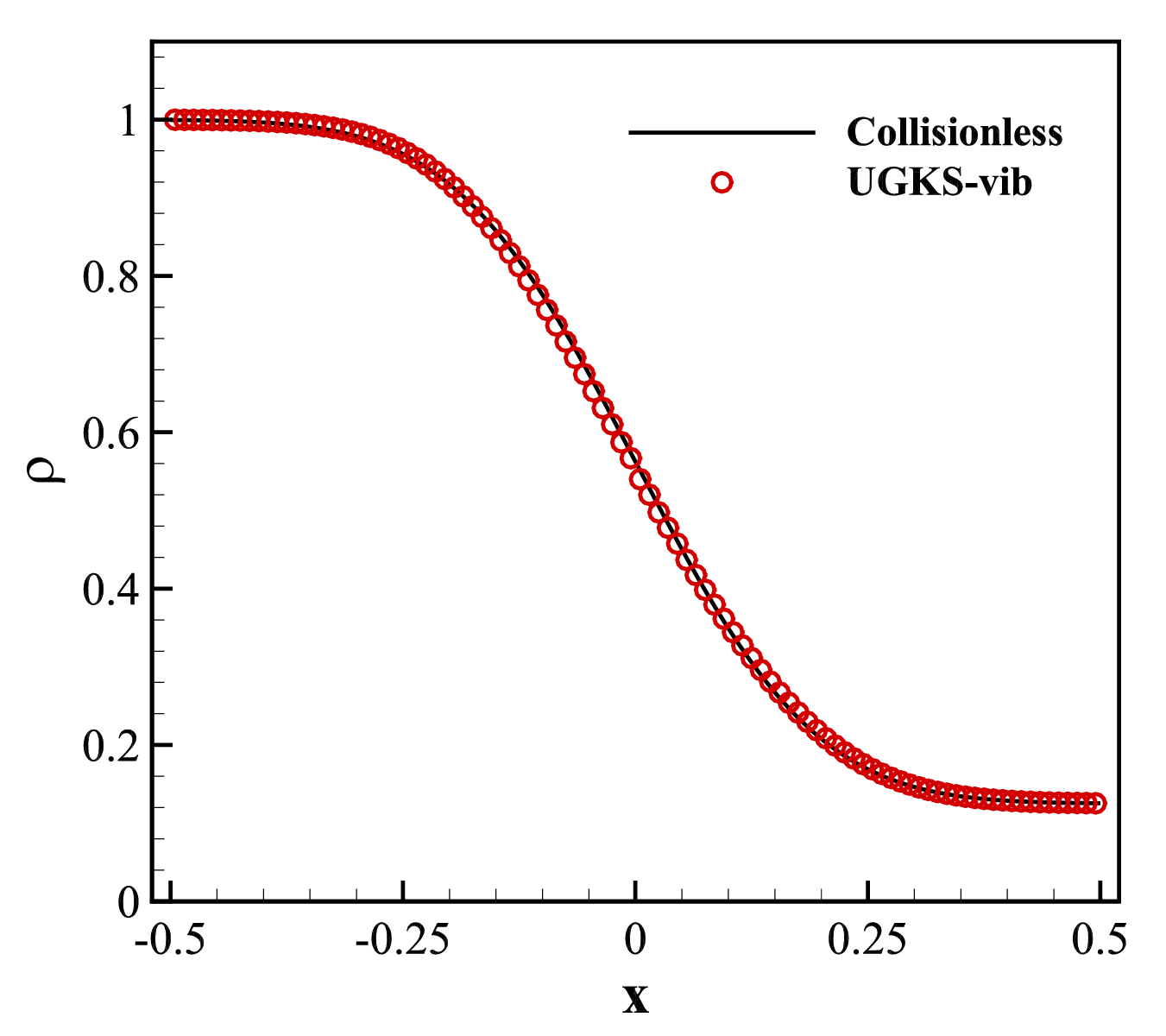}	    
		\end{minipage}
		\label{fig_sod_kn0.12_rho}
	}
	\subfigure[Pressure]{
		\begin{minipage}[!t]{0.45\textwidth}
			\centering
			\includegraphics[width=1.0\textwidth]{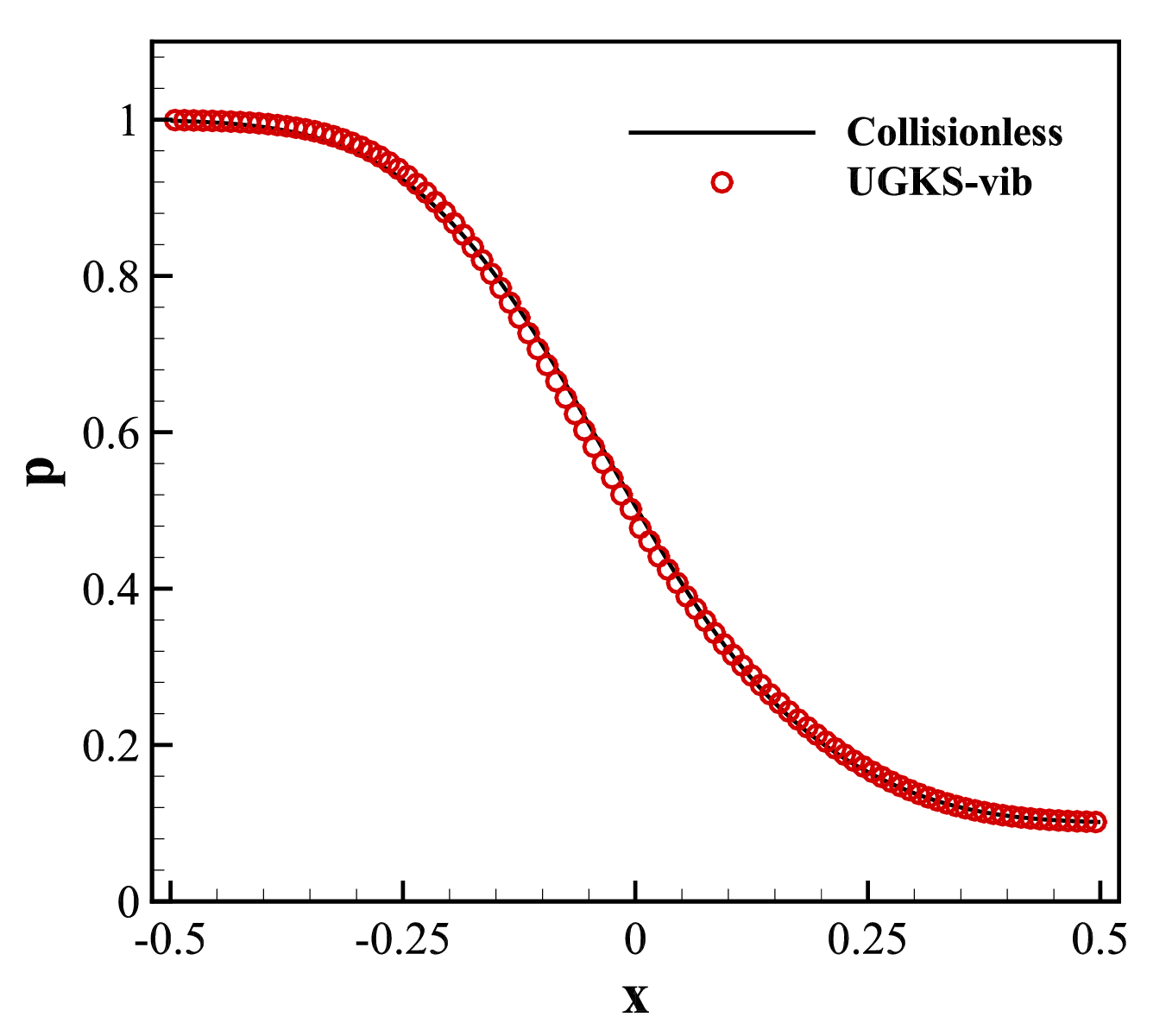}		
		\end{minipage}
		\label{fig_sod_kn0.12_pre}
	}
	\subfigure[Velocity]{
		\begin{minipage}[!t]{0.45\textwidth}
			\centering
			\includegraphics[width=1.0\textwidth]{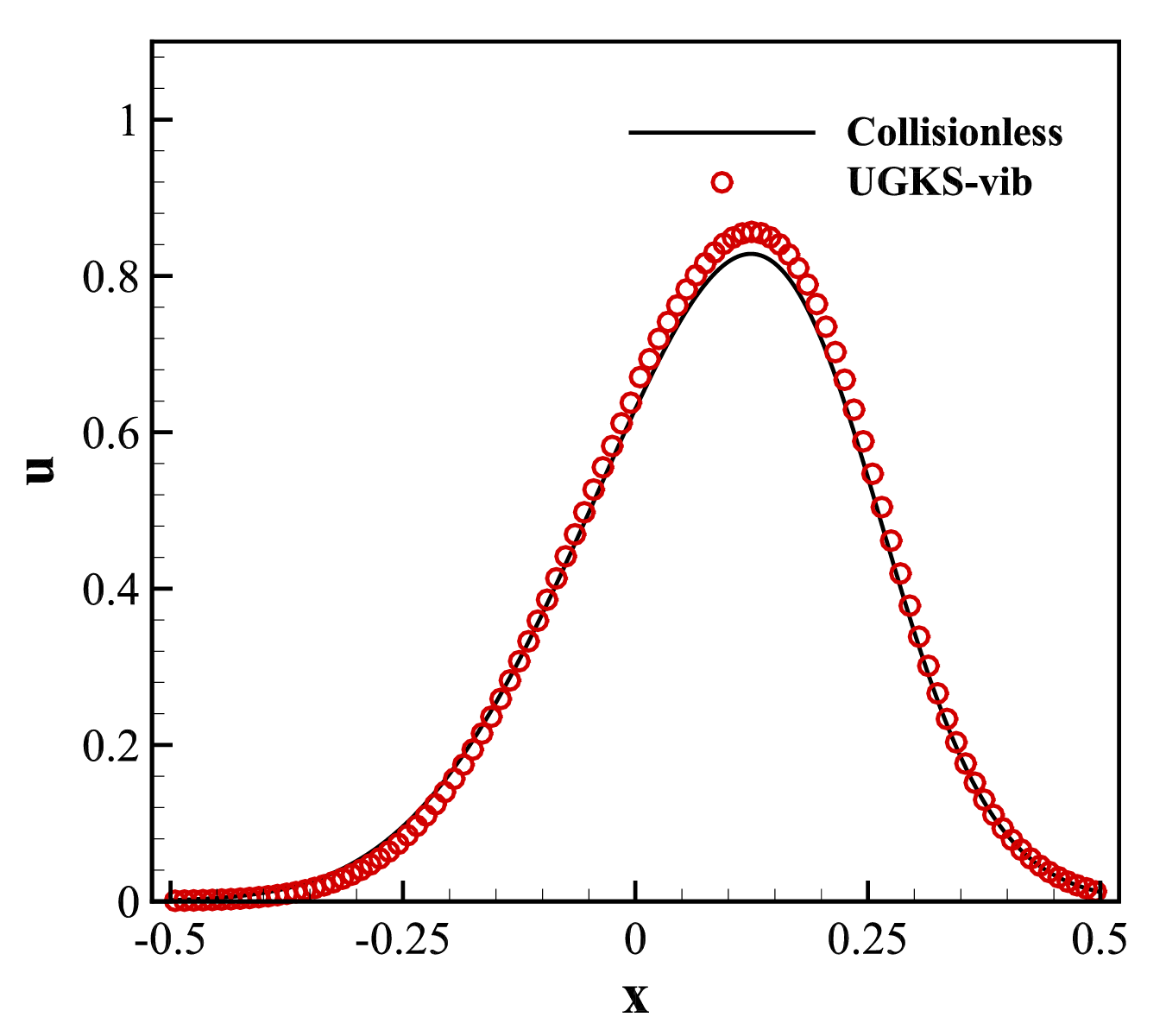}		
		\end{minipage}
		\label{fig_sod_kn0.12_vel}
	}
	\subfigure[Temperature]{
		\begin{minipage}[!t]{0.45\textwidth}
			\centering
			\includegraphics[width=1.0\textwidth]{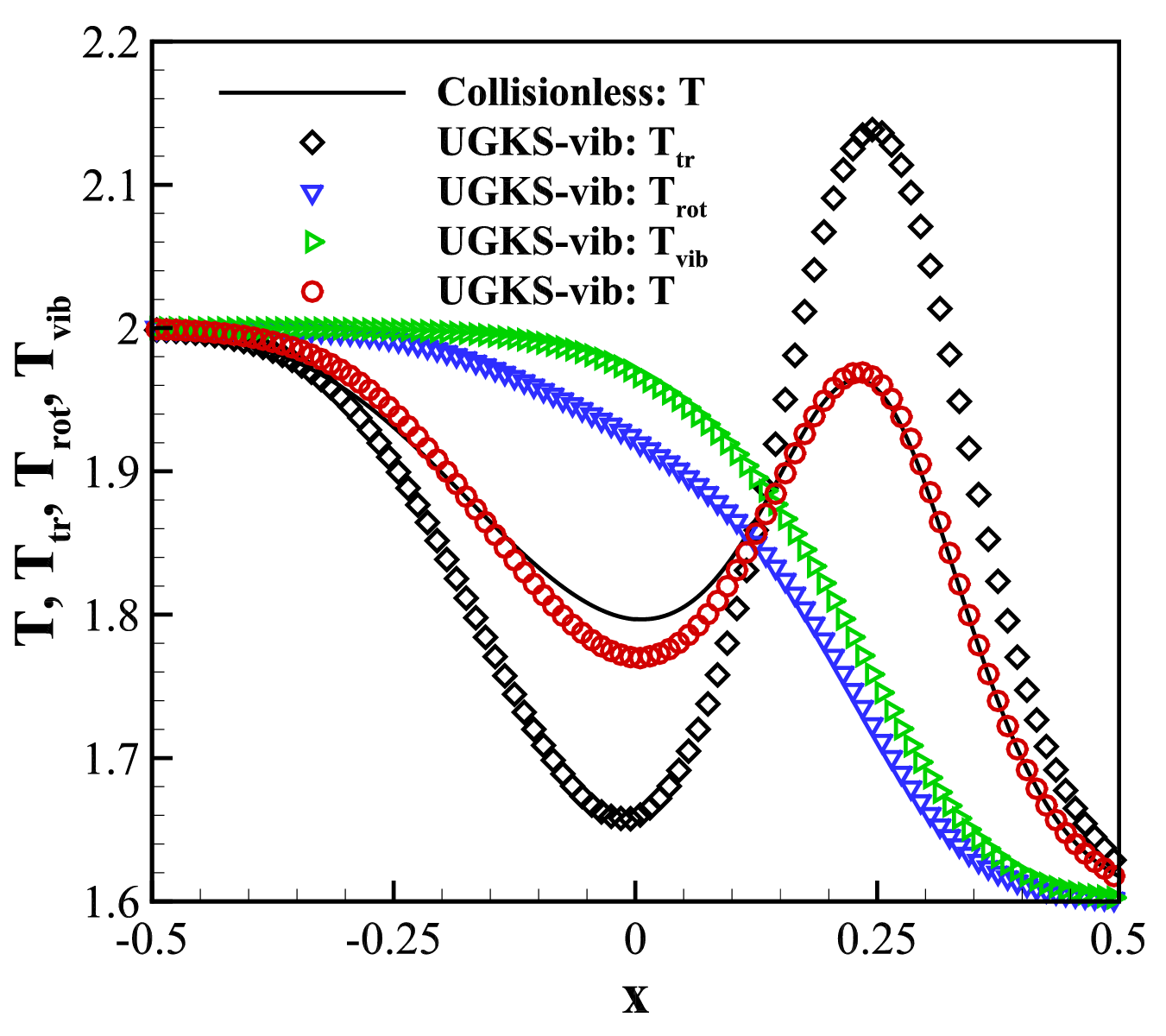}		
		\end{minipage}
		\label{fig_sod_kn0.12_tem}
	}
	\caption{The density, pressure, velocity and temperature profiles of the Sod’s shock tube at $\rm Kn=0.1277$.}
	\label{fig_sod_kn0.12}
\end{figure}
\begin{figure}[!t]
	\centering
	\subfigure[Density]{
		\begin{minipage}[!t]{0.45\textwidth}
			\centering
			\includegraphics[width=1.0\textwidth]{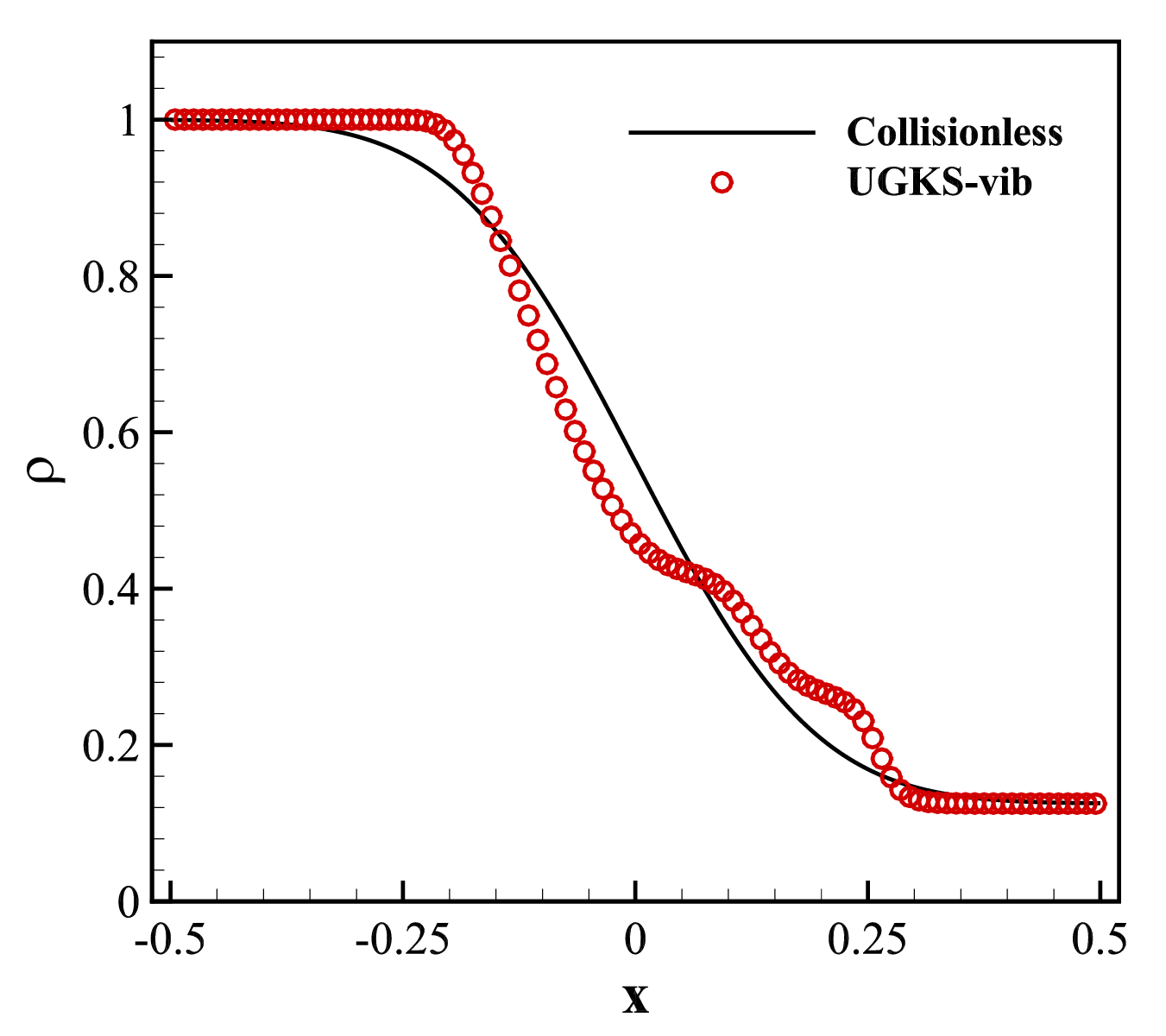}	    
		\end{minipage}
		\label{fig_sod_kn0.0012_rho}
	}
	\subfigure[Pressure]{
		\begin{minipage}[!t]{0.45\textwidth}
			\centering
			\includegraphics[width=1.0\textwidth]{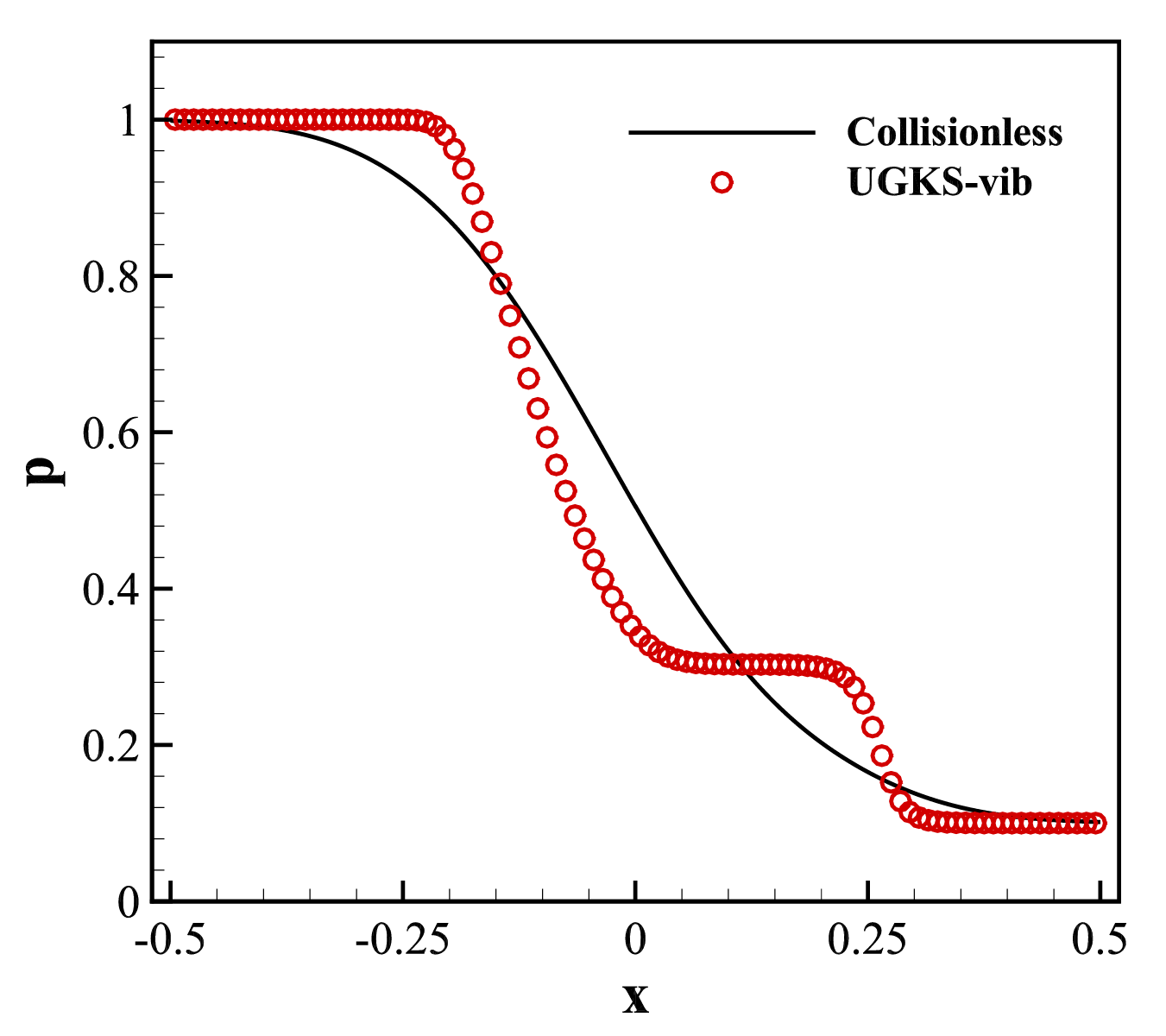}		
		\end{minipage}
		\label{fig_sod_kn0.0012_pre}
	}
	\subfigure[Velocity]{
		\begin{minipage}[!t]{0.45\textwidth}
			\centering
			\includegraphics[width=1.0\textwidth]{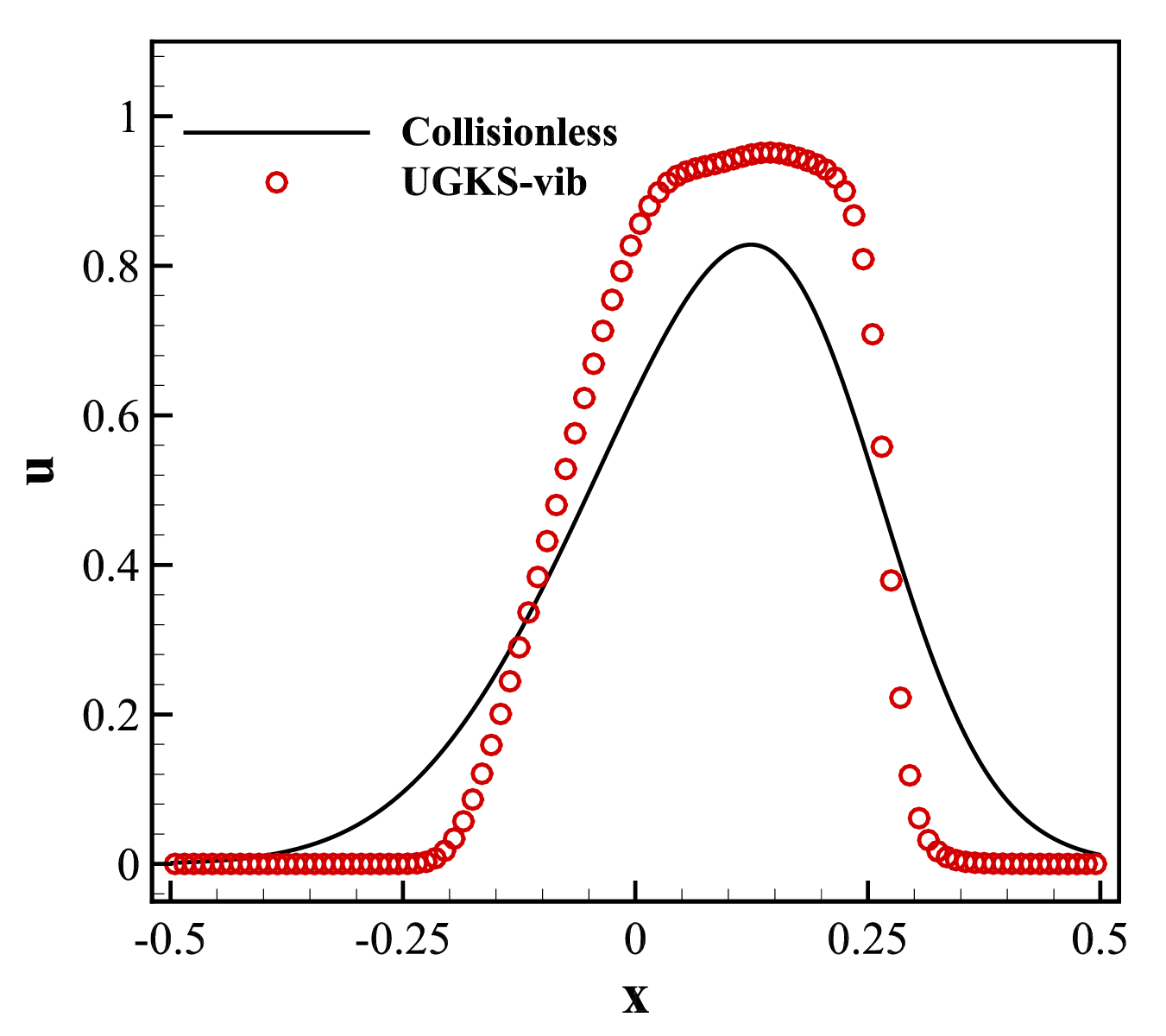}		
		\end{minipage}
		\label{fig_sod_kn0.0012_vel}
	}
	\subfigure[Temperature]{
		\begin{minipage}[!t]{0.45\textwidth}
			\centering
			\includegraphics[width=1.0\textwidth]{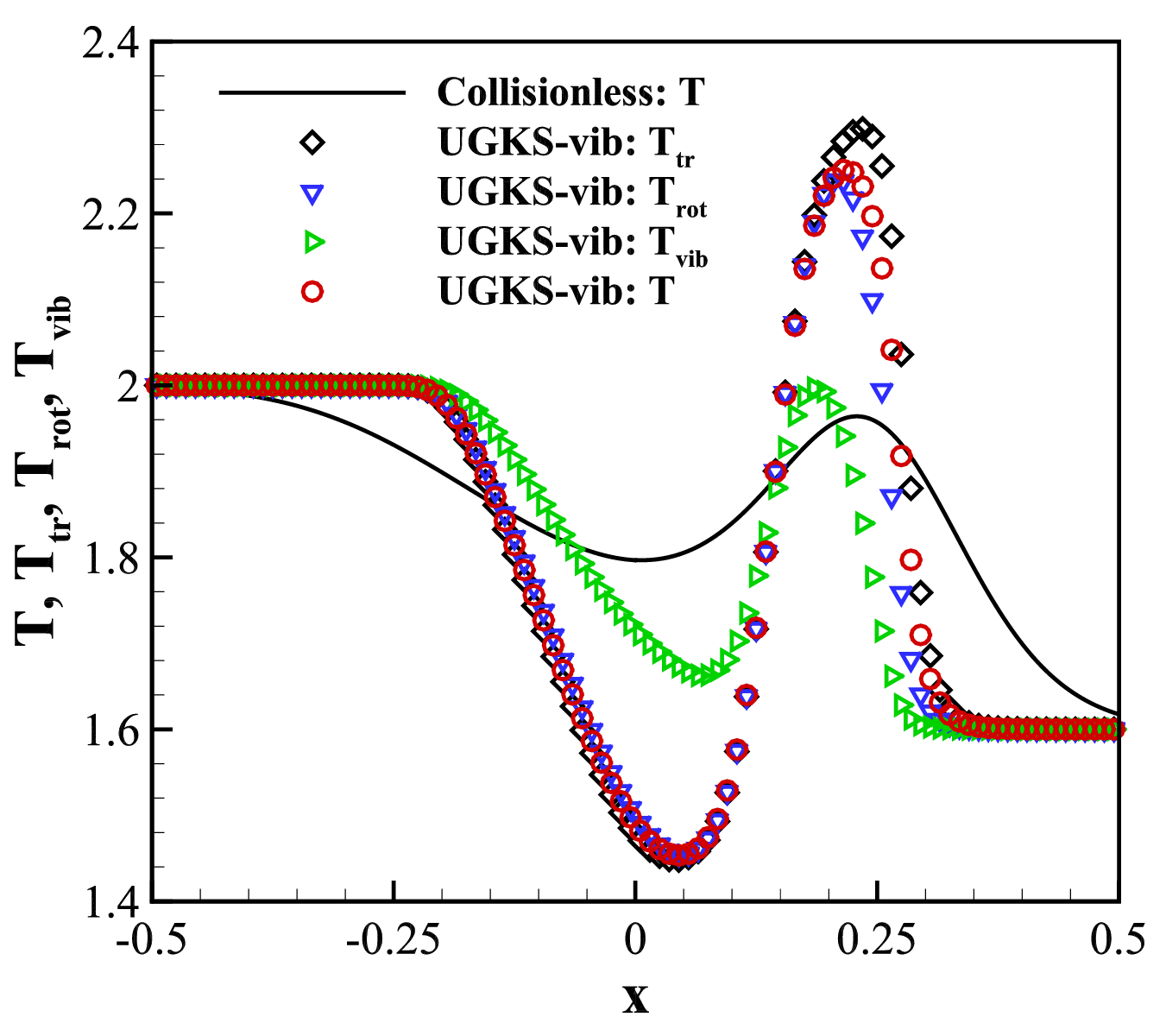}		
		\end{minipage}
		\label{fig_sod_kn0.0012_tem}
	}
	\caption{The density, pressure, velocity and temperature profiles of the Sod’s shock tube at $\rm Kn=1.277 \times 10^{-3}$.}
	\label{fig_sod_kn0.0012}
\end{figure}
\begin{figure}[!t]
	\centering
	\subfigure[Density]{
		\begin{minipage}[!t]{0.45\textwidth}
			\centering
			\includegraphics[width=1.0\textwidth]{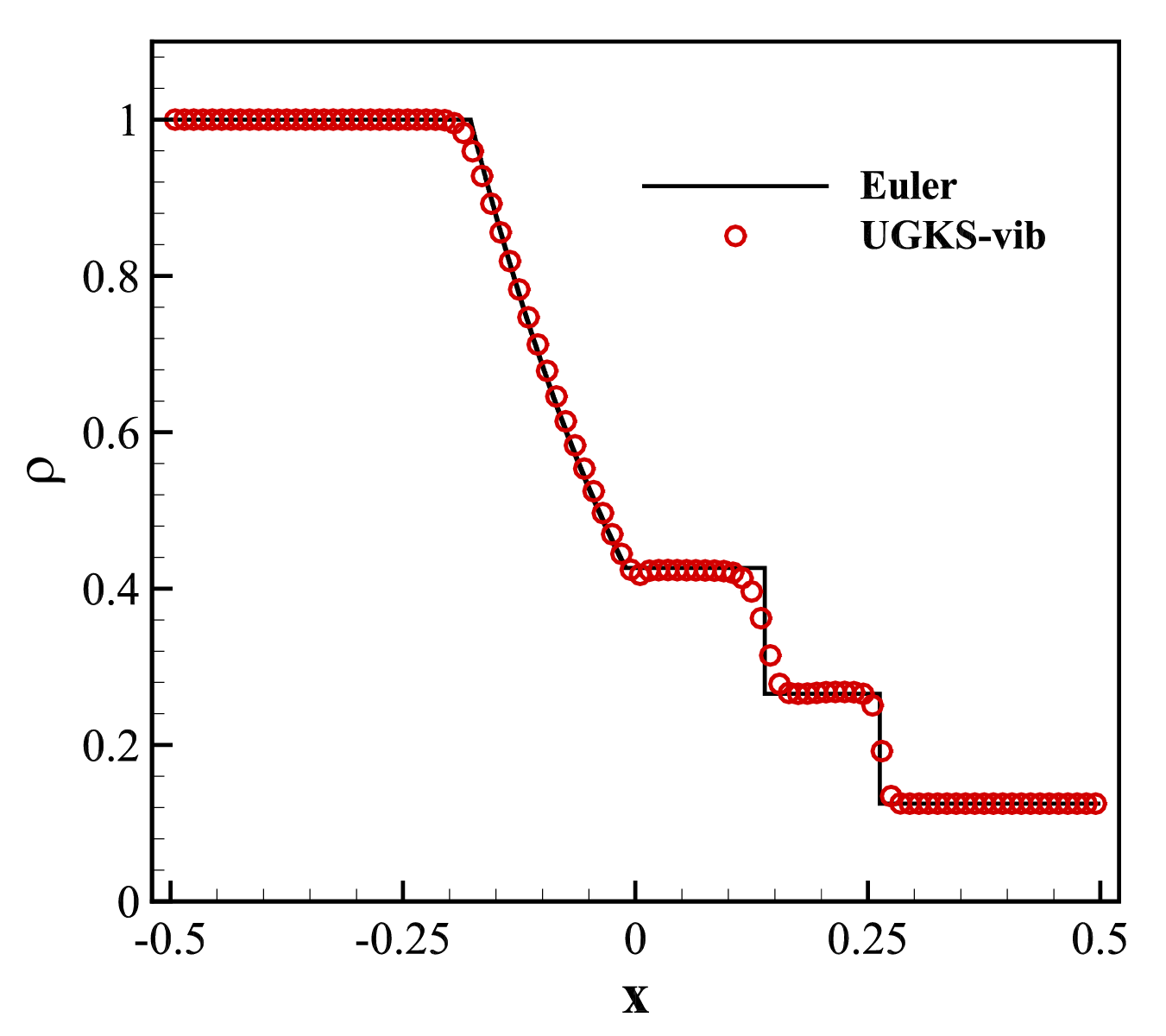}	    
		\end{minipage}
		\label{fig_sod_kn0.00_rho}
	}
	\subfigure[Pressure]{
		\begin{minipage}[!t]{0.45\textwidth}
			\centering
			\includegraphics[width=1.0\textwidth]{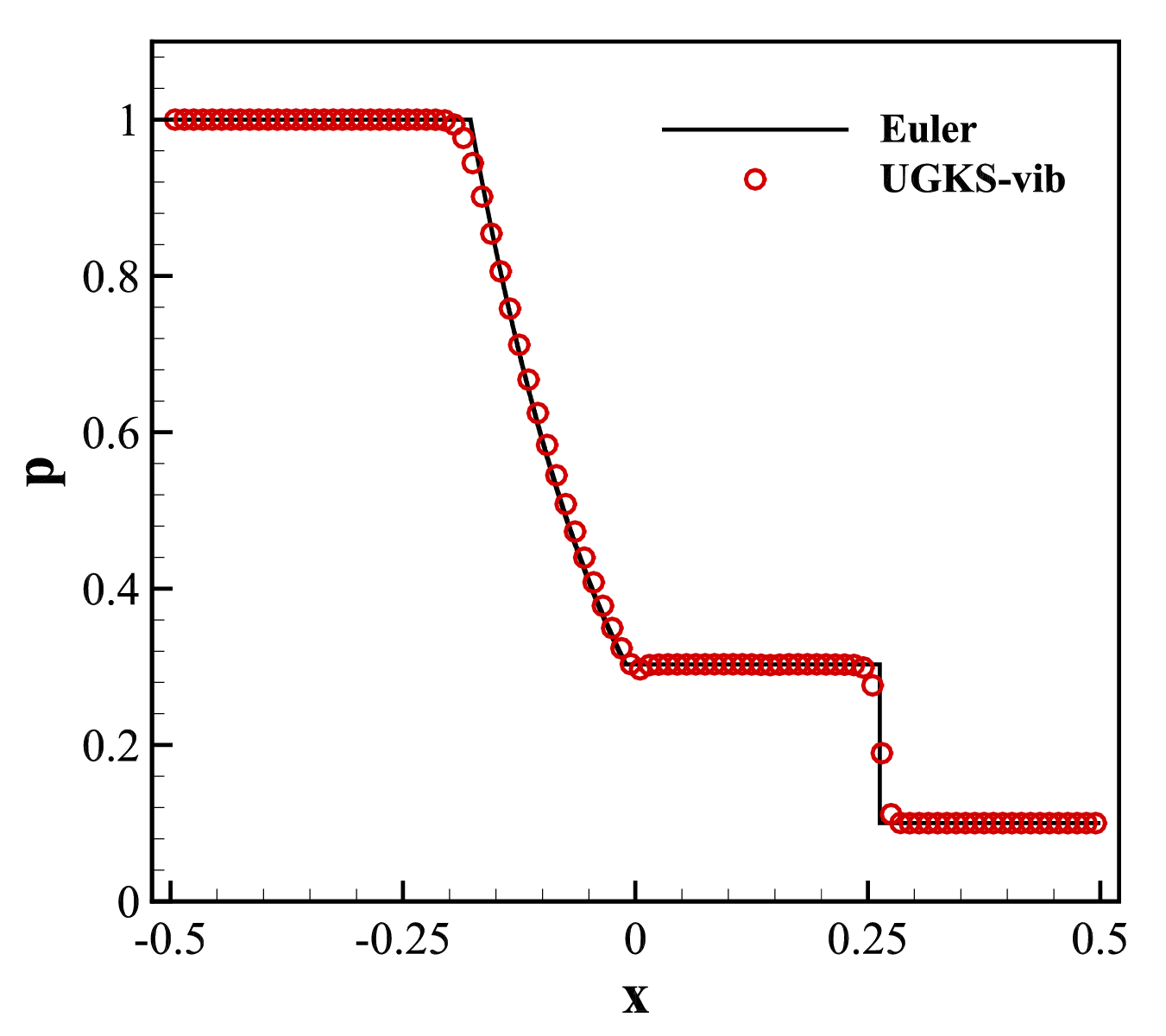}		
		\end{minipage}
		\label{fig_sod_kn0.00_pre}
	}
	\subfigure[Velocity]{
		\begin{minipage}[!t]{0.45\textwidth}
			\centering
			\includegraphics[width=1.0\textwidth]{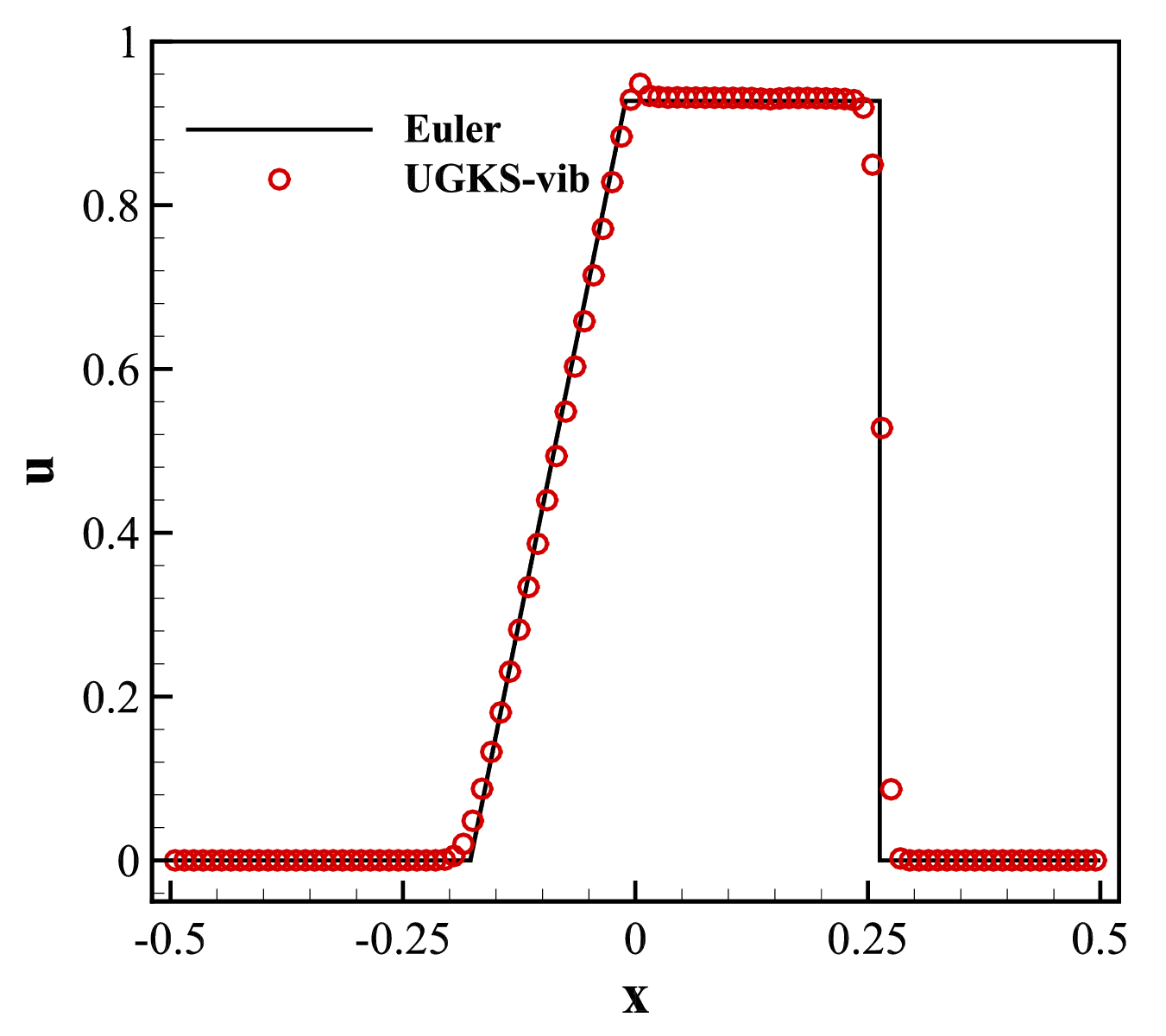}		
		\end{minipage}
		\label{fig_sod_kn0.00_vel}
	}
	\subfigure[Temperature]{
		\begin{minipage}[!t]{0.45\textwidth}
			\centering
			\includegraphics[width=1.0\textwidth]{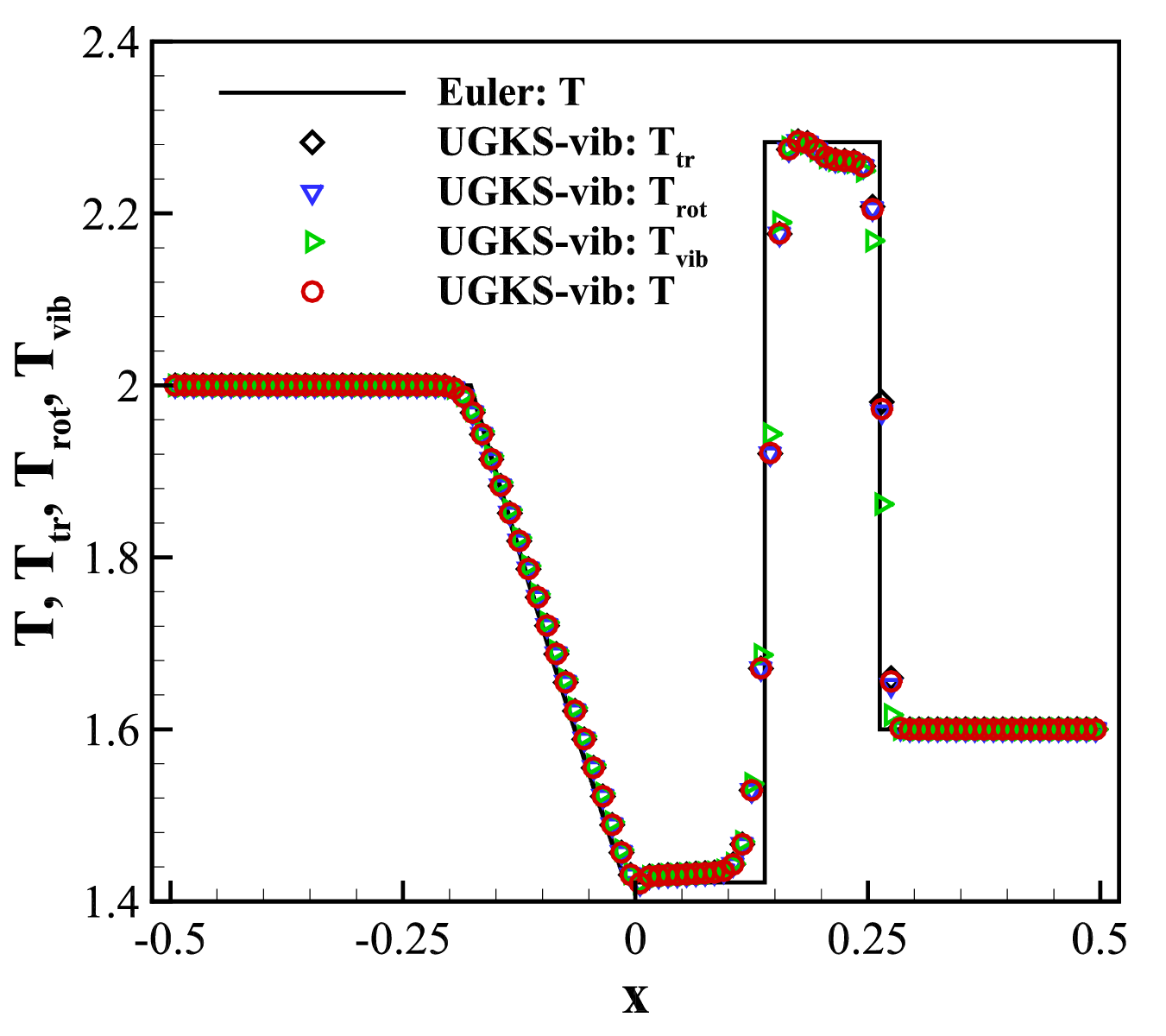}		
		\end{minipage}
		\label{fig_sod_kn0.00_tem}
	}
	\caption{The density, pressure, velocity and temperature profiles of the Sod’s shock tube at $\rm Kn=1.277 \times 10^{-5}$.}
	\label{fig_sod_kn0.00}
\end{figure}

The results of $\mu_1$ change from $10$, $0.1$, $1\times10^{-3}$ to $1\times10^{-5}$ at time $t=0.15$
are plotted in Fig.~\ref{fig_sod_kn12} to Fig.~\ref{fig_sod_kn0.00}. 
The solutions of the collisionless Boltzmann equation and the Euler equations are
also given in free-molecular regime and continuum regime, respectively. 
The density, velocity, pressure and equilibrium temperature curves of $\mu_1=10$ (Fig.~\ref{fig_sod_kn12})
predicted by the present method are in accord with the results of collisionless Boltzmann equation, 
while the results of $\mu_1=1\times10^{-5}$ (Fig.~\ref{fig_sod_kn0.00})
are agreement with those of Euler equations. 
\par
It should be noticed that in the continuum regime, there is enough time for system relaxing to the thermodynamic
equilibrium state because of the high collision frequency. Consequently, the translational, rotational and 
vibrational temperatures are almost the same for $\mu_1=1\times10^{-5}$ (Fig.~\ref{fig_sod_kn0.00_tem}). 
In the rarefied regime, the reduction of particle collisions enables more particles to move farther without 
collision, and propagate the initial information farther, thus showing the obvious thermodynamic 
non-equilibrium effect (since thermodynamic equilibrium needs sufficient collisions). The translational, rotational 
and vibrational temperatures are significantly different at $\mu_1=10$, $\mu_1=0.1$ and $\mu_1=1\times10^{-3}$ 
(Figs.~\ref{fig_sod_kn12_tem},~\ref{fig_sod_kn0.12_tem} and~\ref{fig_sod_kn0.0012_tem}).
\par

\subsection{Shock structure}\label{Shock_Structure}
\par
The planar shock structures are conducted to verify the capacity of the present method for simulating the 
highly non-equilibrium flows. Because the accuracy of the initial and downstream boundary conditions can have 
significant effects on the simulations, one natural requirement is to specify the post-shock equilibrium state.
\par
For diatomic molecules, the specific heat ratio $\gamma$ is variant across the shock wave 
because of the excitation of vibrational degrees of freedom under high temperature. The Rankine–Hugoniot 
relations with constant specific heat ratio $\gamma$ can not provide the correct post-shock state. Therefore, 
the generalized Rankine–Hugoniot relations~\citep{ZhaoWang_2017,ChunpeiCai_2008} assuming all temperature 
relaxation processes are completed (the temperatures of different inertial energies are the same) is used to 
determine the post-shock state, and the relations between pre-shock (denoted by subscript 1) and 
post-shock (denoted by subscript 2) states are as follows:
\begin{equation}\label{gRH_relation_pre}
	\frac{{{p_2}}}{{{p_1}}} = \frac{{1 + {\gamma _1}M_1^2}}{{1 + {\gamma _2}M_2^2}},
\end{equation}
\begin{equation}\label{gRH_relation_tem}
	\frac{{{T_2}}}{{{T_1}}} = \frac{{\left[ {{{{\gamma _1}} \mathord{\left/
	{\vphantom {{{\gamma _1}} {\left( {{\gamma _1} - 1} \right)}}} \right.
	\kern-\nulldelimiterspace} {\left( {{\gamma _1} - 1} \right)}}} \right] + \left( {{{{\gamma _1}} 
					\mathord{\left/{\vphantom {{{\gamma _1}} 2}} \right.
	\kern-\nulldelimiterspace} 2}} \right)M_1^2}}{{\left[ {{{{\gamma _2}} \mathord{\left/
			{\vphantom {{{\gamma _2}} {\left( {{\gamma _2} - 1} \right)}}} \right.
	\kern-\nulldelimiterspace} {\left( {{\gamma _2} - 1} \right)}}} \right] + \left( {{{{\gamma _2}} 
	\mathord{\left/{\vphantom {{{\gamma _2}} 2}} \right.\kern-\nulldelimiterspace} 2}} \right)M_2^2}},
\end{equation}
\begin{equation}\label{gRH_relation_vel}
	\frac{{{u_2}}}{{{u_1}}} = \sqrt {\frac{{{\gamma _2}}}{{{\gamma _1}}}} \frac{{{M_2}}}{{{M_1}}}\sqrt 
	{\frac{{\left[ {{{{\gamma 
		_1}} \mathord{\left/{\vphantom {{{\gamma _1}} {\left( {{\gamma _1} - 1} \right)}}} \right.
	\kern-\nulldelimiterspace} {\left( {{\gamma _1} - 1} \right)}}} \right] + \left( {{{{\gamma _1}} 
		\mathord{\left/{\vphantom {{{\gamma _1}} 2}} \right.
		\kern-\nulldelimiterspace} 2}} \right)M_1^2}}{{\left[ {{{{\gamma _2}} \mathord{\left/
		{\vphantom {{{\gamma _2}} {\left( {{\gamma _2} - 1} \right)}}} \right.
		\kern-\nulldelimiterspace} {\left( {{\gamma _2} - 1} \right)}}} \right] + \left( {{{{\gamma _2}} 
		\mathord{\left/	{\vphantom {{{\gamma _2}} 2}} \right.
		\kern-\nulldelimiterspace} 2}} \right)M_2^2}}},
\end{equation}
\begin{equation}\label{gRH_relation_mach}
	\frac{{{{\left( {1 + {\gamma _1}M_1^2} \right)}^2}}}{{\left\{ {\left[ {{{{\gamma _1}} \mathord{\left/
	{\vphantom {{{\gamma _1}} {\left( {{\gamma _1} - 1} \right)}}} \right.
			\kern-\nulldelimiterspace} {\left( {{\gamma _1} - 1} \right)}}} \right] + \left( {{{{\gamma _1}} 
			\mathord{\left/{\vphantom {{{\gamma _1}} 2}} \right.
		\kern-\nulldelimiterspace} 2}} \right)M_1^2} \right\}{\gamma _1}M_1^2}} = \frac{{{{\left( {1 + {\gamma 
		_2}M_2^2} \right)}^2}}}{{\left\{ {\left[ {{{{\gamma _2}} \mathord{\left/
		{\vphantom {{{\gamma _2}} {\left( {{\gamma _2} - 1} \right)}}} \right.
			\kern-\nulldelimiterspace} {\left( {{\gamma _2} - 1} \right)}}} \right] + \left( {{{{\gamma _2}} 
		\mathord{\left/{\vphantom {{{\gamma _2}} 2}} \right.
					\kern-\nulldelimiterspace} 2}} \right)M_2^2} \right\}{\gamma _2}M_2^2}}.	
\end{equation}
In addition, the specific heat ratio is determined as follows:
\begin{equation}\label{specific_heat_ratio}
	\gamma = \frac{{{K_{tr}}+{K_{rot}}+{K_{vib}}(T)+2}}{{{K_{tr}}+{K_{rot}}+{K_{vib}}(T)}}.
\end{equation}
\par
Noted that the parameters of post-shock can not be expressed explicitly using the parameters of pre-shock,
so the following iteration process is used to obtain the post-shock state.\\
\textbf{(a)} Calculate the $M_2$ according to Eq.~\eqref{gRH_relation_mach} with a specified 
$\gamma _2^{(n)}$;\\
\textbf{(b)} Calculate the temperature $T_2$ according to Eq.~\eqref{gRH_relation_tem} using 
$M_2$ and $\gamma _2^{(n)}$;\\
\textbf{(c)} Solve the $\gamma _2^{(n+1)}$ using Eq.~\eqref{specific_heat_ratio}.
\par
We performed two numerical simulations of nitrogen shock structure
with $\rm Ma=10$ and $\rm Ma=15$. The viscosity in the present simulation 
is calculated with the VHS model ($\omega = 0.74$). 
The pre-shock density $\rho_1 = 1.7413 \times 10^{-2}kg/m^3$, 
temperature $T_1 = 226.149K$ are same for those two cases. In the calculations, the dimensionless quantities 
are used. The reference length, density, temperature, and velocity are set as 
$L_{ref}=\lambda_1$, $\rho_{ref}=\rho_1$, $T_{ref}=T_1$ and $U_{ref}=\sqrt{2RT_{ref}}$, 
where the mean free path $\lambda_1$ is computed by Eq.~\eqref{mfp_mu_T}. Corresponding, the dimensionless 
parameters of pre-shock equilibrium state and post-shock equilibrium state are shown 
in Table~\ref{tbl:ss_parameters}.
\par
The computational domain is set as $\left[{-100{\lambda _1}}, {100{\lambda _1}}\right]$, 
and a uniform mesh with 400 cells is used so that the mesh space is $\Delta x = 0.5{\lambda _1}$.
The uniform discrete velocity space is determined by the Newton-Cotes quadrature with 301 points distributed 
in $\left[{-21,21}\right]$ for $\rm Ma=10$ and 401 points distributed in $\left[{-30,30}\right]$ for $\rm Ma=15$. 
The rotational and vibrational collision numbers are $Z_{rot}=4$ and $Z_{vib}=50$ for $\rm Ma=10$, 
and $Z_{rot}=5$ and $Z_{vib}=25$ for $\rm Ma=15$ to be consistent with the relaxation rate in 
Ref~\citep{ChunpeiCai_2008}.
\par
\begin{figure}[!t]
	\centering
	\subfigure[Density]{
		\begin{minipage}[!t]{0.45\textwidth}
			\centering
			\includegraphics[width=1.0\textwidth]{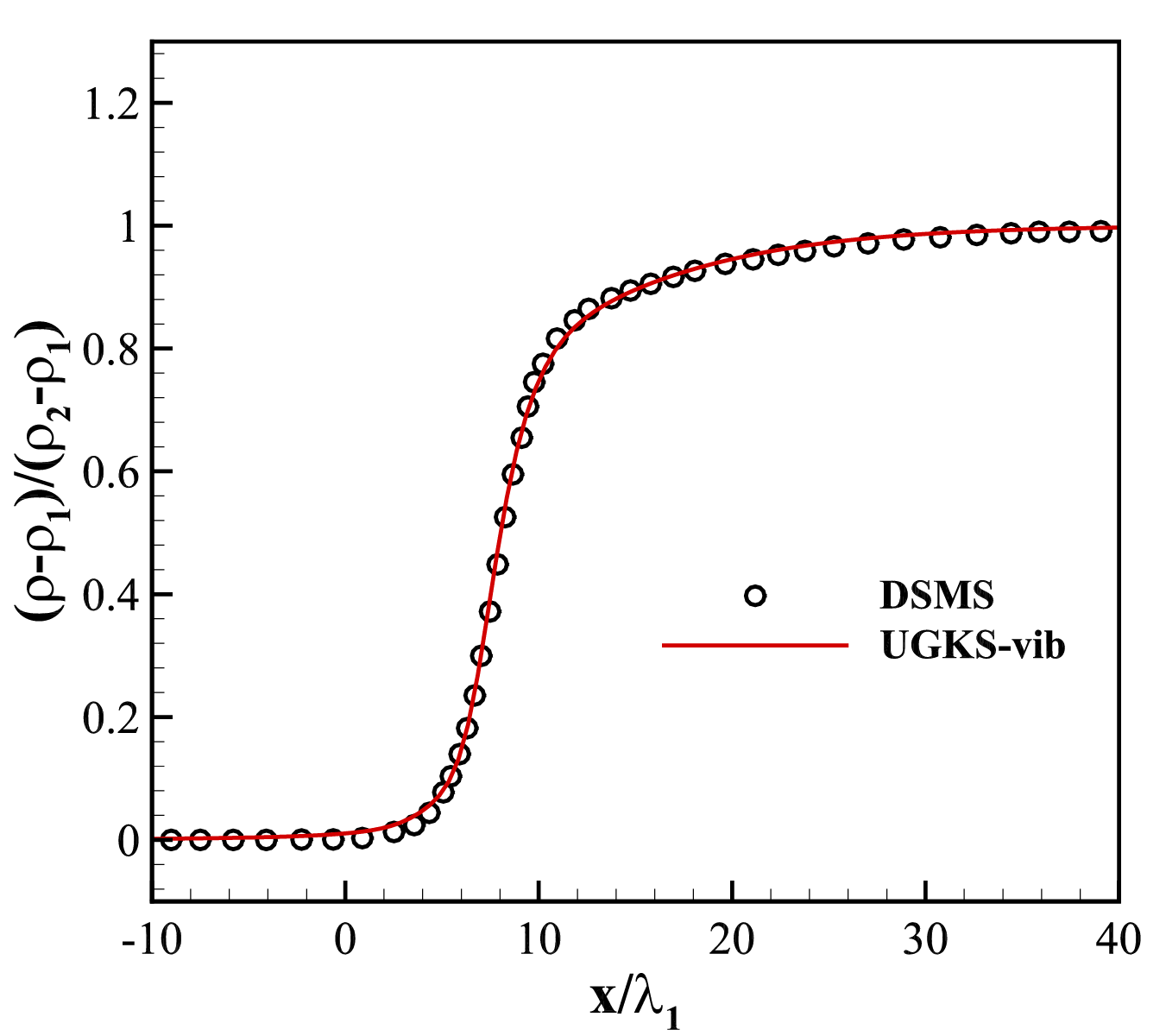}	    
		\end{minipage}
		\label{fig_ss_ma10_rho}
	}
	\subfigure[Normal translational temperature and parallel translational temperature]{
		\begin{minipage}[!t]{0.45\textwidth}
			\centering
			\includegraphics[width=1.0\textwidth]{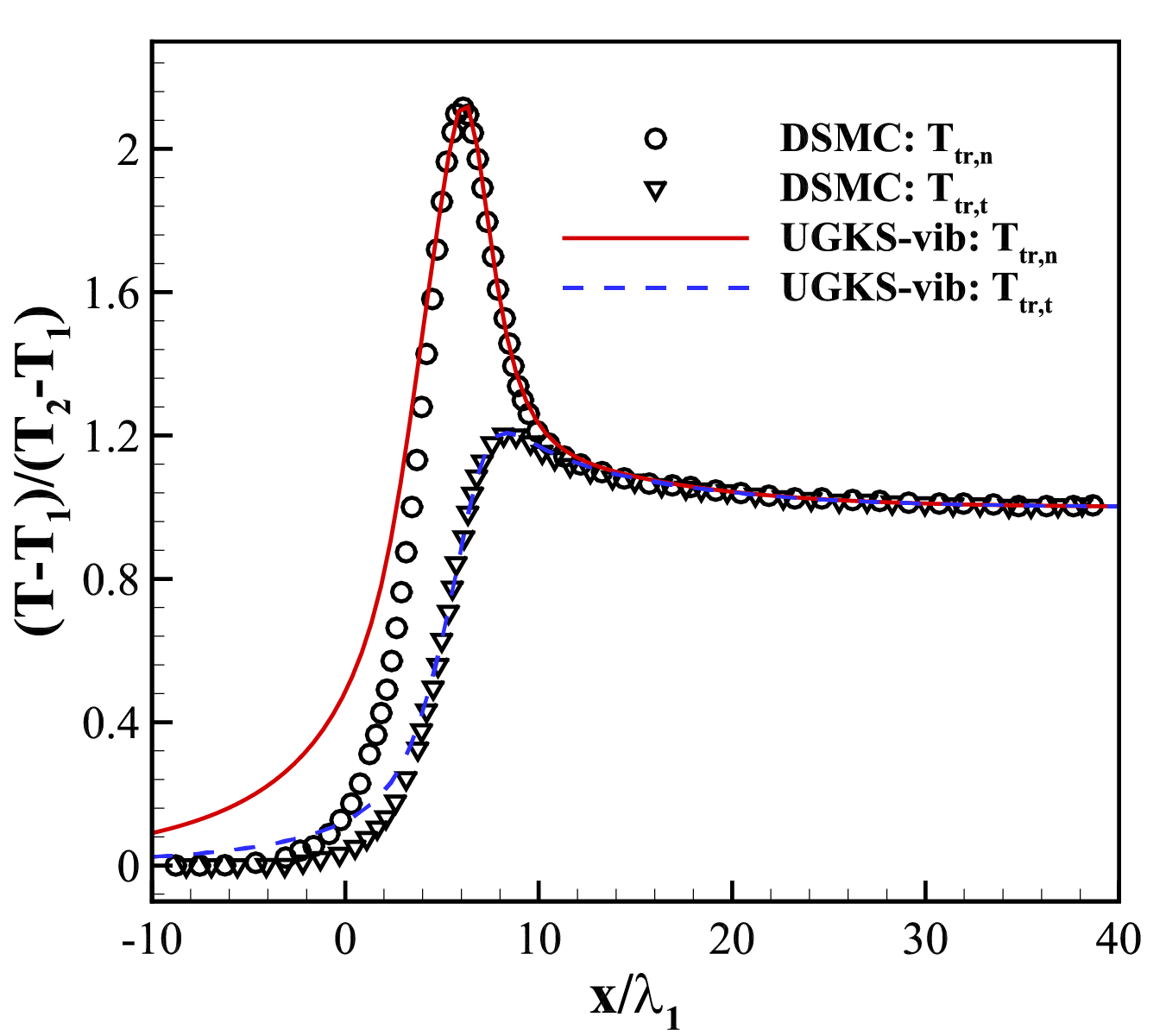}		
		\end{minipage}
		\label{fig_ss_ma10_temtra}
	}
	\subfigure[Rotational temperature and vibrational temperature]{
		\begin{minipage}[!t]{0.45\textwidth}
			\centering
			\includegraphics[width=1.0\textwidth]{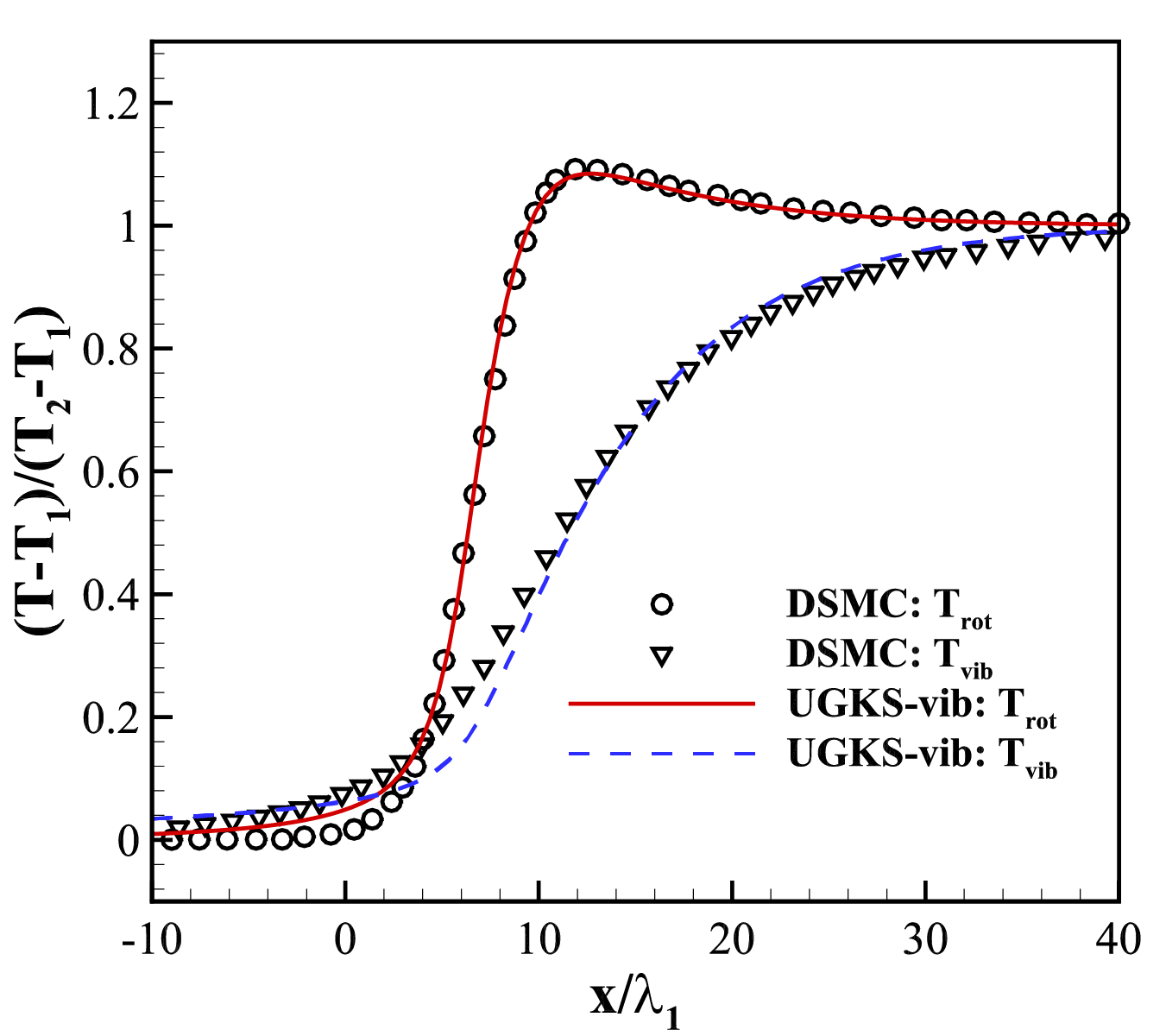}		
		\end{minipage}
		\label{fig_ss_ma10_temrot_temvib}
	}
	\subfigure[Vibrational degrees of freedom and specific heat ratio $\gamma$]{
		\begin{minipage}[!t]{0.45\textwidth}
			\centering
			\includegraphics[width=1.0\textwidth]{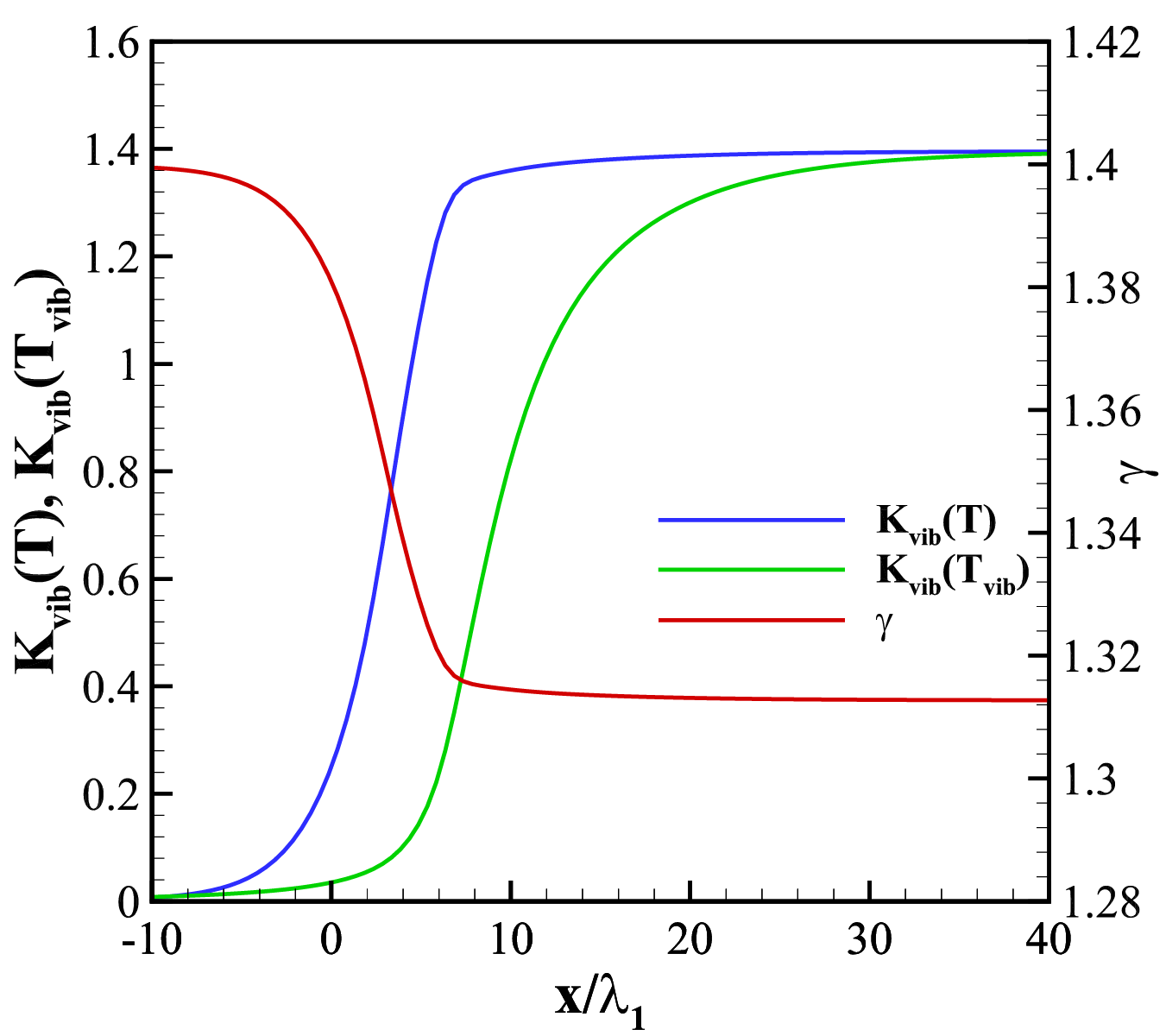}		
		\end{minipage}
		\label{fig_ss_ma10_lambda_kvib}
	}
	\caption{Nitrogen gas shock structure of variable hard sphere molecule at $\rm Ma=10$.}
	\label{fig_ss_ma10}
\end{figure}
\begin{figure}[!t]
	\centering
	\subfigure[Density]{
		\begin{minipage}[!t]{0.45\textwidth}
			\centering
			\includegraphics[width=1.0\textwidth]{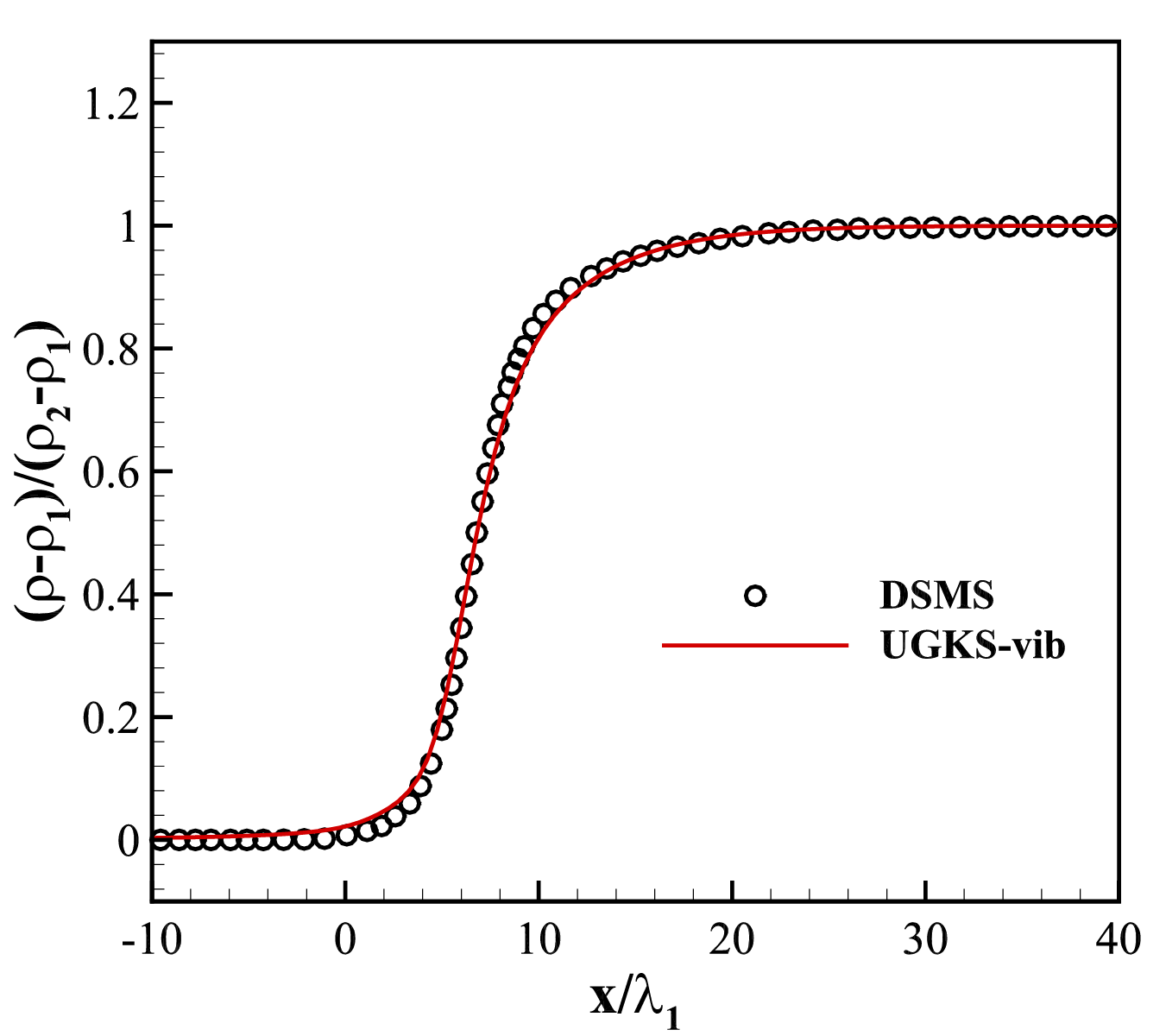}	    
		\end{minipage}
		\label{fig_ss_ma15_rho}
	}
	\subfigure[Normal translational temperature and parallel translational temperature]{
		\begin{minipage}[!t]{0.45\textwidth}
			\centering
			\includegraphics[width=1.0\textwidth]{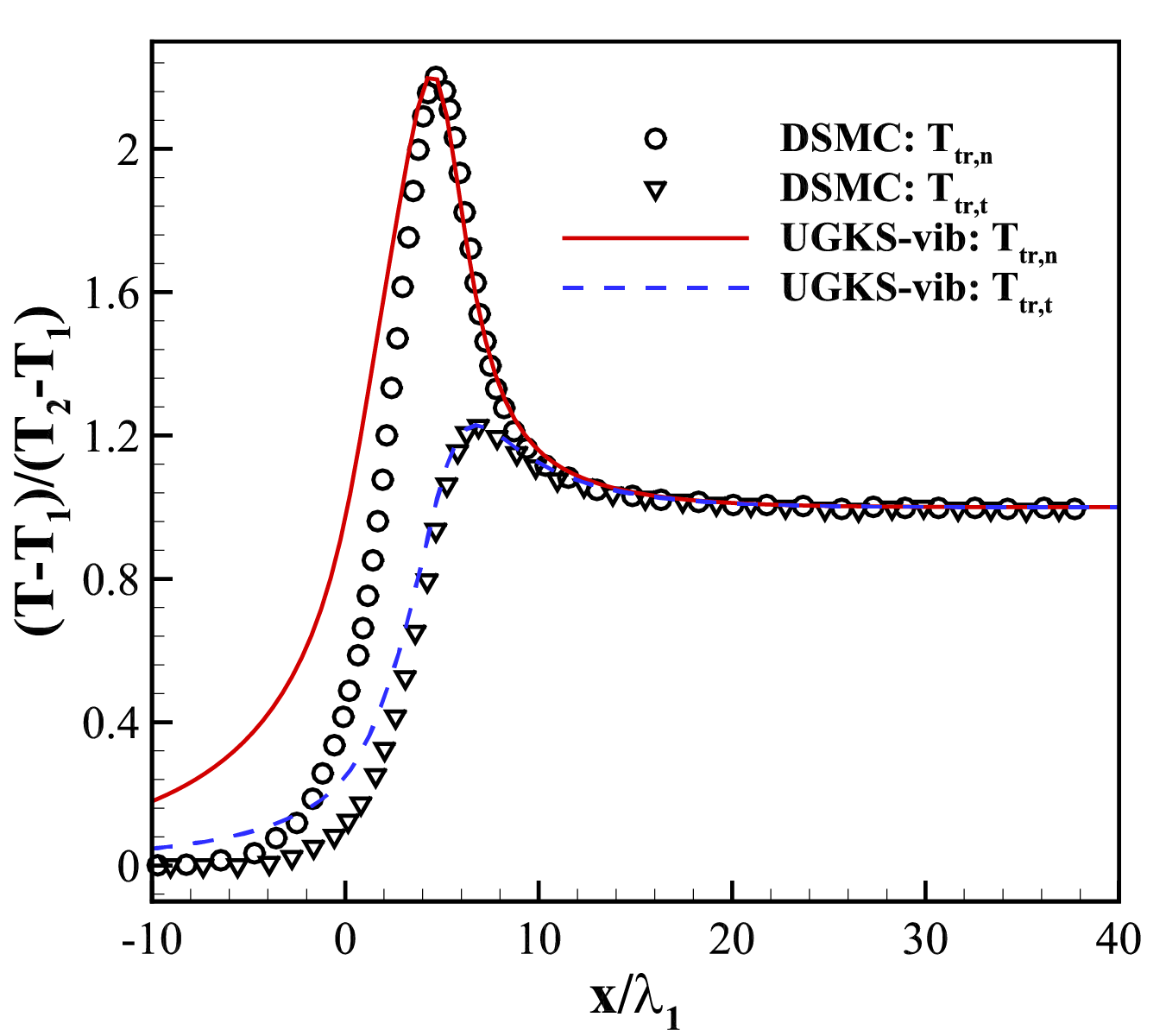}		
		\end{minipage}
		\label{fig_ss_ma15_temtra}
	}
	\subfigure[Rotational temperature and vibrational temperature]{
		\begin{minipage}[!t]{0.45\textwidth}
			\centering
			\includegraphics[width=1.0\textwidth]{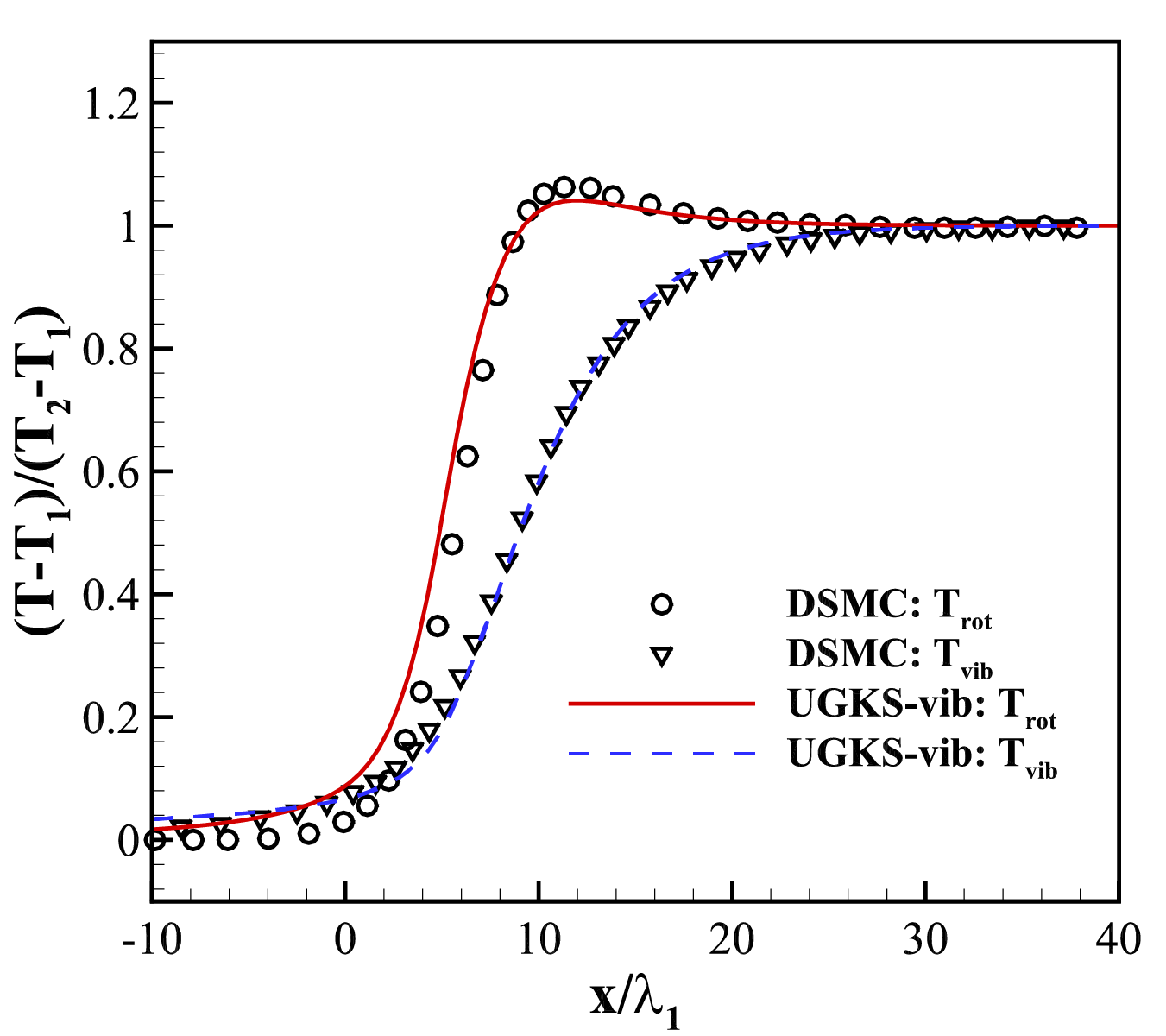}		
		\end{minipage}
		\label{fig_ss_ma15_temrot_temvib}
	}
	\subfigure[Vibrational degrees of freedom and specific heat ratio $\gamma$]{
		\begin{minipage}[!t]{0.45\textwidth}
			\centering
			\includegraphics[width=1.0\textwidth]{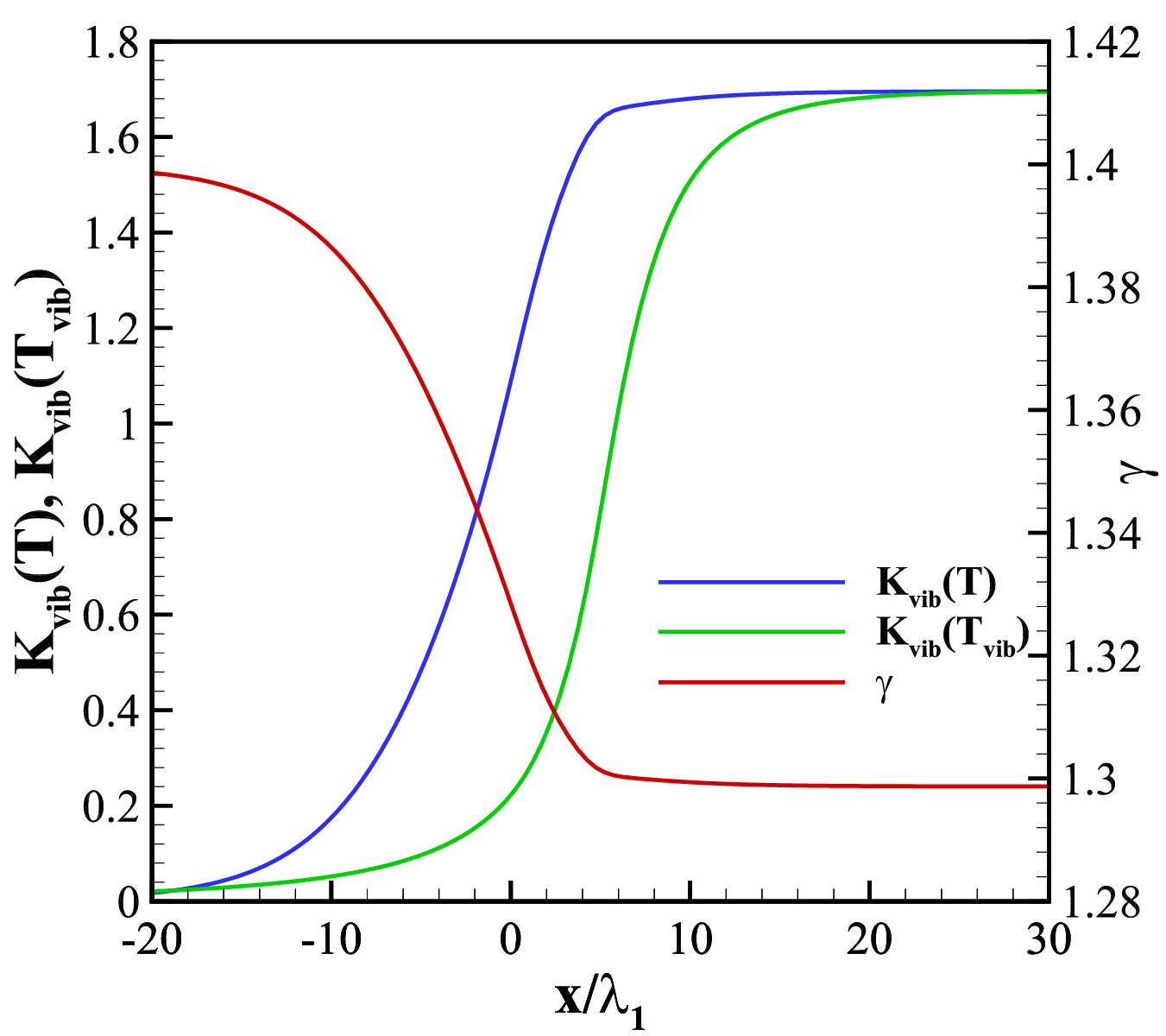}		
		\end{minipage}
		\label{fig_ss_ma15_lambda_kvib}
	}
	\caption{Nitrogen gas shock structure of variable hard sphere molecule at $\rm Ma=15$.}
	\label{fig_ss_ma15}
\end{figure}

Fig.~\ref{fig_ss_ma10} and Fig.~\ref{fig_ss_ma15} illustrate the comparisons of the density and temperature 
distributions between the present and DSMC’s results~\citep{ChunpeiCai_2008} for $\rm Ma=10$ and $\rm Ma=15$, 
respectively. Besides, the distributions of vibrational degrees of freedom and specific heat ratio $\gamma$
are also plotted in Fig.~\ref{fig_ss_ma10_lambda_kvib} ($\rm Ma=10$)
and Fig.~\ref{fig_ss_ma15_lambda_kvib} ($\rm Ma=15$).  
It can be noted that the density distributions of the present results for $\rm Ma=10$ and $\rm Ma=15$
are both agreement with the results of DSMC. 
The present temperature distributions also agree well with the DSMC’s results in the downstream field,
while the present translational and rotational temperatures are generally higher than the DSMC’s
results in the upstream field. This may due to the fact that the relaxation time is independent of the 
molecular velocity in the BGK-type model equations, which lead to overestimation of the temperature in the 
upstream flow~\citep{RuifengYuan_2020,ChangLiu_2016}. Recently, a modification of the relaxation time 
according to the particle velocity have been implemented in the unified gas-kinetic wave-particle (UGKWP) 
method to fix this problem~\citep{XiaocongXu_2021}, which can be extended to the present method.
\par

\subsection{Hypersonic flow around a circular cylinder}\label{Cylinder}
\par  
In the case of high-speed molecular gas flows past bodies in a rarefied gas environment, 
in addition to the translational non-equilibrium such as the bimodal distributions 
and prominent surface slip phenomena, the thermodynamic non-equilibrium related to the difference between 
the translational, rotational and vibrational temperatures both in the shock wave zone and in the boundary 
layer will arise. Therefore, in order to verify the reliability of the present UGKS with simplified multi-scale 
numerical flux involving internal molecular energies to simulate such high-speed flows, the hypersonic flow 
around a circular cylinder is implemented and the numerical results are compared with those from 
DSMC computed by the DS2V software~\citep{JunLinWu_2021}. 
\par
In the present study, the two numerical simulations of hypersonic nitrogen gas flows around a 
circular cylinder with the same free-stream Knudsen number ($\rm Kn_{\infty}=0.01$), 
but different free-stream Mach number ($\rm Ma_{\infty}=5$ and $\rm Ma_{\infty}=20$) are performed.   
The radius of the cylinder is $R_c = 0.5m$. The free-stream density and temperature 
are $\rho_{\infty}=6.9592\times10^{-6}kg/m^3$ and $T_{\infty}=500 K$, 
respectively. The VHS molecular model with $\omega = 0.74$ is applied and the free-stream
mean free path $\lambda_{mfp,\infty}$ is around $0.01m$. 
The hot wall boundary condition is chosen, the temperatures of the wall surface are $T_w=500 K$ for 
$\rm Ma_{\infty}=5$ and $T_w=2000 K$ for $\rm Ma_{\infty}=20$. The Parker formula~\citep{JG_Parker_1959} is 
used to calculate the rotational collision number, and the vibrational collision numbers are $Z_{vib}=30$ 
for $\rm Ma_{\infty}=5$ and $Z_{vib}=35$ for $\rm Ma_{\infty}=20$, respectively.
\par
In the calculations, the dimensionless quantities normalized by the reference length $L_{ref}=2R_c$, 
density $\rho_{ref}=\rho_{\infty}$, temperature $T_{ref}=T_{\infty}$, and 
velocity $U_{ref}=\sqrt{2RT_{ref}}$ are introduced. Corresponding, the dimensionless gas flow parameters at 
infinity and the wall surface temperature of those two cases are shown in Table~\ref{tb2:cylinder_parameters}. 
For the boundary conditions, the diffuse reflection boundary condition with full thermodynamic
accommodation at the wall surface along with the inlet, outlet boundaries discussed in 
subsection~\ref{Boundary_Conditions} are applied.
\par
\begin{figure}[!t]
	\centering
	\subfigure[Physical space mesh ($180 \times 88$ cells)]{
		\begin{minipage}[!t]{0.45\textwidth}
			\centering
			\includegraphics[width=1.0\textwidth]{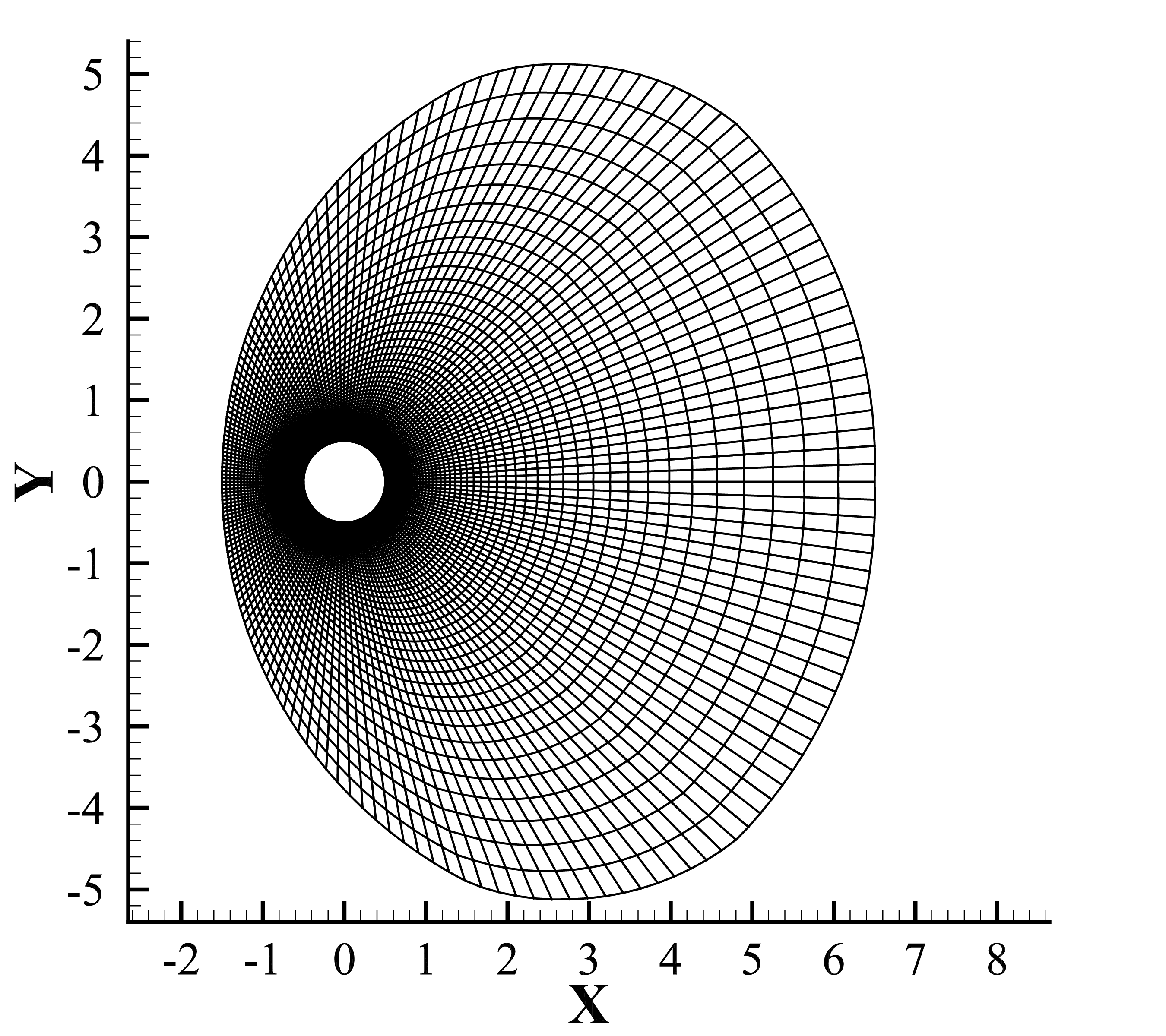}	    
		\end{minipage}
		\label{fig_cylinder_ma5_macmesh}
	}
	\subfigure[Unstructured discrete velocity space mesh ($1931$ cells)]{
		\begin{minipage}[!t]{0.45\textwidth}
			\centering
			\includegraphics[width=1.0\textwidth]{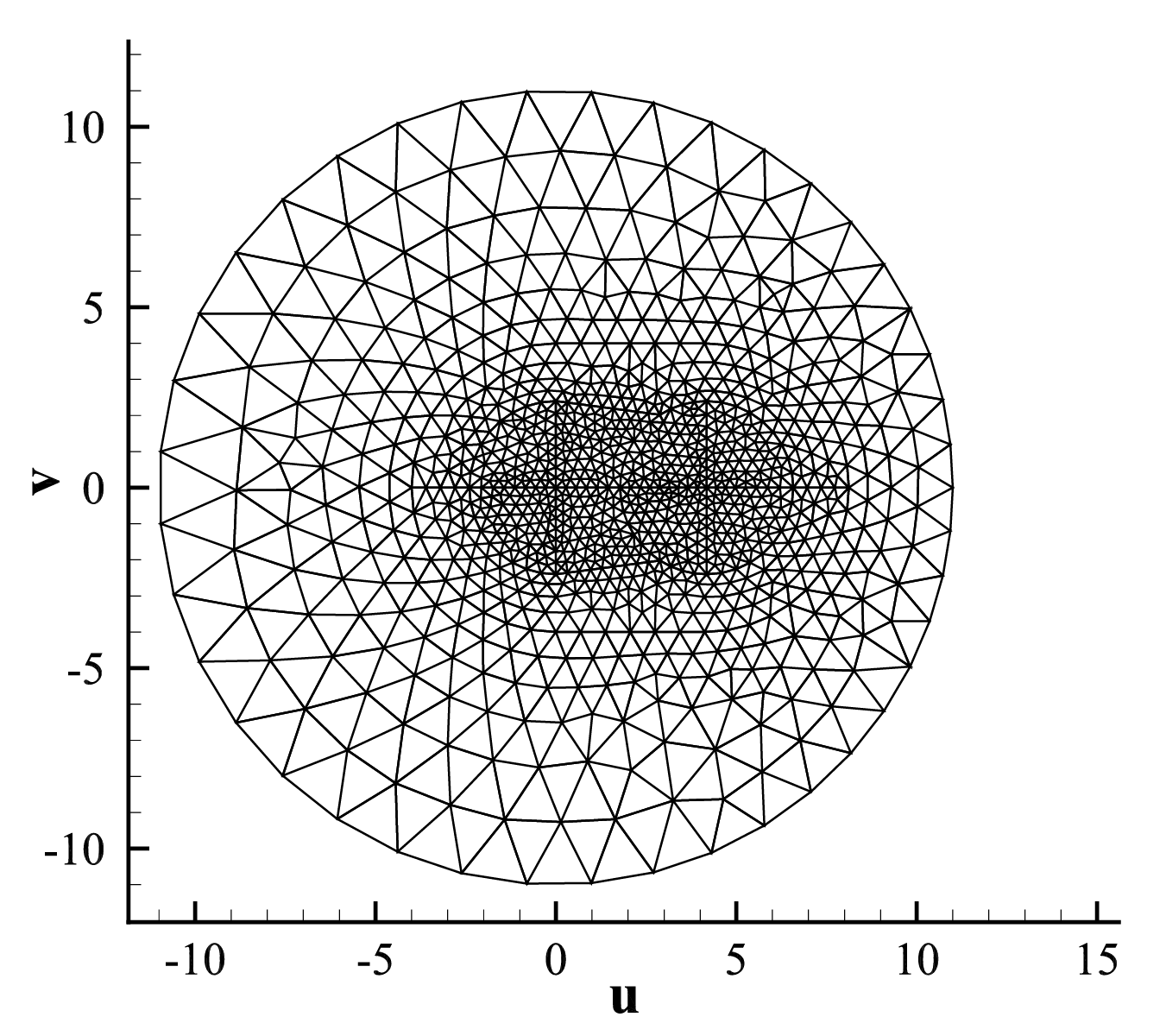}		
		\end{minipage}
		\label{fig_cylinder_ma5_micmesh}
	}
	\caption{The physical space mesh and unstructured discrete velocity space mesh for the cylinder at $\rm Ma=5$.}
	\label{fig_cylinder_ma5_mesh}
\end{figure}

For the case of $\rm Ma=5$, the 15840 elements are used in the physical space with $180 \times 88$ mesh cells, 
as shown in Fig.~\ref{fig_cylinder_ma5_macmesh}, in which the height of the first layer 
near the wall is approximately 0.001. A 1931-cell unstructured DVS is adopted, 
as shown in Fig.~\ref{fig_cylinder_ma5_micmesh}. For the unstructured DVS, the discrete velocity around the 
point $(0,0)$ (solid wall) and $(4.1814,0)$ (the free-stream velocity is 4.1814) are refined 
because the temperatures of the wall surface and free-stream are the lowest throughout the flow field. 
The discrete velocity space range is approximately set as a circle of radius $6\sqrt{RT_0}$, where the 
total temperature $T_0$ is estimated as $T_0 = {[{1+{Ma_{\infty}^2}({\gamma} -1)/2}]{T_{\infty}}}$.
\par
\begin{figure}[!t]
	\centering
	\subfigure[Pressure]{
		\begin{minipage}[!t]{0.45\textwidth}
			\centering
			\includegraphics[width=1.0\textwidth]{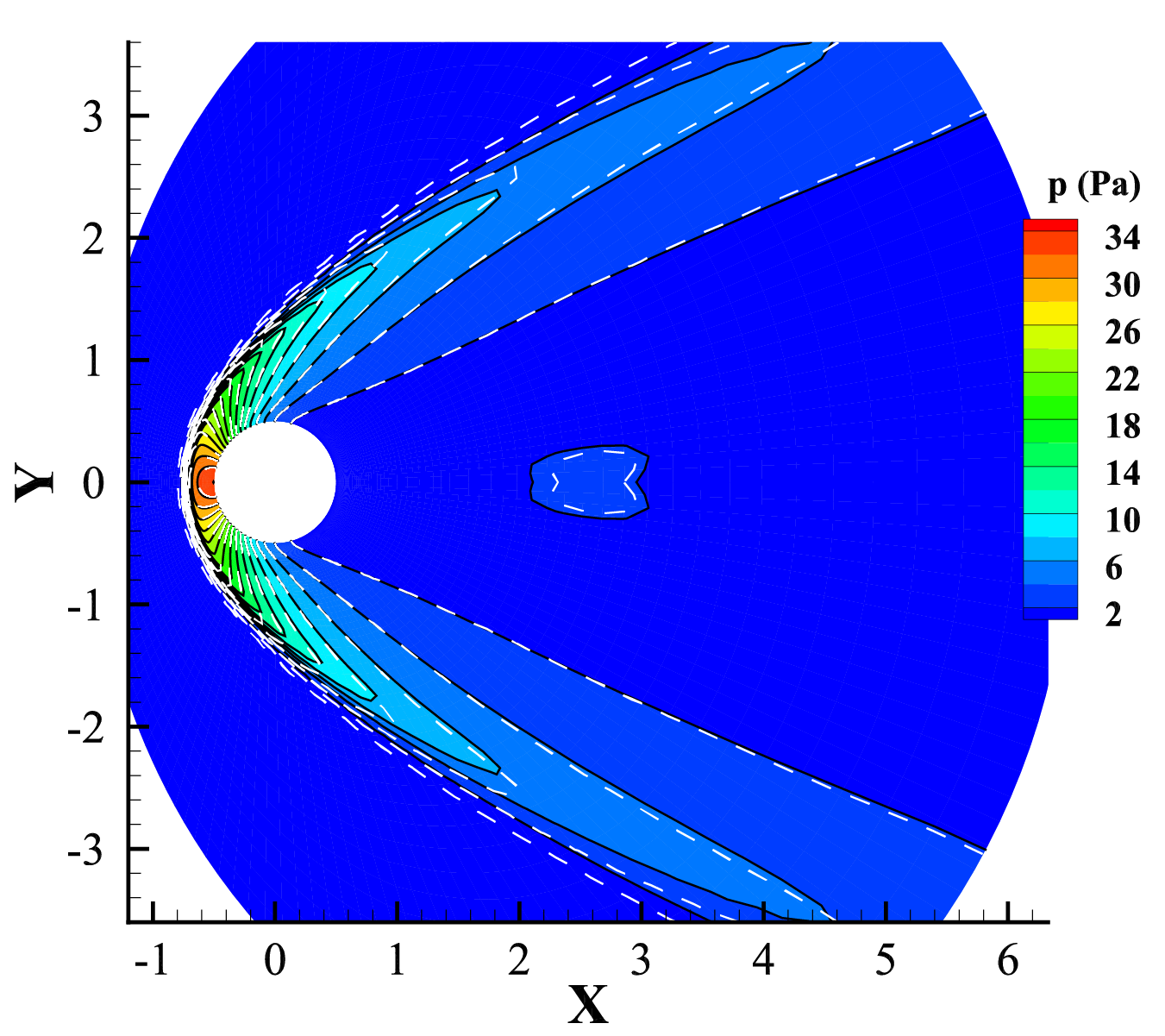}	    
		\end{minipage}
		\label{fig_cylinder_ma5_field_pressure}
	}
	\subfigure[Mach number]{
		\begin{minipage}[!t]{0.45\textwidth}
			\centering
			\includegraphics[width=1.0\textwidth]{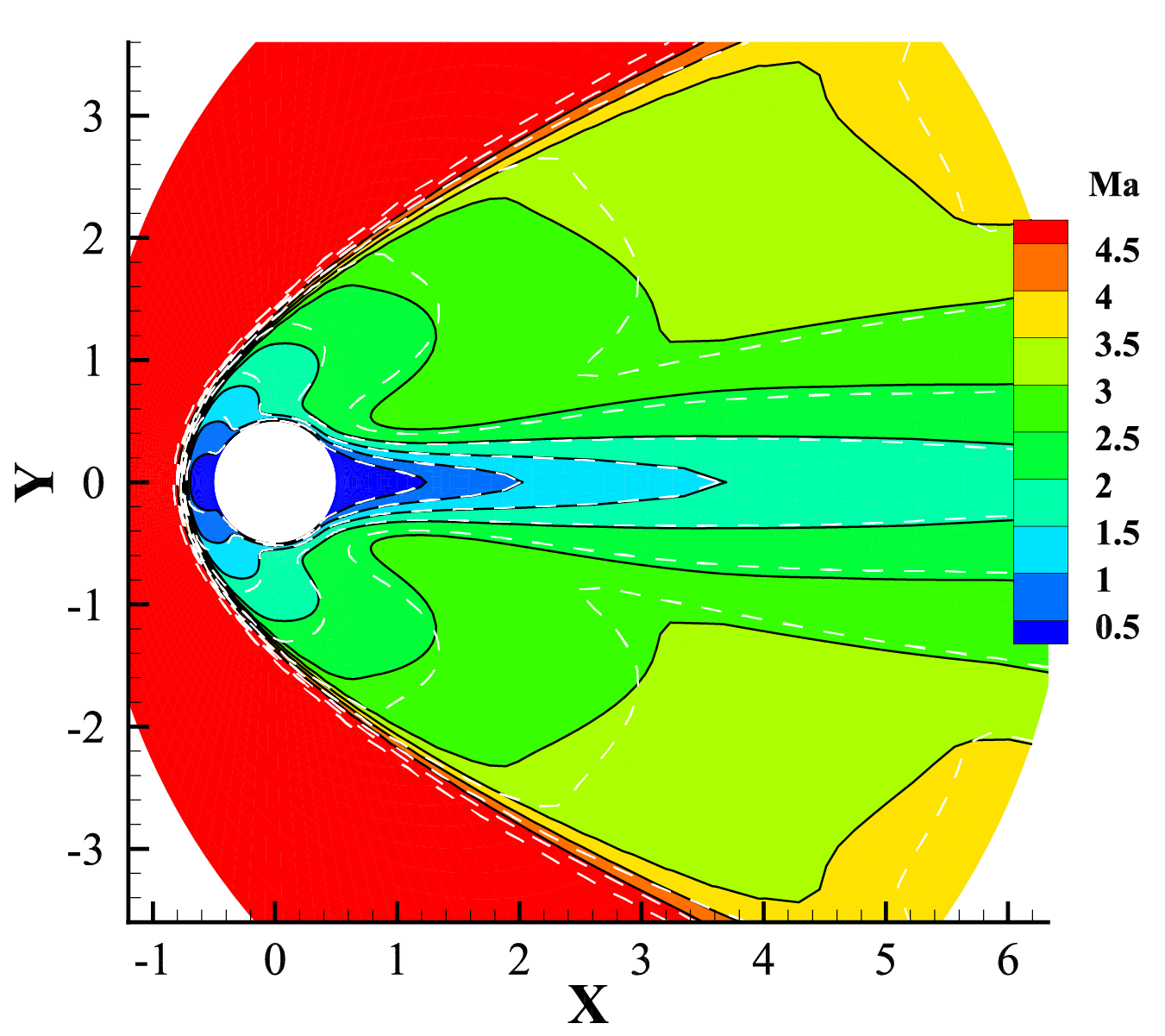}		
		\end{minipage}
		\label{fig_cylinder_ma5_field_mach}
	}
	\subfigure[Equilibrium temperature]{
		\begin{minipage}[!t]{0.45\textwidth}
			\centering
			\includegraphics[width=1.0\textwidth]{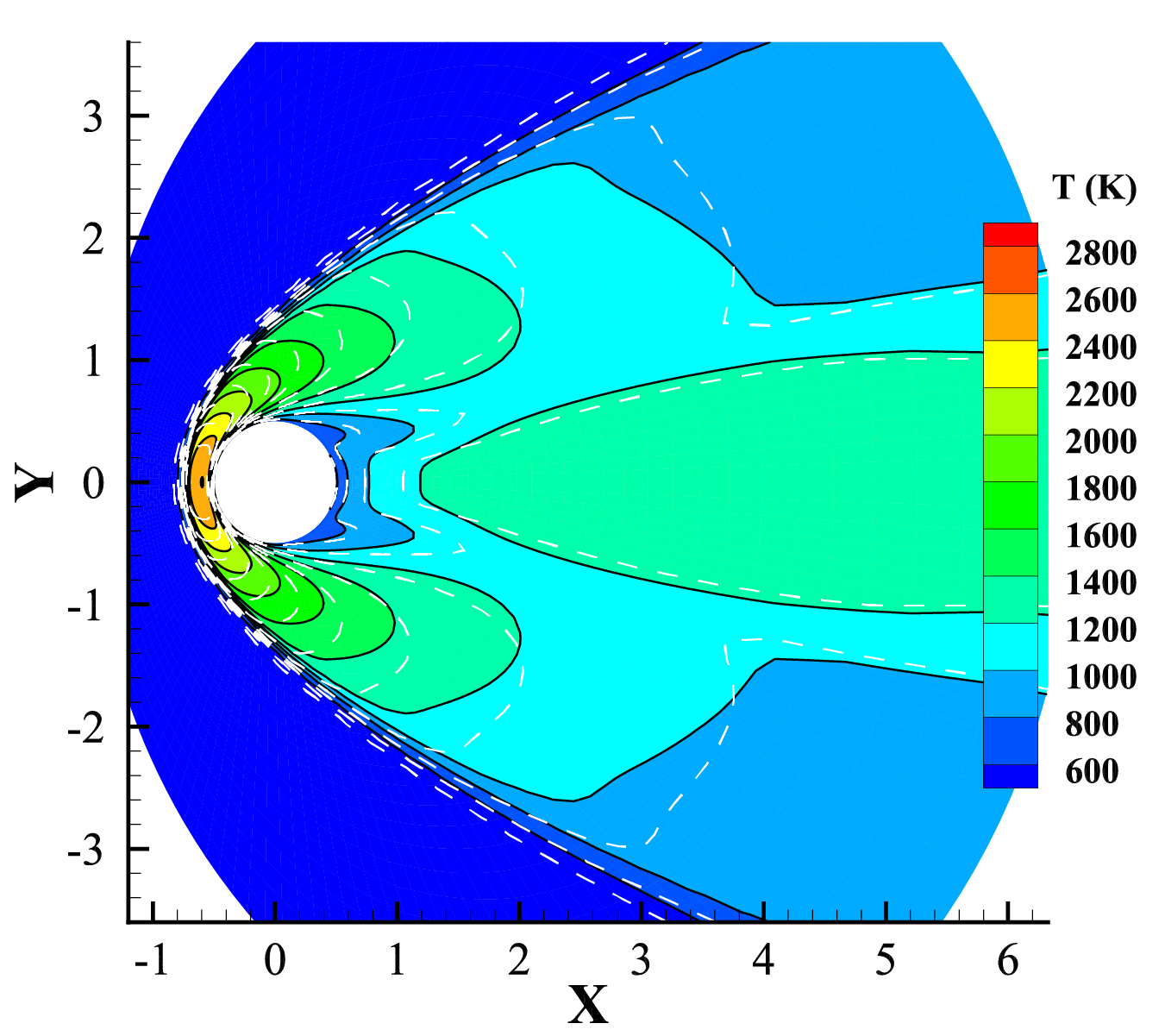}	    
		\end{minipage}
		\label{fig_cylinder_ma5_field_tem}
	}
	\subfigure[Translational temperature]{
		\begin{minipage}[!t]{0.45\textwidth}
			\centering
			\includegraphics[width=1.0\textwidth]{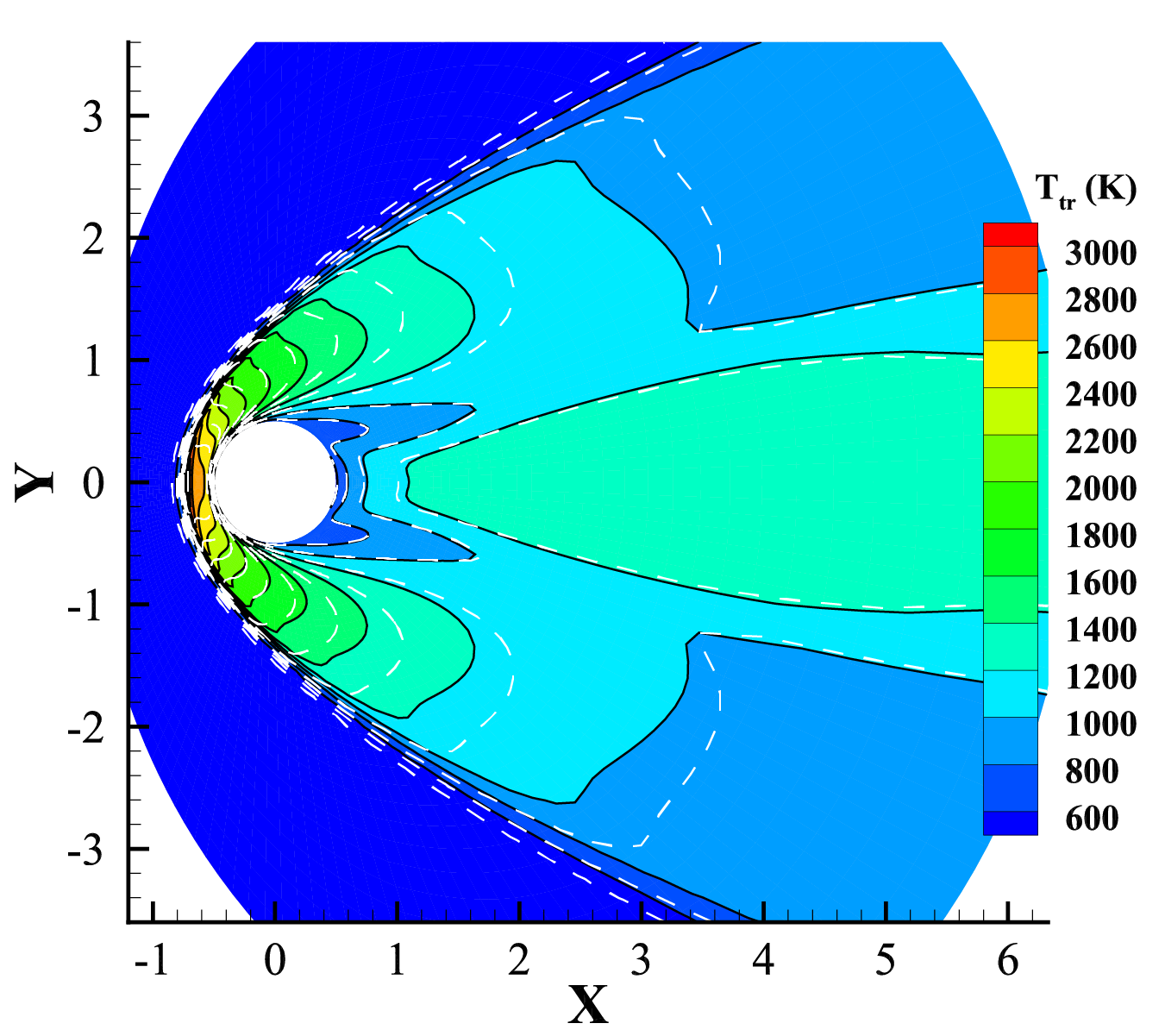}		
		\end{minipage}
		\label{fig_cylinder_ma5_field_temtra}
	}
	\subfigure[Rotational temperature]{
		\begin{minipage}[!t]{0.45\textwidth}
			\centering
			\includegraphics[width=1.0\textwidth]{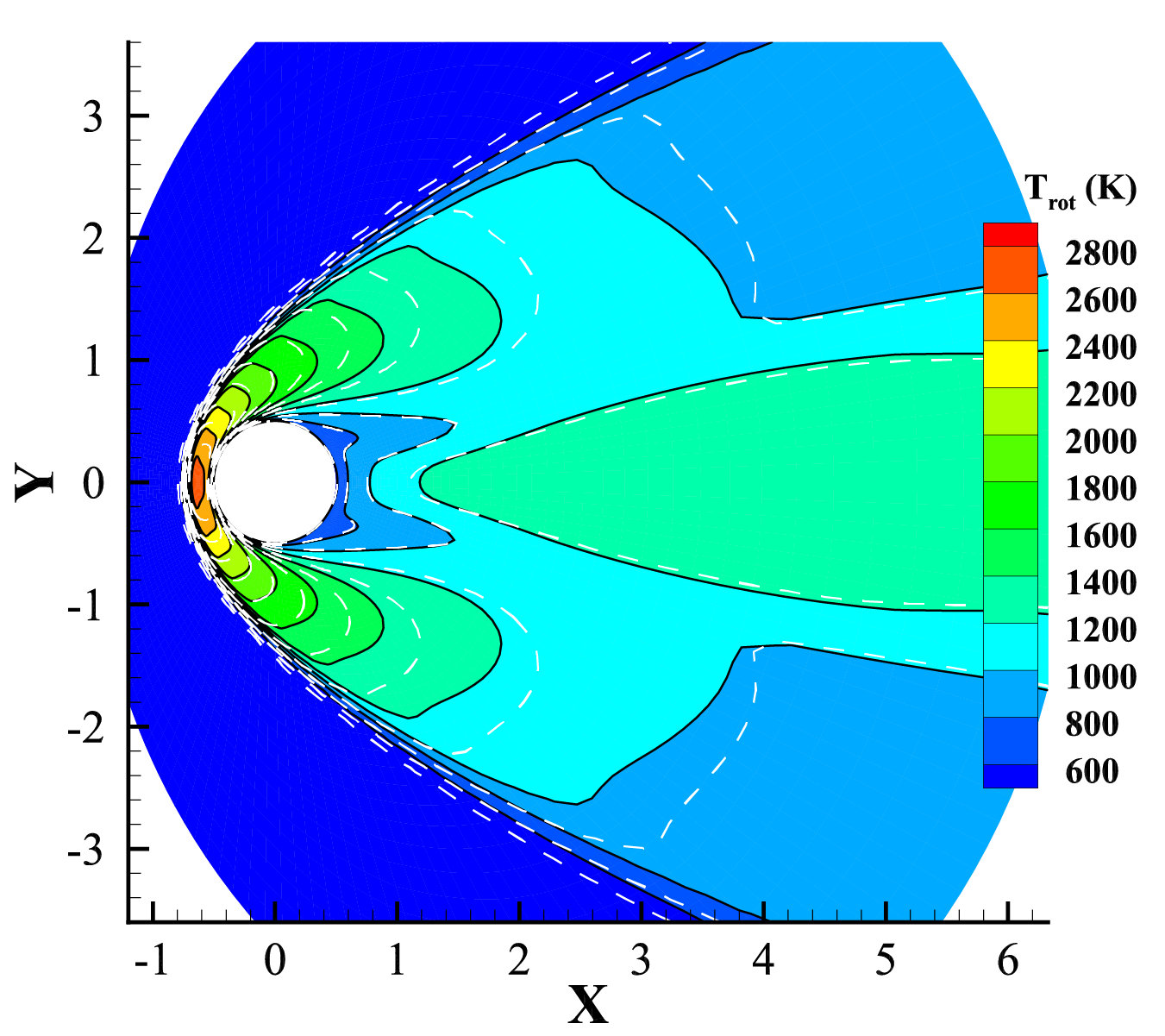}		
		\end{minipage}
		\label{fig_cylinder_ma5_field_temrot}
	}
	\subfigure[Vibrational temperature]{
		\begin{minipage}[!t]{0.45\textwidth}
			\centering
			\includegraphics[width=1.0\textwidth]{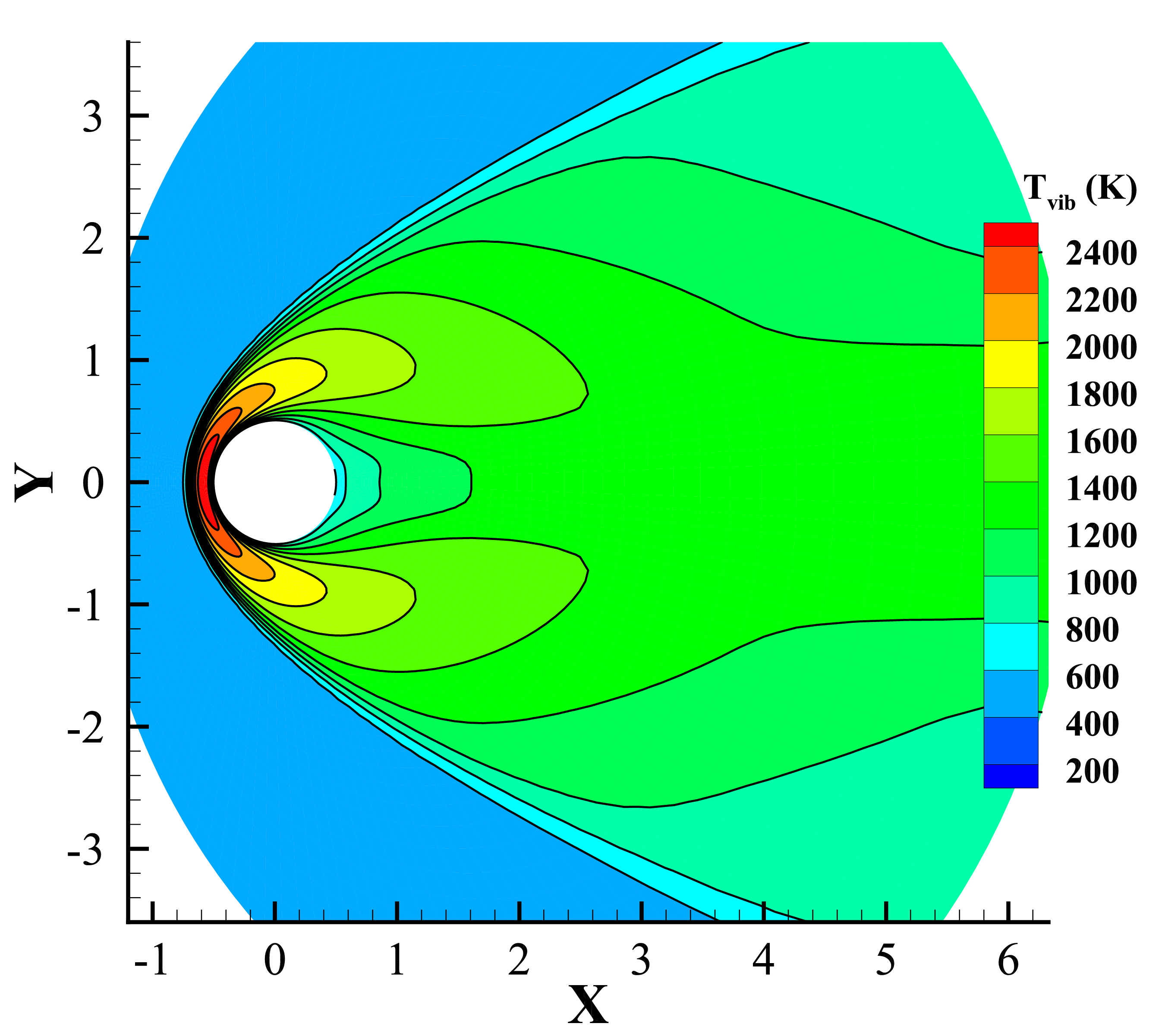}		
		\end{minipage}
		\label{fig_cylinder_ma5_field_temvib}
	}
	\caption{The contours of macroscopic flow variables around the cylinder at $\rm Ma=5$. The dash line: 
		UGKS with Rykov model, color band: present.}
	\label{fig_cylinder_ma5_field}
\end{figure}
\begin{figure}[!t]
	\centering
	\subfigure[Density]{
		\begin{minipage}[!t]{0.45\textwidth}
			\centering
			\includegraphics[width=1.0\textwidth]{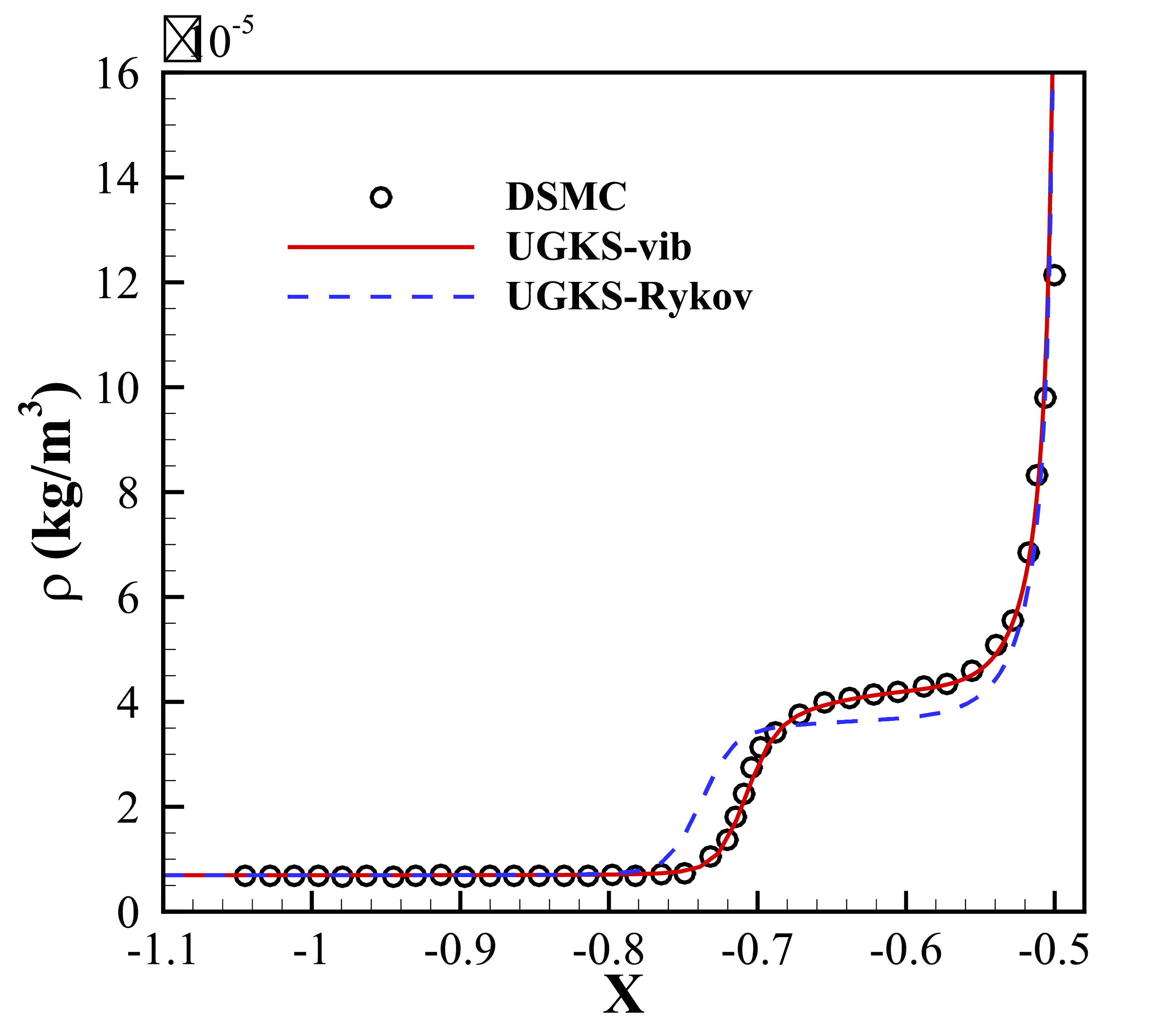}	    
		\end{minipage}
		\label{fig_cylinder_ma5_stagnationline_density}
	}
	\subfigure[Translational, rotational and vibrational temperatures]{
		\begin{minipage}[!t]{0.45\textwidth}
			\centering
			\includegraphics[width=1.0\textwidth]{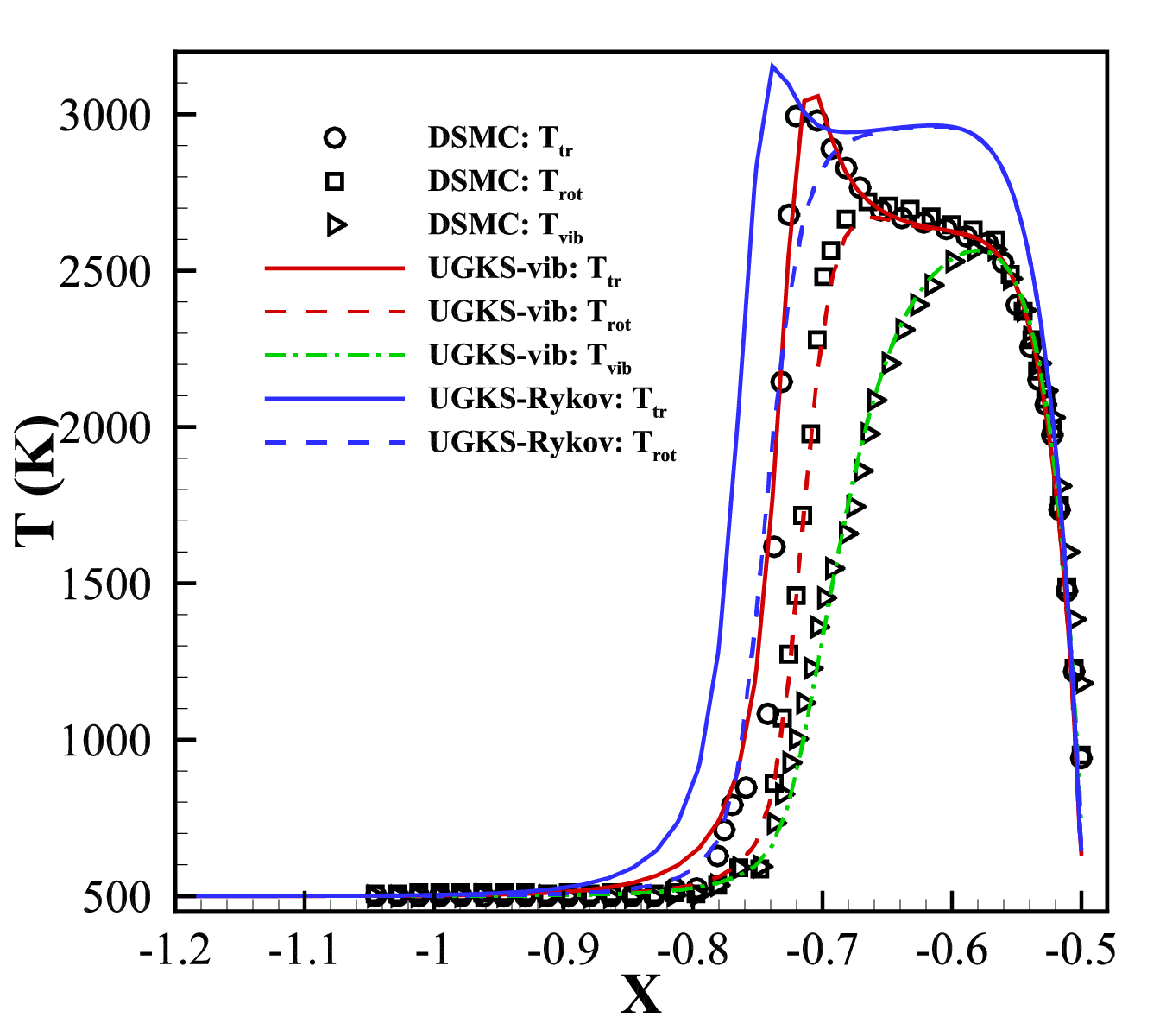}		
		\end{minipage}
		\label{fig_cylinder_ma5_stagnationline_tem}
	}
	\caption{Comparison of the density and the translational, rotational and vibrational temperatures 
		along the forward stagnation line for the cylinder at $\rm Ma=5$.}
	\label{fig_cylinder_ma5_stagnationline}
\end{figure}
\begin{figure}[!t]
	\centering
	\subfigure[Pressure]{
		\begin{minipage}[!t]{0.45\textwidth}
			\centering
			\includegraphics[width=1.0\textwidth]{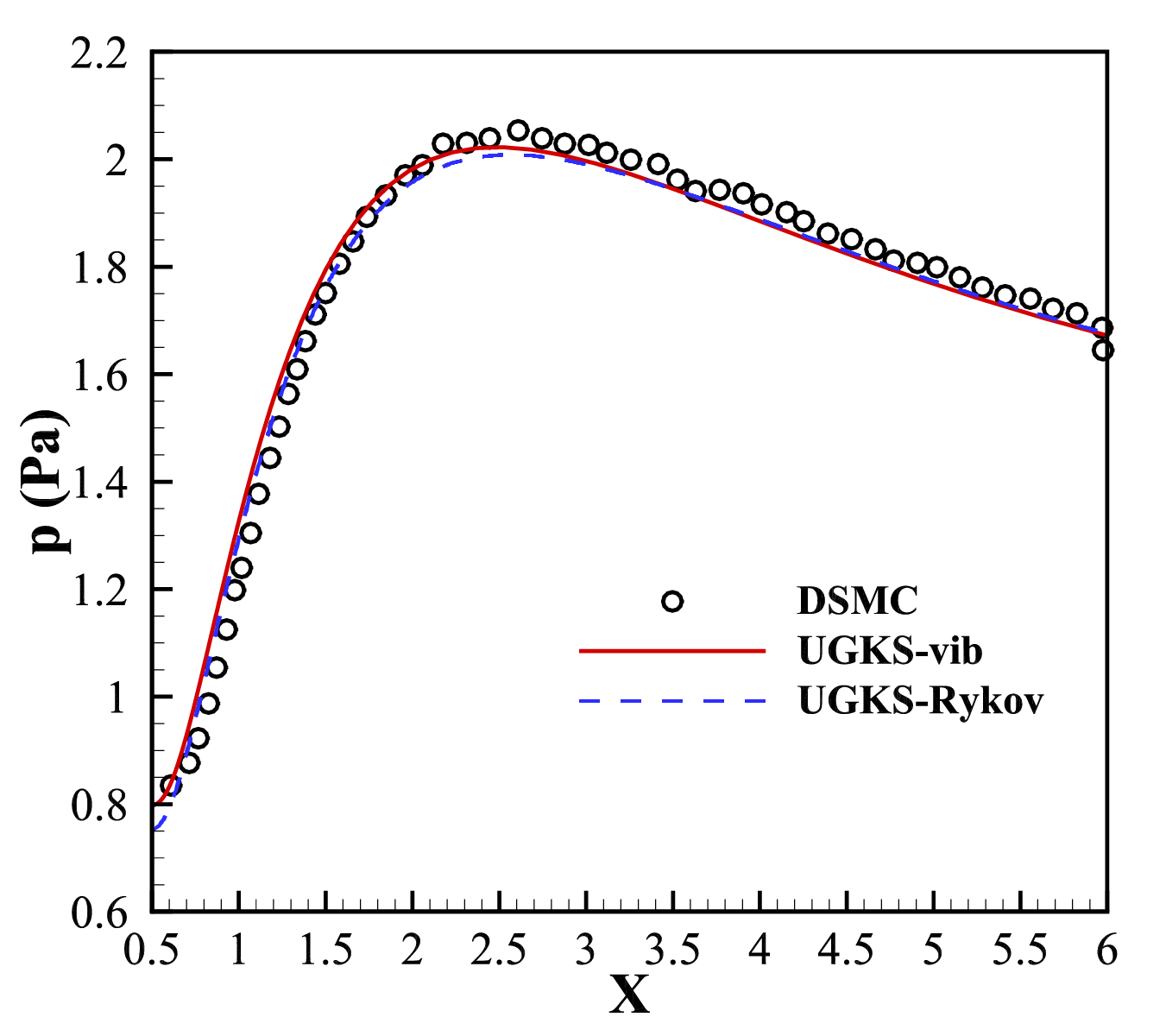}	    
		\end{minipage}
		\label{fig_cylinder_ma5_leewardLine_pressure}
	}
	\subfigure[Rotational temperature and vibrational temperature]{
		\begin{minipage}[!t]{0.45\textwidth}
			\centering
			\includegraphics[width=1.0\textwidth]{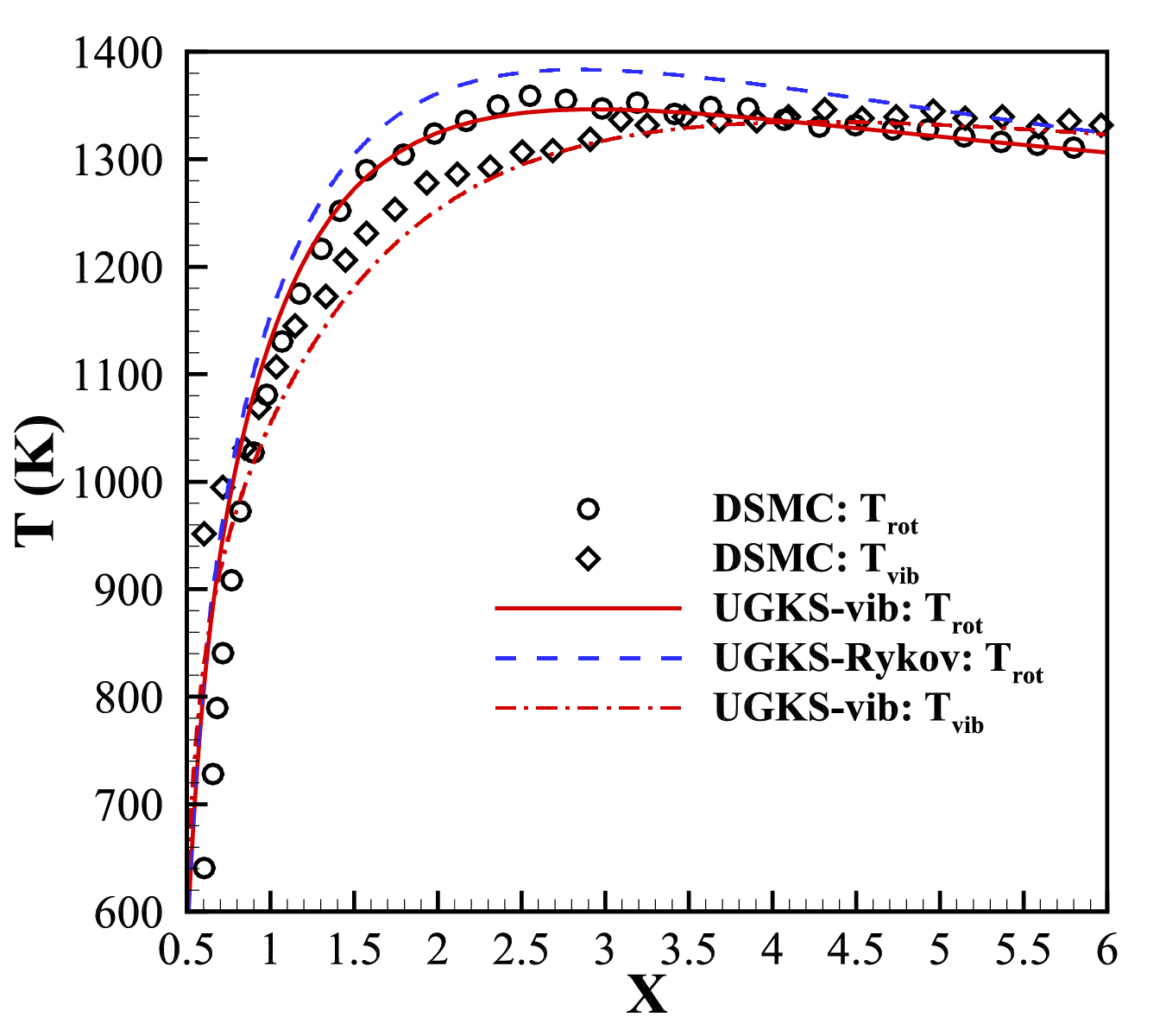}		
		\end{minipage}
		\label{fig_cylinder_ma5_leewardLine_temrot_temvib}
	}
	\caption{Comparison of the pressure, rotational temperature and vibrational temperature 
		along the backward stagnation line for the cylinder at $\rm Ma=5$.}
	\label{fig_cylinder_ma5_leewardLine}
\end{figure}

Fig.~\ref{fig_cylinder_ma5_field} shows the contours of macroscopic flow variables at $\rm Ma=5$, including the 
pressure, Mach number, equilibrium temperature, and translational, rotational and vibrational temperatures.
Fig.~\ref{fig_cylinder_ma5_stagnationline} and Fig.~\ref{fig_cylinder_ma5_leewardLine} show the distributions of 
density and temperatures along the forward stagnation line, and the distributions of pressure and temperatures along 
the backward stagnation line, respectively. The present results of density and translational, rotational 
and vibrational temperatures along the forward stagnation line are in good agreement with those from DSMC, except the 
translational temperature raises up earlier than that in DSMC due to the common defect of the relaxation-type kinetic 
models which have a single relaxation time for particles with different velocity. The present pressure, rotational and 
vibrational temperatures along the backward stagnation line are in reasonable agreement with the results of DSMC. 
It can be seen that the thermodynamic non-equilibrium effect is extremely apparent in the shock wave zone from 
fig.~\ref{fig_cylinder_ma5_stagnationline}. Besides, the results of density and temperatures along the forward 
stagnation line from the UGKS-Rykov method~\citep{ShaLiu_2014} show some deviations compared with the results of 
present method and DSMC. The excitation of molecular vibrational degrees of freedom will result in the reduction 
of shock detachment distance.
\par
\begin{figure}[!t]
	\centering
	\subfigure[Pressure]{
		\begin{minipage}[!t]{0.45\textwidth}
			\centering
			\includegraphics[width=1.0\textwidth]{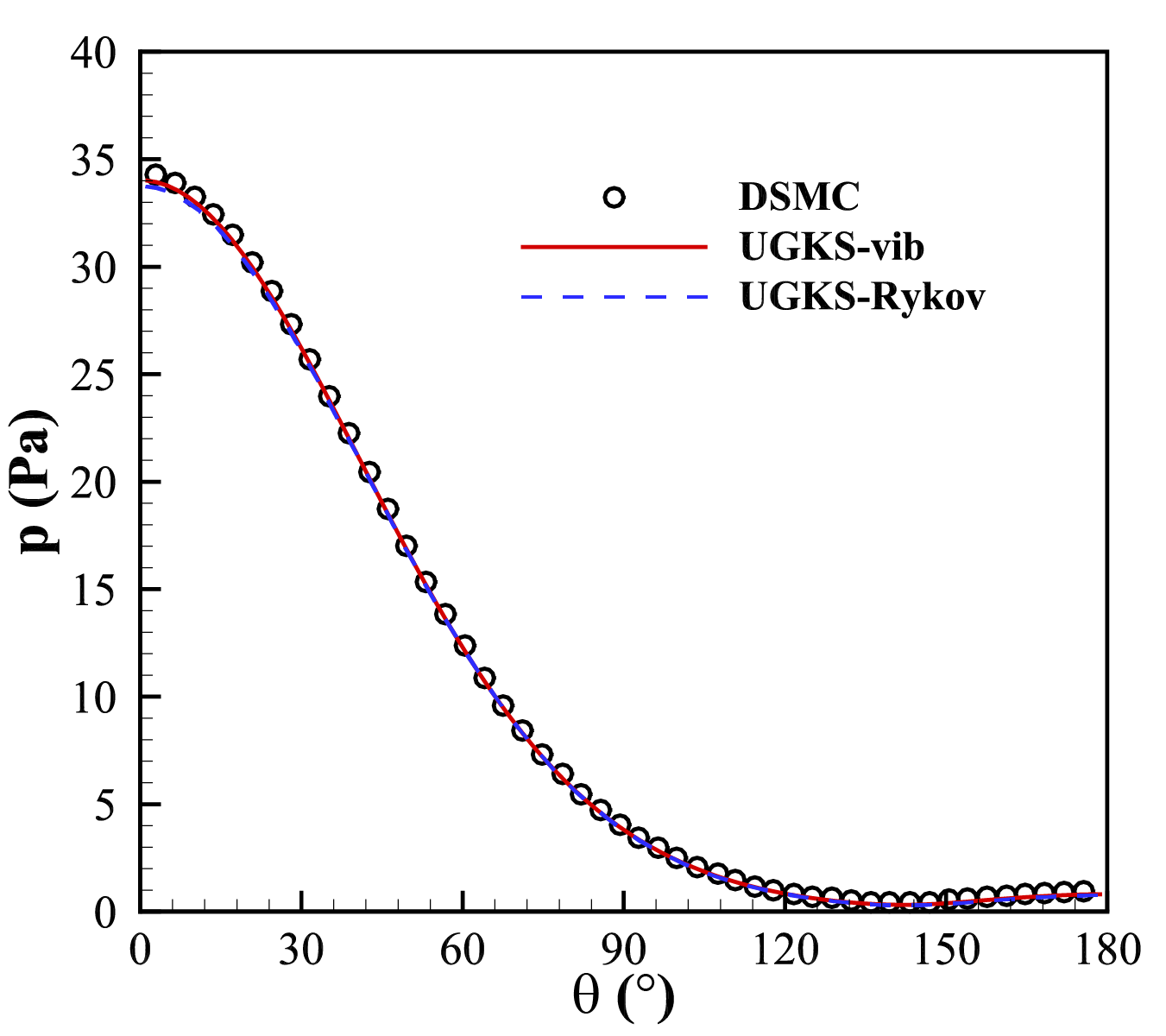}	    
		\end{minipage}
		\label{fig_cylinder_ma5_surfaceline_pressure}
	}
	\subfigure[Heat flux]{
		\begin{minipage}[!t]{0.45\textwidth}
			\centering
			\includegraphics[width=1.0\textwidth]{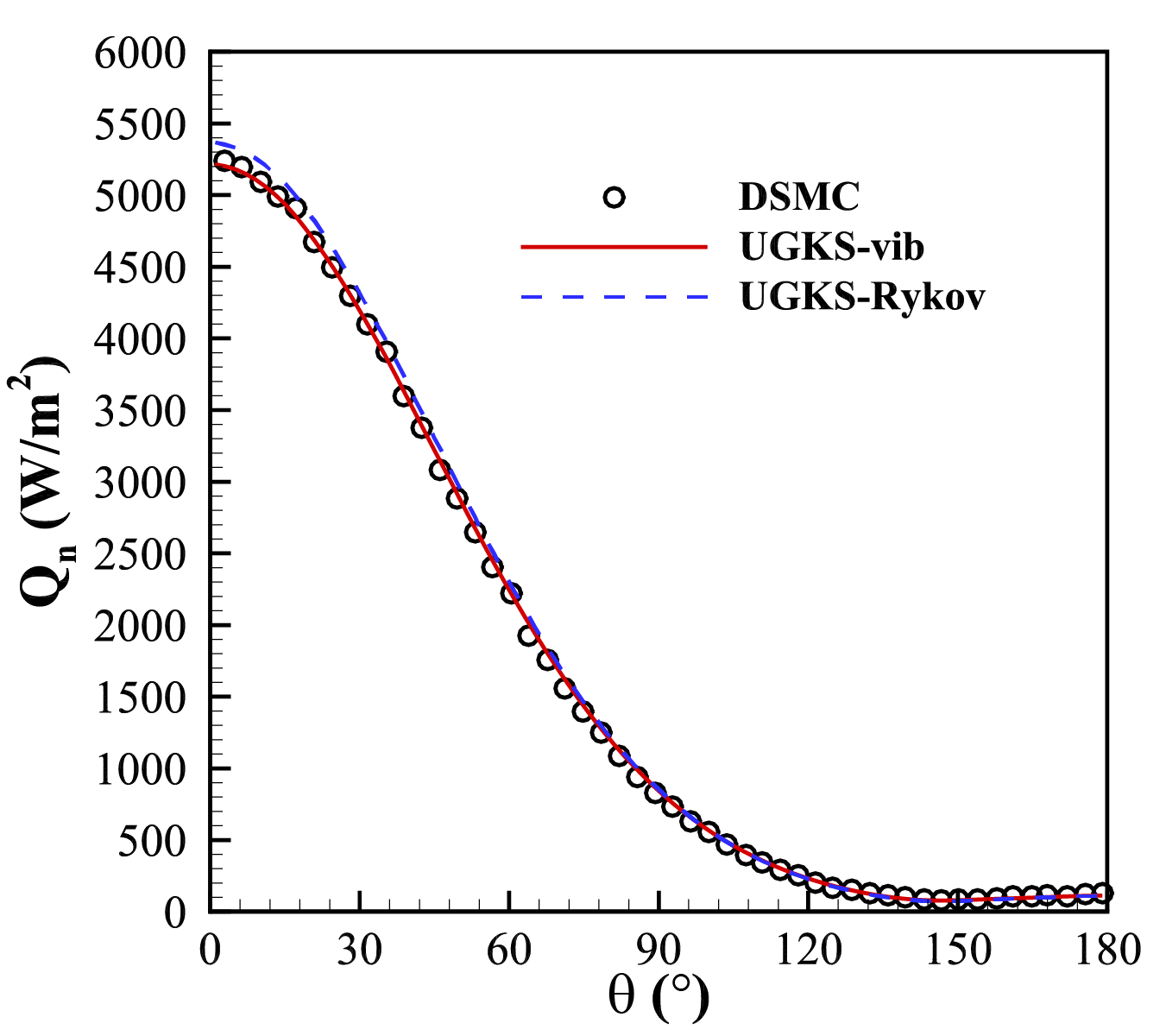}	    
		\end{minipage}
		\label{fig_cylinder_ma5_surfaceline_heatflux}
	}
	\subfigure[Translational temperature]{
		\begin{minipage}[!t]{0.45\textwidth}
			\centering
			\includegraphics[width=1.0\textwidth]{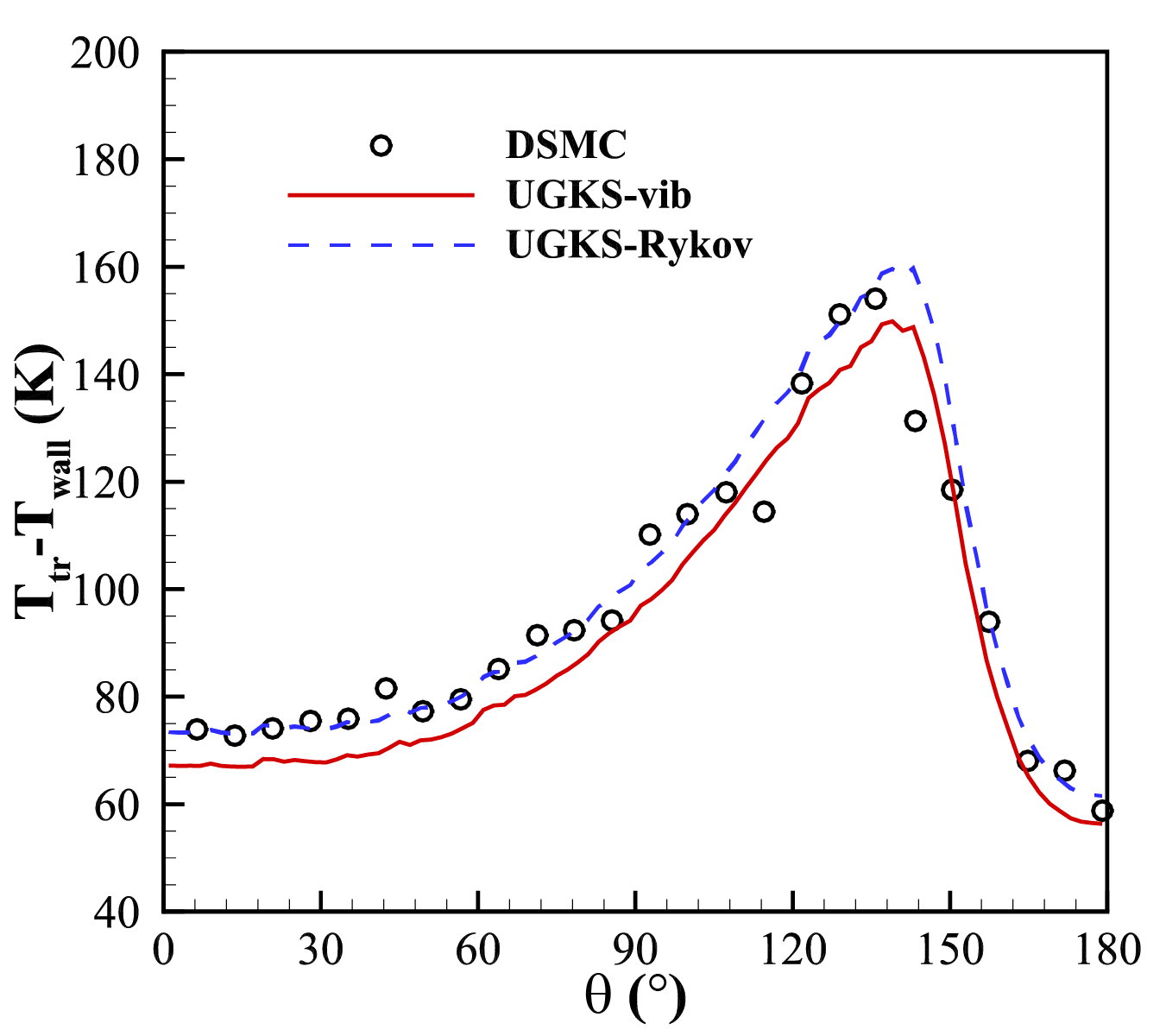}		
		\end{minipage}
		\label{fig_cylinder_ma5_surfaceline_temtra}
	}
	\subfigure[Rotational temperature]{
		\begin{minipage}[!t]{0.45\textwidth}
			\centering
			\includegraphics[width=1.0\textwidth]{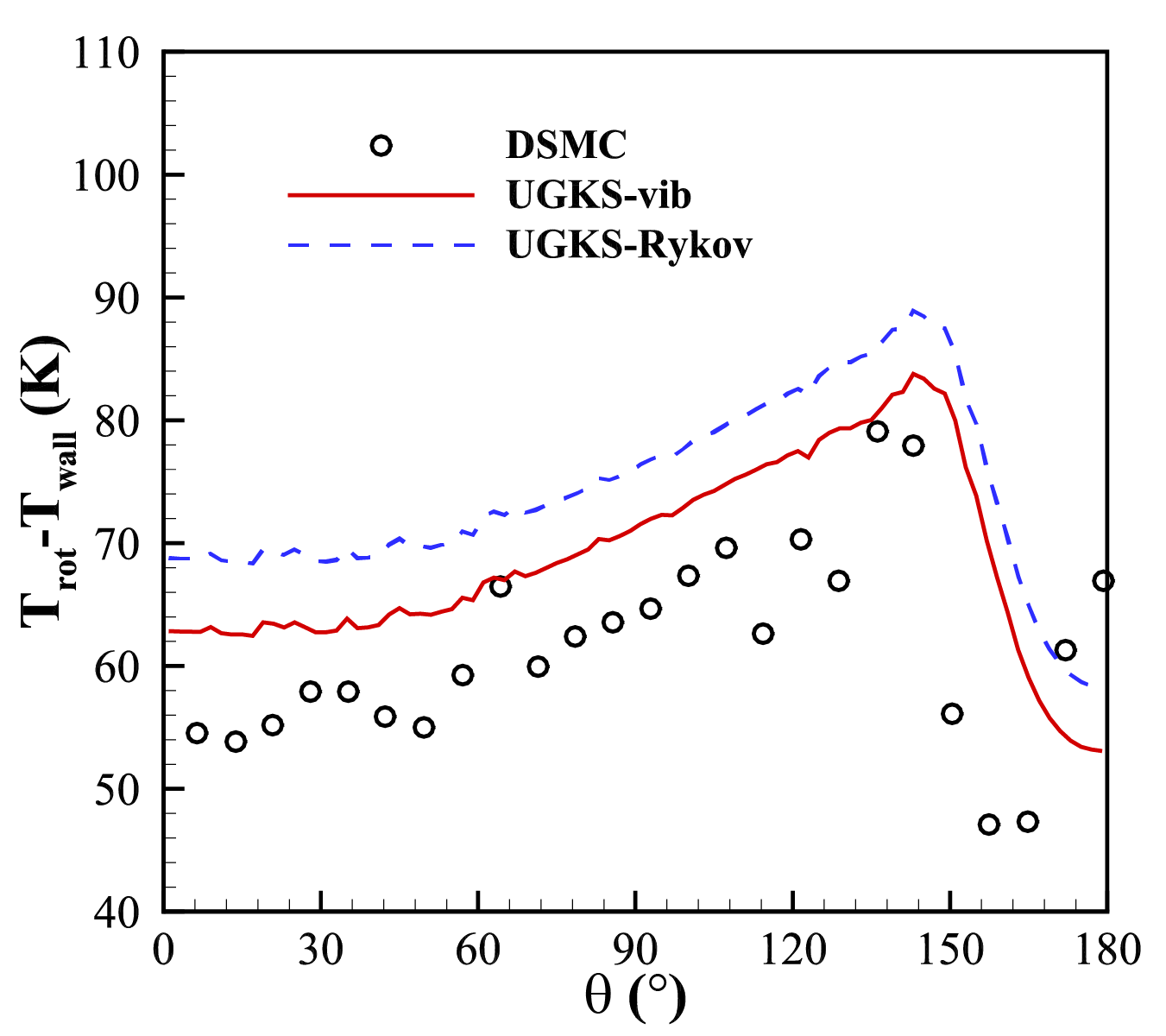}		
		\end{minipage}
		\label{fig_cylinder_ma5_surfaceline_temrot}
	}
	\caption{Distributions of the pressure, heat flux, translational and rotational temperatures
		on the wall surface for the cylinder at $\rm Ma=5$.}
	\label{fig_cylinder_ma5_surfaceline}
\end{figure}

The detailed pressure, heat flux, translational and rotational temperatures on the wall surface compared with the results 
of DSMC~\citep{JunLinWu_2021} and UGKS-Rykov method~\citep{ShaLiu_2014} are shown in Fig.~\ref{fig_cylinder_ma5_surfaceline}. 
It can be seen that the distribution of pressure on the wall surface from present method and UGKS-Rykov method agree well 
with the result of DSMC. The distribution of heat flux solved by present method also agree well with the result of DSMC, 
while the stagnation heat flux calculated by UGKS-Rykov method is generally higher than the result of DSMC. The current UGKS 
with simplified multi-scale numerical flux involving the excitation of molecular vibrational degrees has a significant effect 
on improving the result of heat flux on the wall surface. It can be seen that the translational and rotational temperature 
jumps on the wall surface are also qualitatively consistent with the benchmark solutions. 
\par
\begin{figure}[!t]
	\centering
	\subfigure[Physical space mesh ($200 \times 124$ cells)]{
		\begin{minipage}[!t]{0.45\textwidth}
			\centering
			\includegraphics[width=1.0\textwidth]{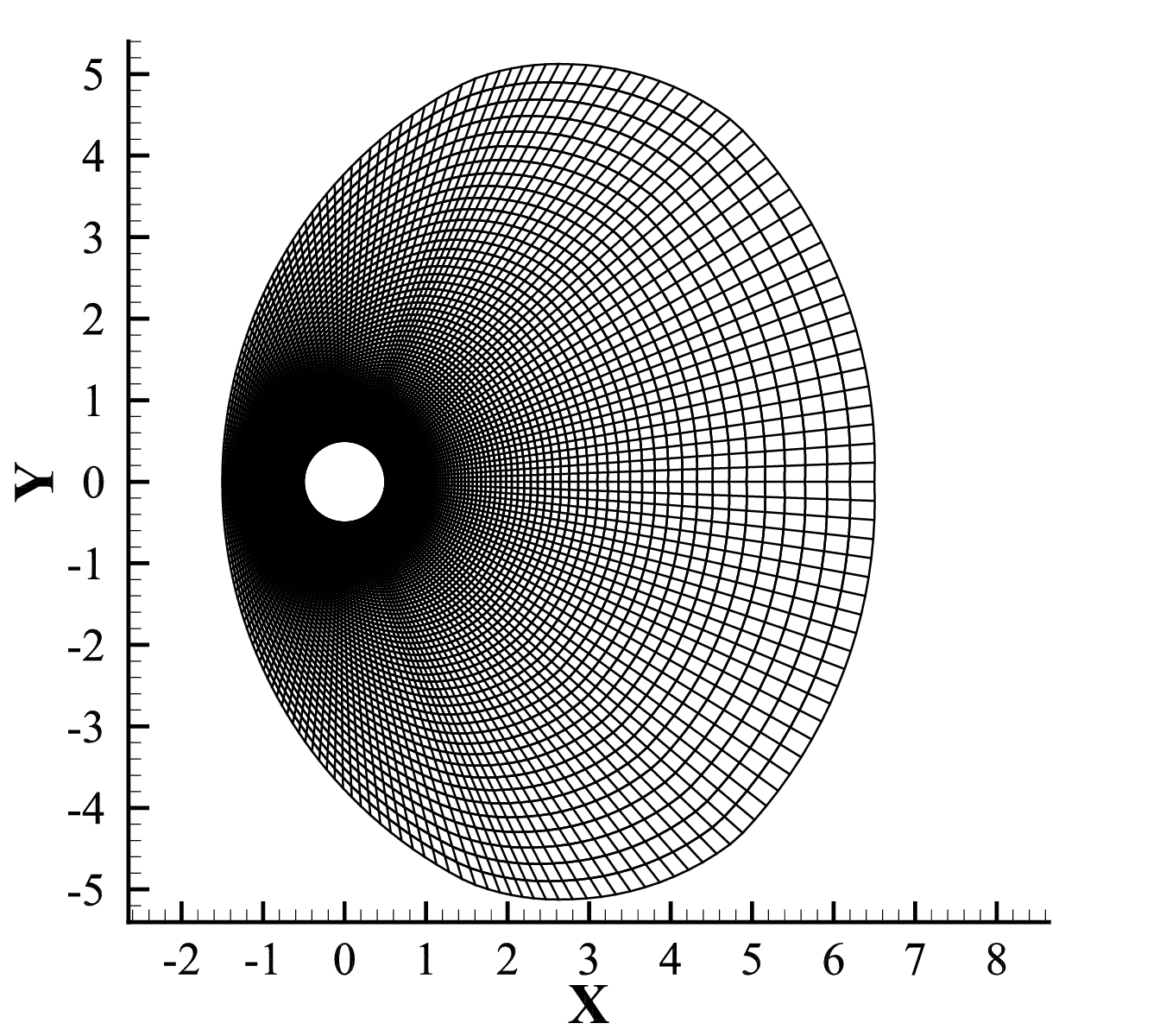}	    
		\end{minipage}
		\label{fig_cylinder_ma20_macmesh}
	}
	\subfigure[Unstructured discrete velocity space mesh ($2735$ cells)]{
		\begin{minipage}[!t]{0.45\textwidth}
			\centering
			\includegraphics[width=1.0\textwidth]{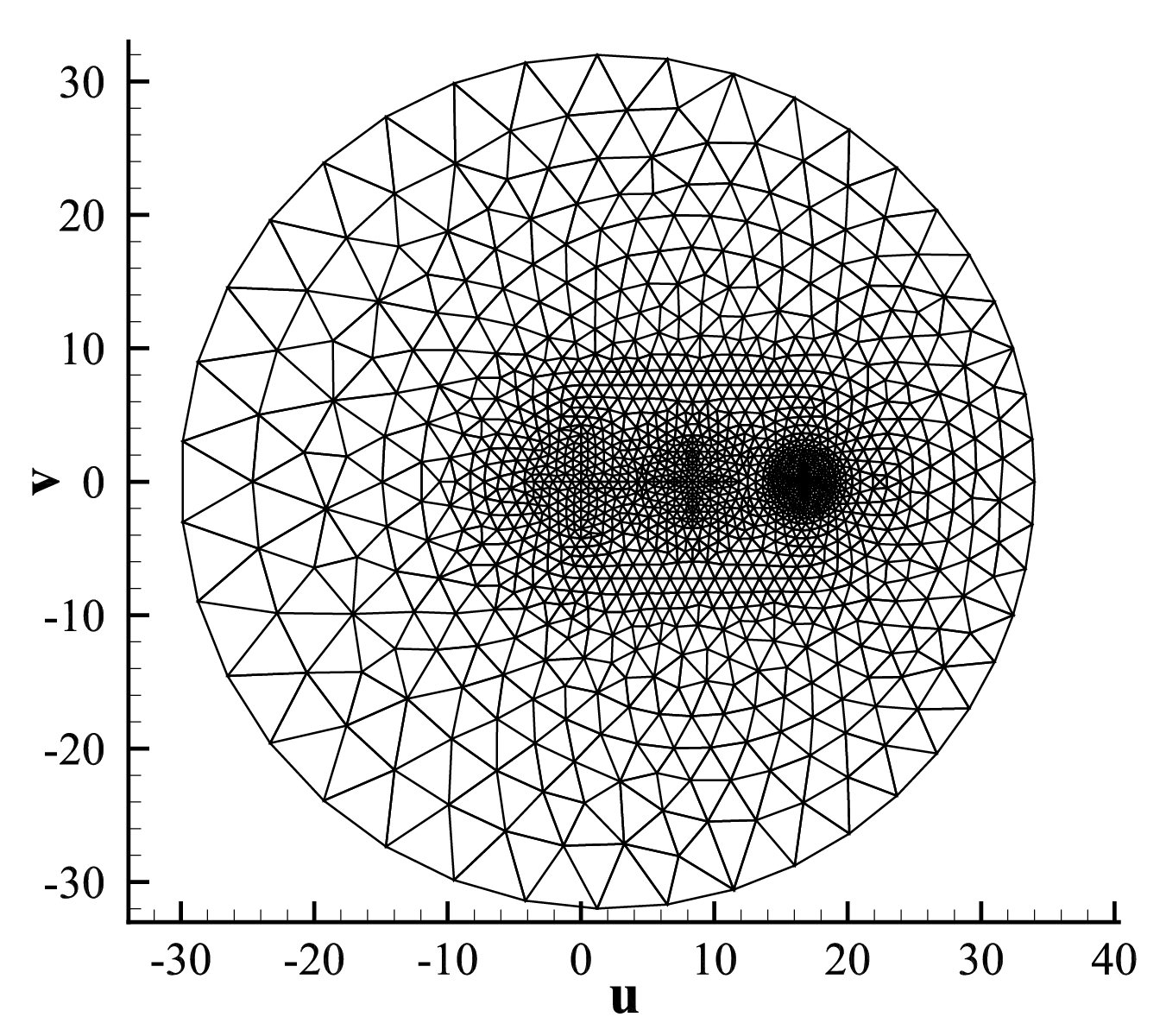}		
		\end{minipage}
		\label{fig_cylinder_ma20_micmesh}
	}
	\caption{The physical space mesh and unstructured discrete velocity space mesh for the cylinder at $\rm Ma=20$.}
	\label{fig_cylinder_ma20_mesh}
\end{figure}

For the case of $\rm Ma=20$, the physical domain is discretized by a mesh with $200 \times 124$ cells, 
as shown in Fig.~\ref{fig_cylinder_ma20_macmesh}, in which the height of the first layer near the wall 
is smaller than 0.001. A 2735-cell unstructured DVS is applied, as shown in Fig.~\ref{fig_cylinder_ma20_micmesh}. 
\par
\begin{figure}[!t]
	\centering
	\subfigure[Pressure]{
		\begin{minipage}[!t]{0.45\textwidth}
			\centering
			\includegraphics[width=1.0\textwidth]{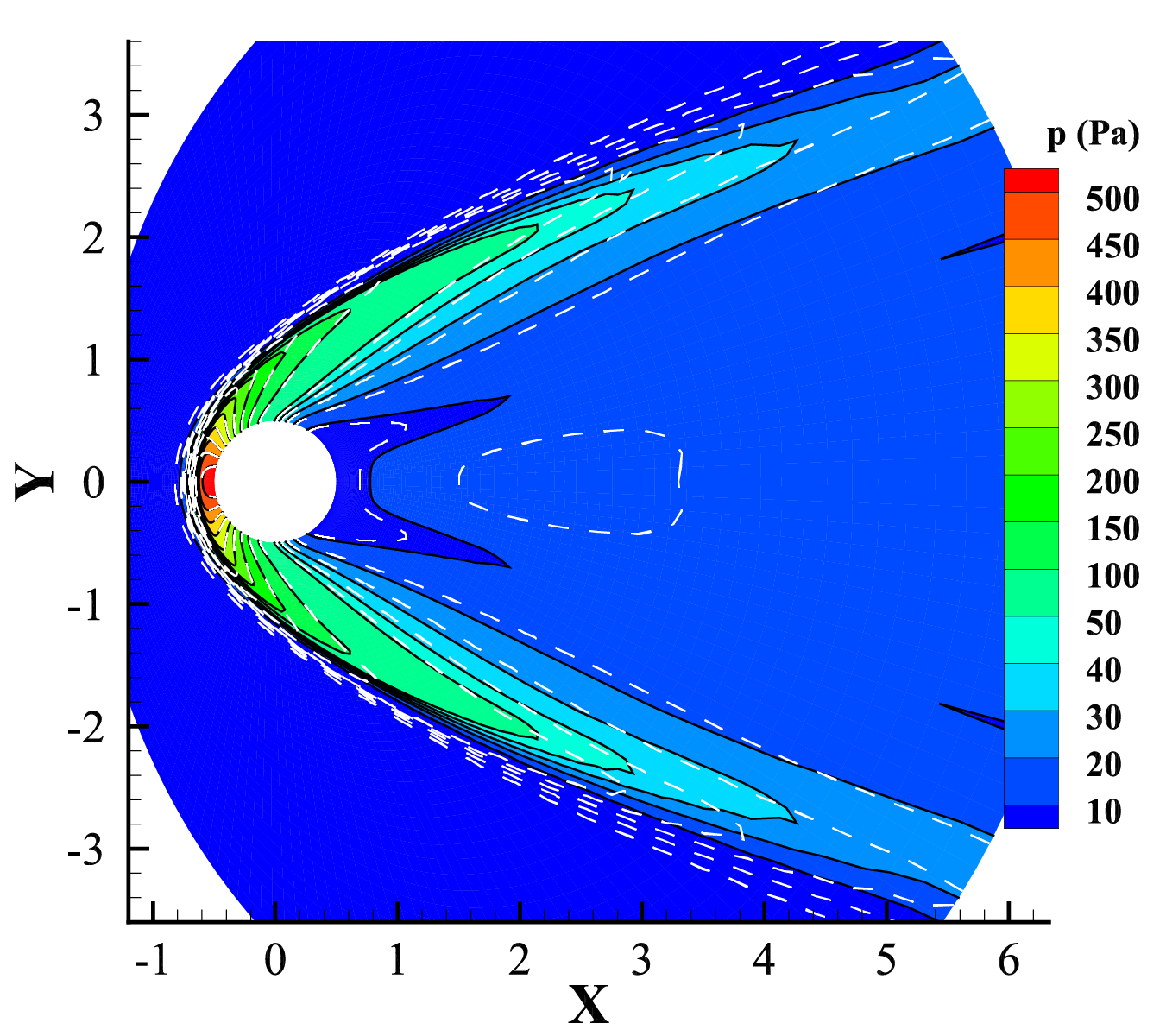}	    
		\end{minipage}
		\label{fig_cylinder_ma20_field_pressure}
	}
	\subfigure[Mach number]{
		\begin{minipage}[!t]{0.45\textwidth}
			\centering
			\includegraphics[width=1.0\textwidth]{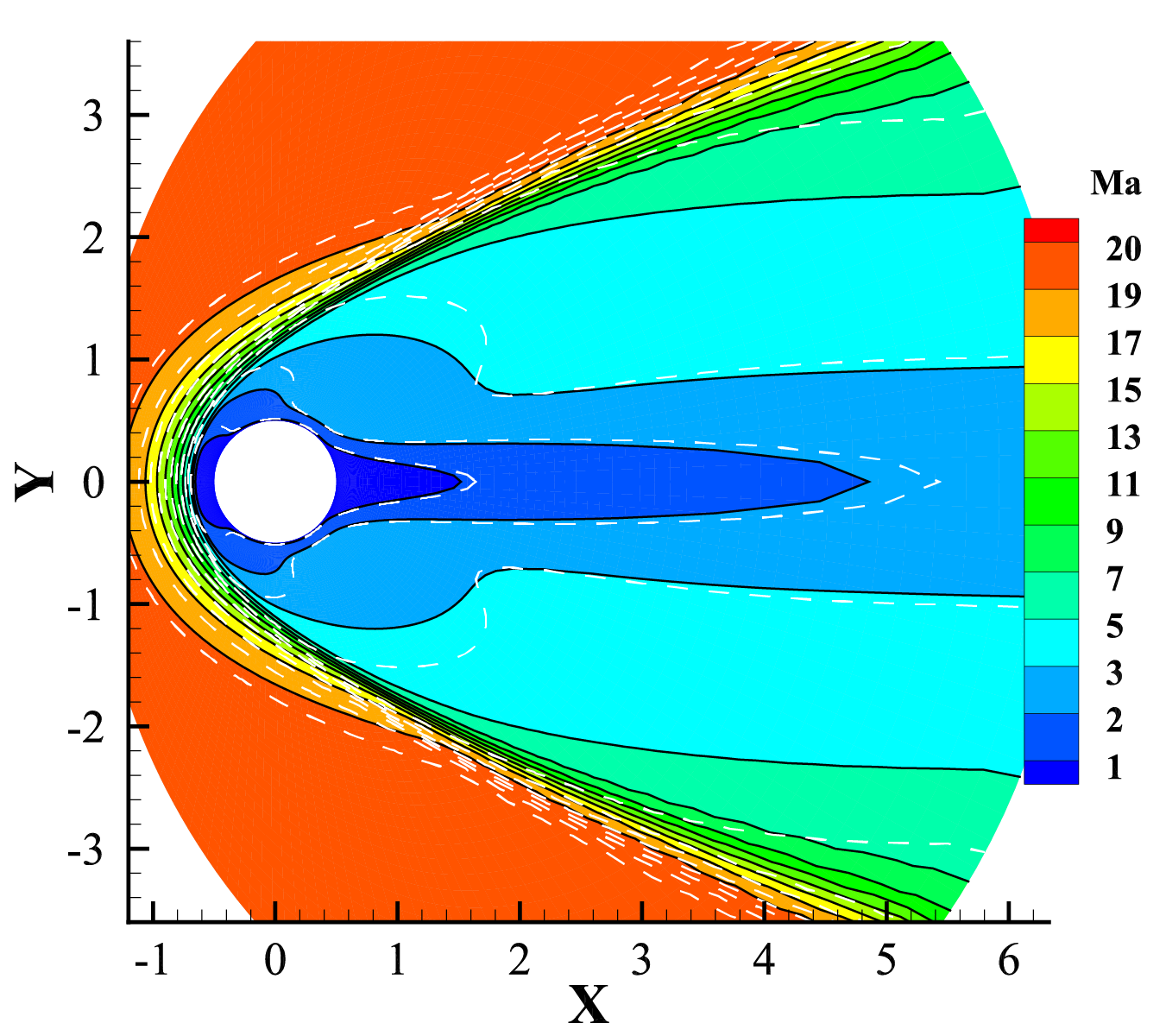}		
		\end{minipage}
		\label{fig_cylinder_ma20_field_mach}
	}
	\subfigure[Equilibrium temperature]{
		\begin{minipage}[!t]{0.45\textwidth}
			\centering
			\includegraphics[width=1.0\textwidth]{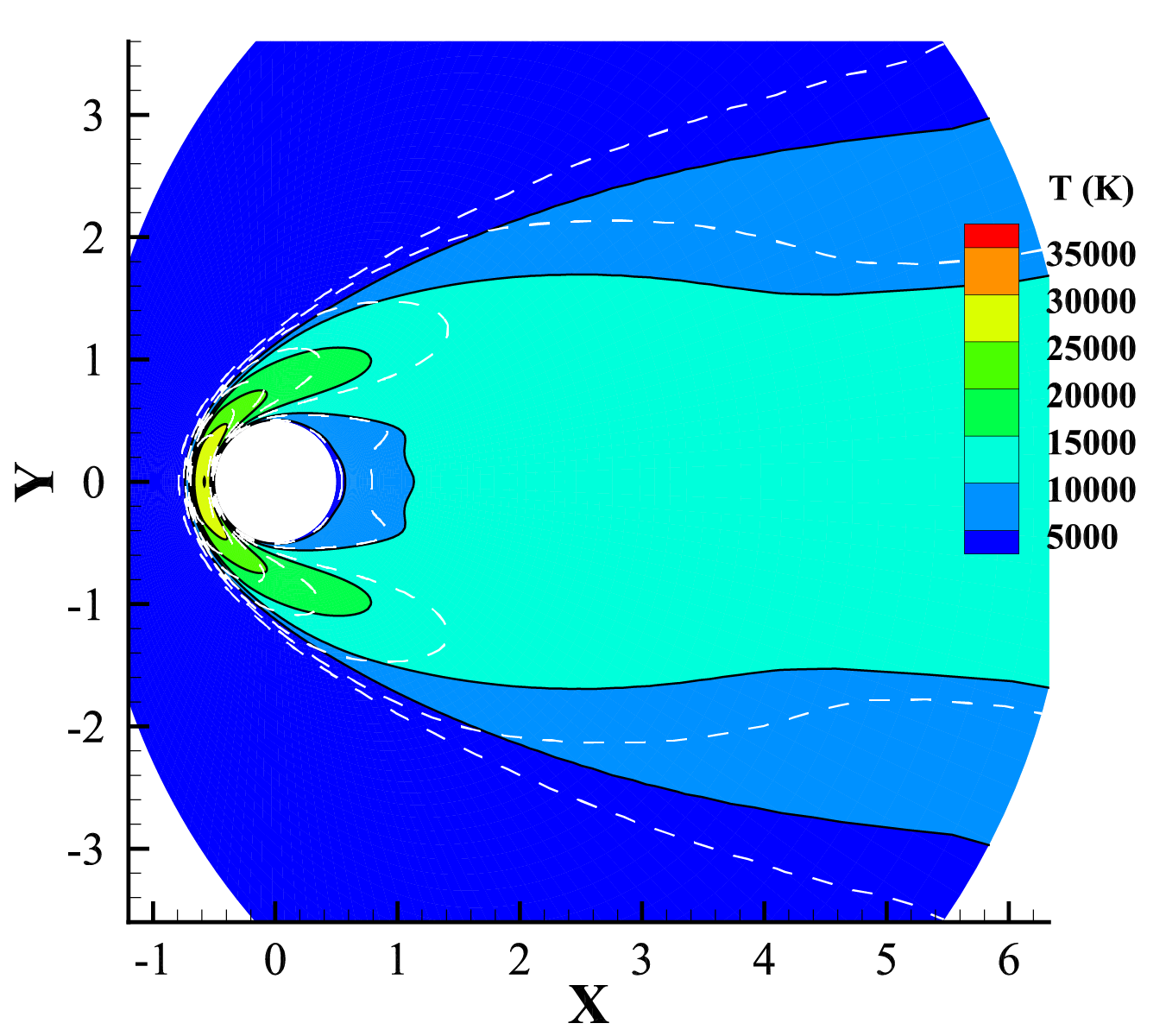}	    
		\end{minipage}
		\label{fig_cylinder_ma20_field_tem}
	}
	\subfigure[Translational temperature]{
		\begin{minipage}[!t]{0.45\textwidth}
			\centering
			\includegraphics[width=1.0\textwidth]{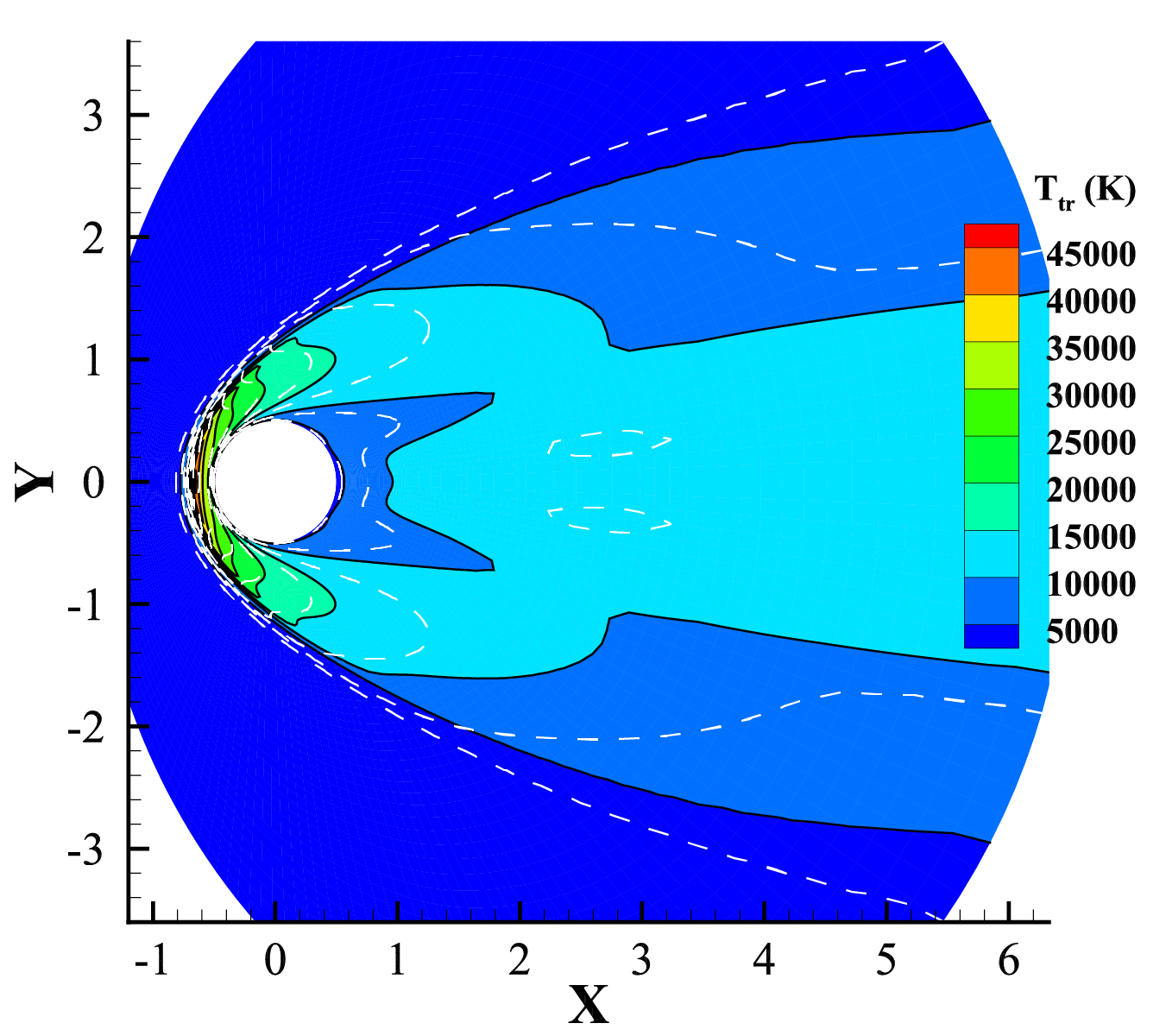}		
		\end{minipage}
		\label{fig_cylinder_ma20_field_temtra}
	}
	\subfigure[Rotational temperature]{
		\begin{minipage}[!t]{0.45\textwidth}
			\centering
			\includegraphics[width=1.0\textwidth]{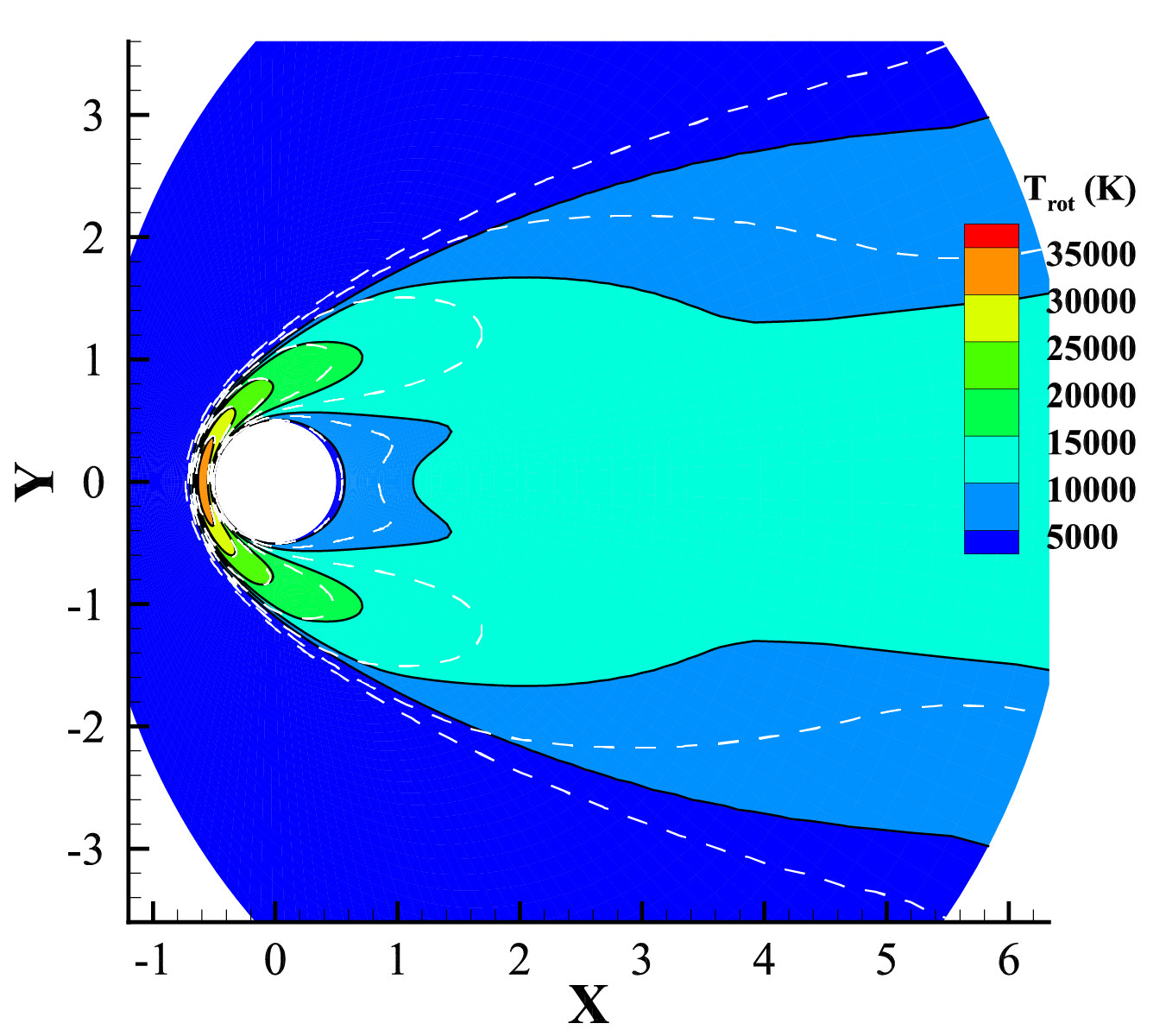}		
		\end{minipage}
		\label{fig_cylinder_ma20_field_temrot}
	}
	\subfigure[Vibrational temperature]{
		\begin{minipage}[!t]{0.45\textwidth}
			\centering
			\includegraphics[width=1.0\textwidth]{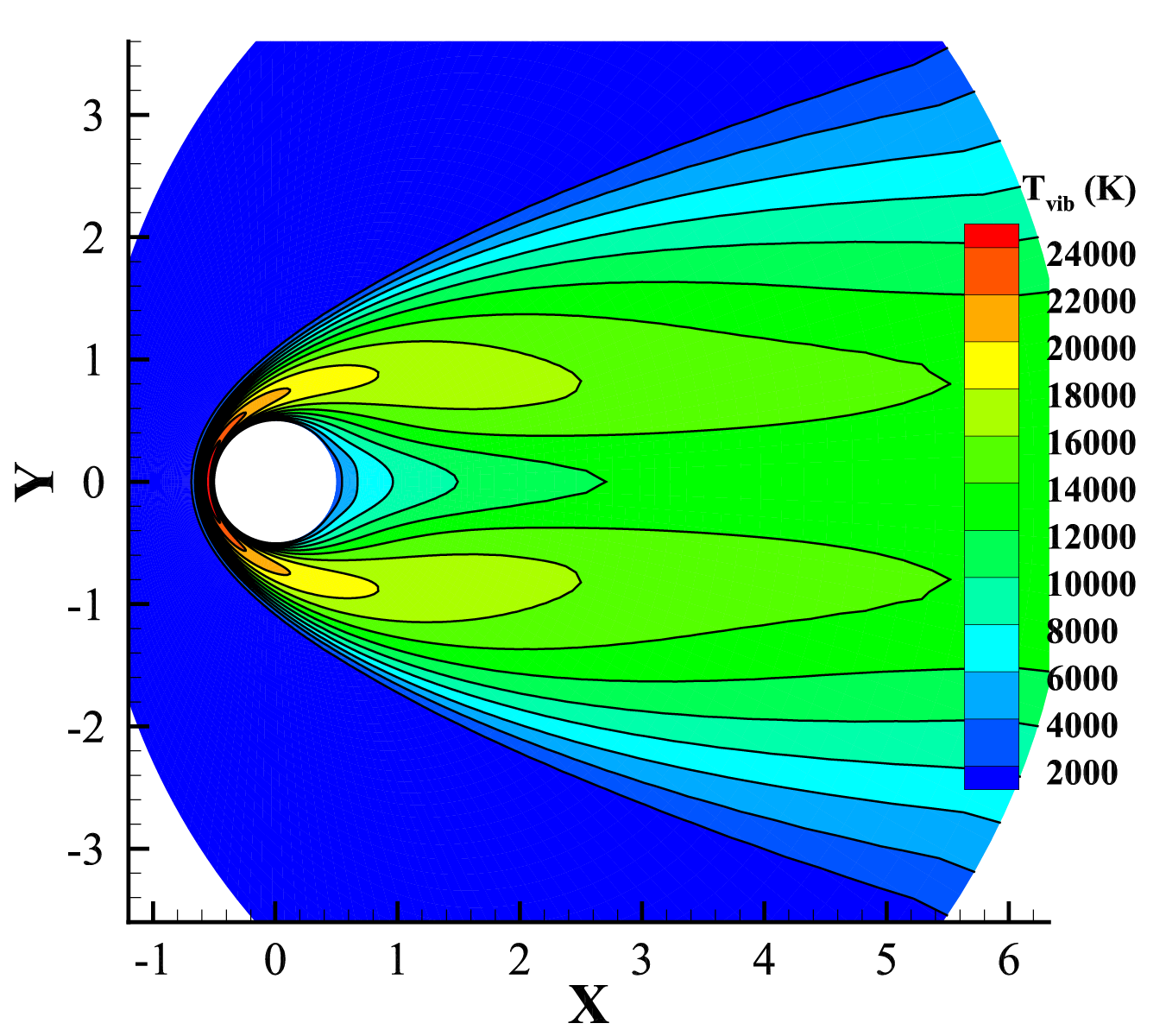}		
		\end{minipage}
		\label{fig_cylinder_ma20_field_temvib}
	}
	\caption{The contours of macroscopic flow variables around the cylinder at $\rm Ma=20$. The dash line: 
		UGKS with Rykov model, color band: present.}
	\label{fig_cylinder_ma20_field}
\end{figure}
\begin{figure}[!t]
	\centering
	\subfigure[Pressure]{
		\begin{minipage}[!t]{0.45\textwidth}
			\centering
			\includegraphics[width=1.0\textwidth]{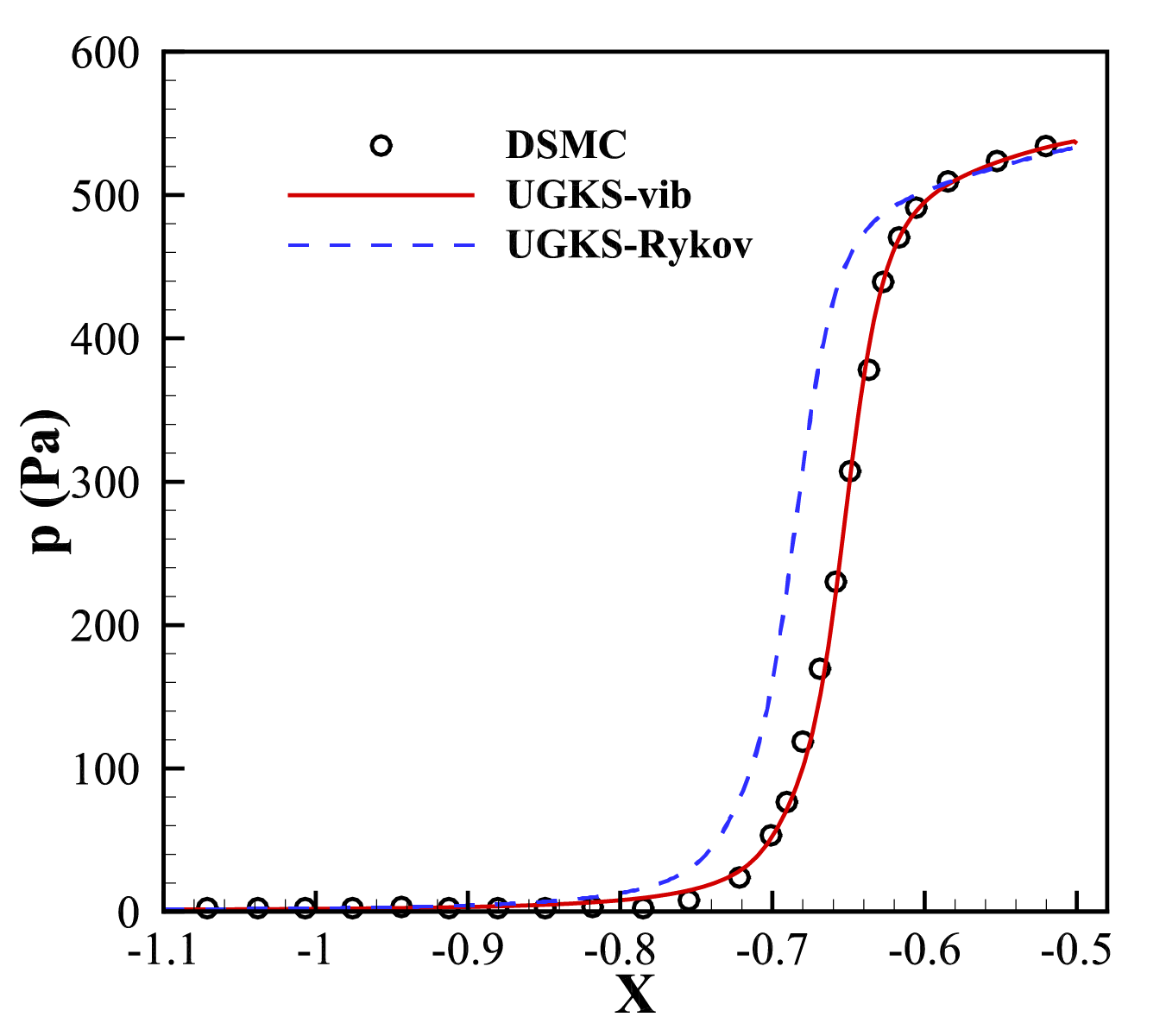}	    
		\end{minipage}
		\label{fig_cylinder_ma20_stagnationline_pressure}
	}
	\subfigure[Translational, rotational and vibrational temperatures]{
		\begin{minipage}[!t]{0.45\textwidth}
			\centering
			\includegraphics[width=1.0\textwidth]{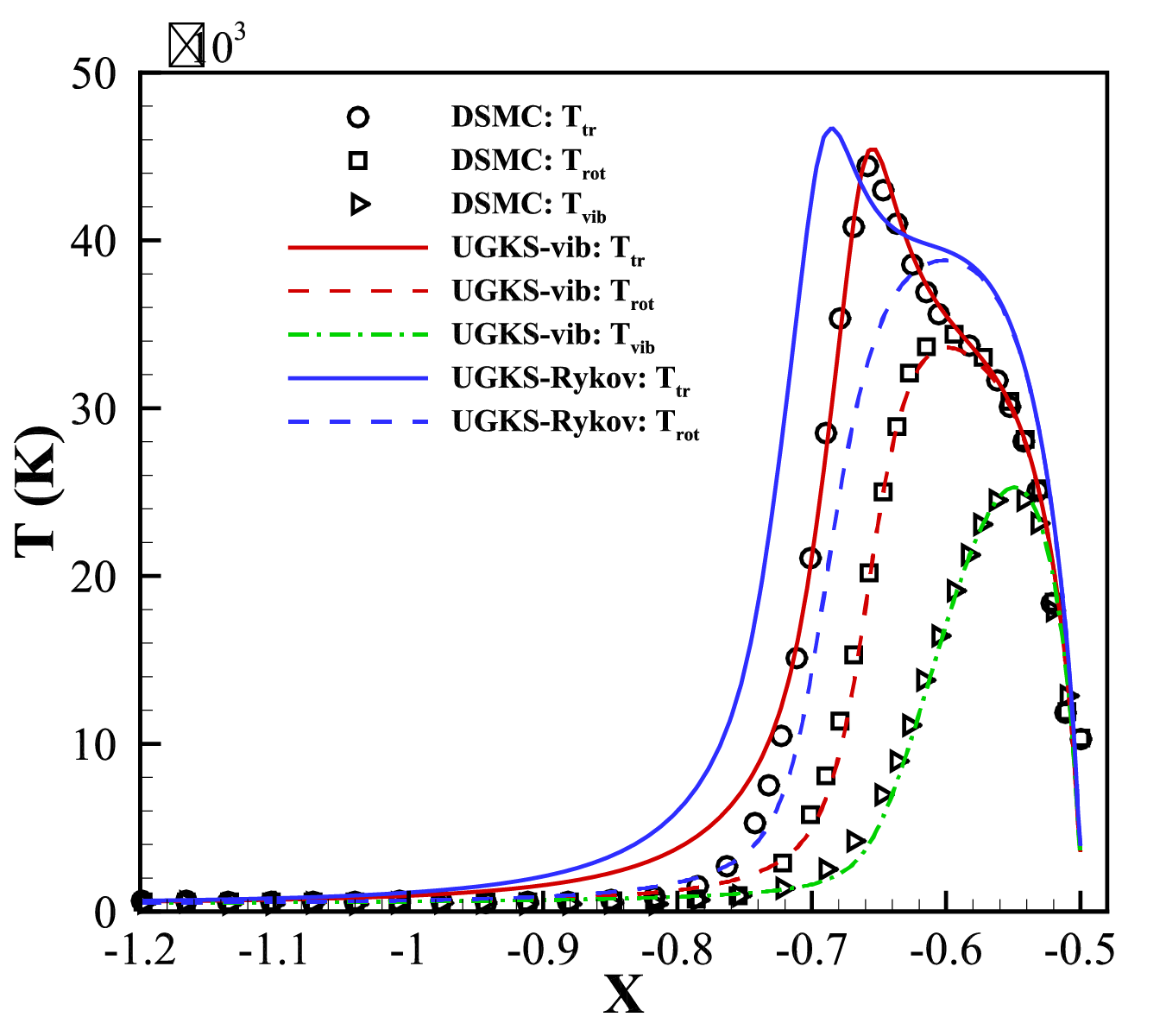}		
		\end{minipage}
		\label{fig_cylinder_ma20_stagnationline_tem}
	}
	\caption{Comparison of the pressure and the translational, rotational and vibrational temperatures 
		along the forward stagnation line for the cylinder at $\rm Ma=20$.}
	\label{fig_cylinder_ma20_stagnationline}
\end{figure}

Fig.~\ref{fig_cylinder_ma20_field} illustrates the contours of the pressure, Mach number, equilibrium temperature, 
translational, rotational and vibrational temperatures obtained from present method and UGKS-Rykov 
method~\citep{ShaLiu_2014}. Fig.~\ref{fig_cylinder_ma20_stagnationline} shows the distributions of pressure and 
translational, rotational and vibrational temperatures along the forward stagnation line. It can be seen that the 
distribution of pressure along the forward stagnation line calculated by present method are consistent with the 
result of DSMC, and the results of translational, rotational and vibrational temperatures are also in good 
agreement with those from DSMC, except the translational temperature is generally higher than the DSMC’s result. 
However, the results of UGKS-Rykov method~\citep{ShaLiu_2014} are significantly different from those of DSMC, 
and the shock detachment distance solved by the UGKS-Rykov method is large than that calculated by present 
method and DSMC in Fig.~\ref{fig_cylinder_ma20_stagnationline}.
\par
\begin{figure}[!t]
	\centering
	\subfigure[Pressure]{
		\begin{minipage}[!t]{0.45\textwidth}
			\centering
			\includegraphics[width=1.0\textwidth]{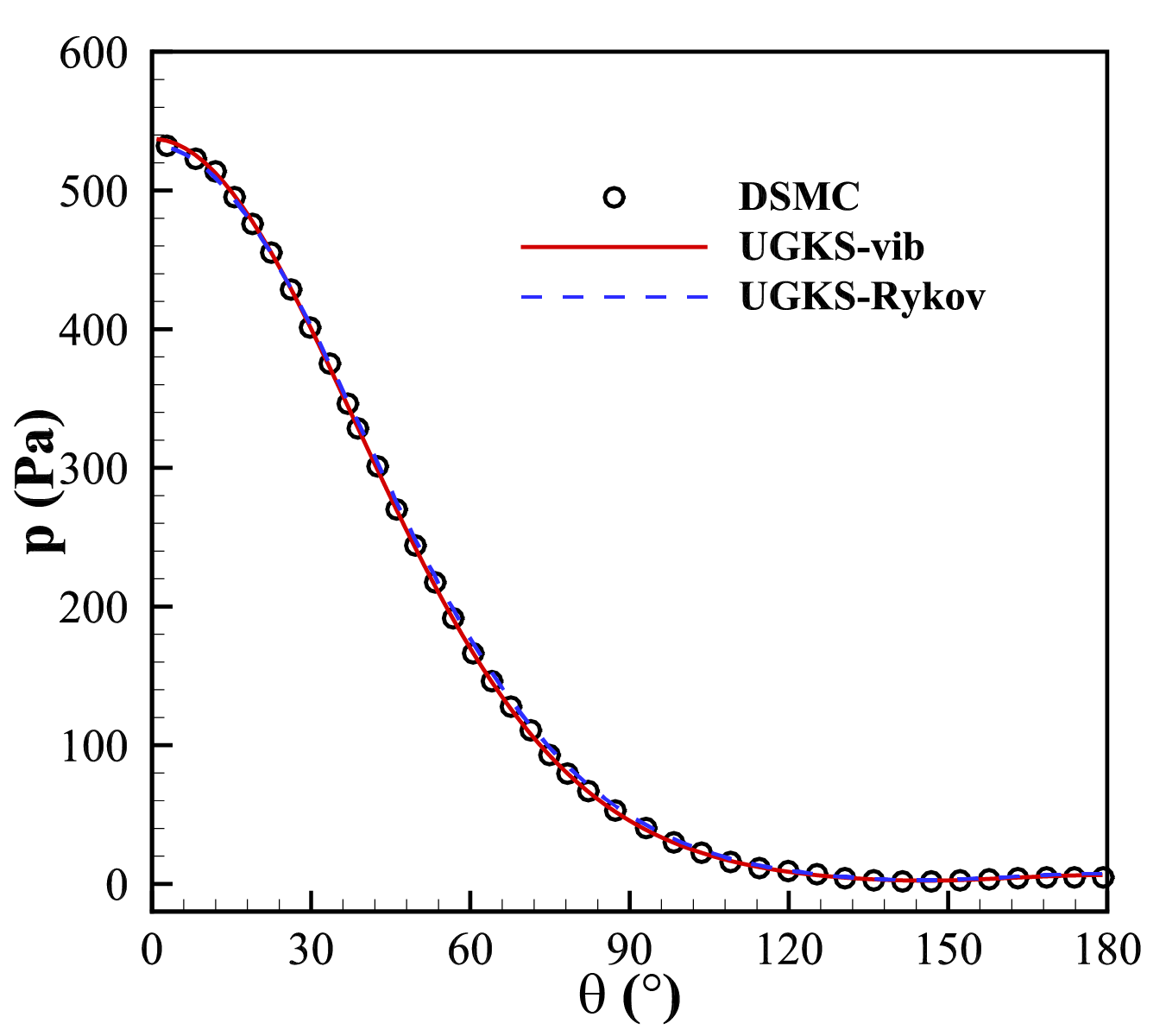}	    
		\end{minipage}
		\label{fig_cylinder_ma20_surfaceline_pressure}
	}
	\subfigure[Heat flux]{
		\begin{minipage}[!t]{0.45\textwidth}
			\centering
			\includegraphics[width=1.0\textwidth]{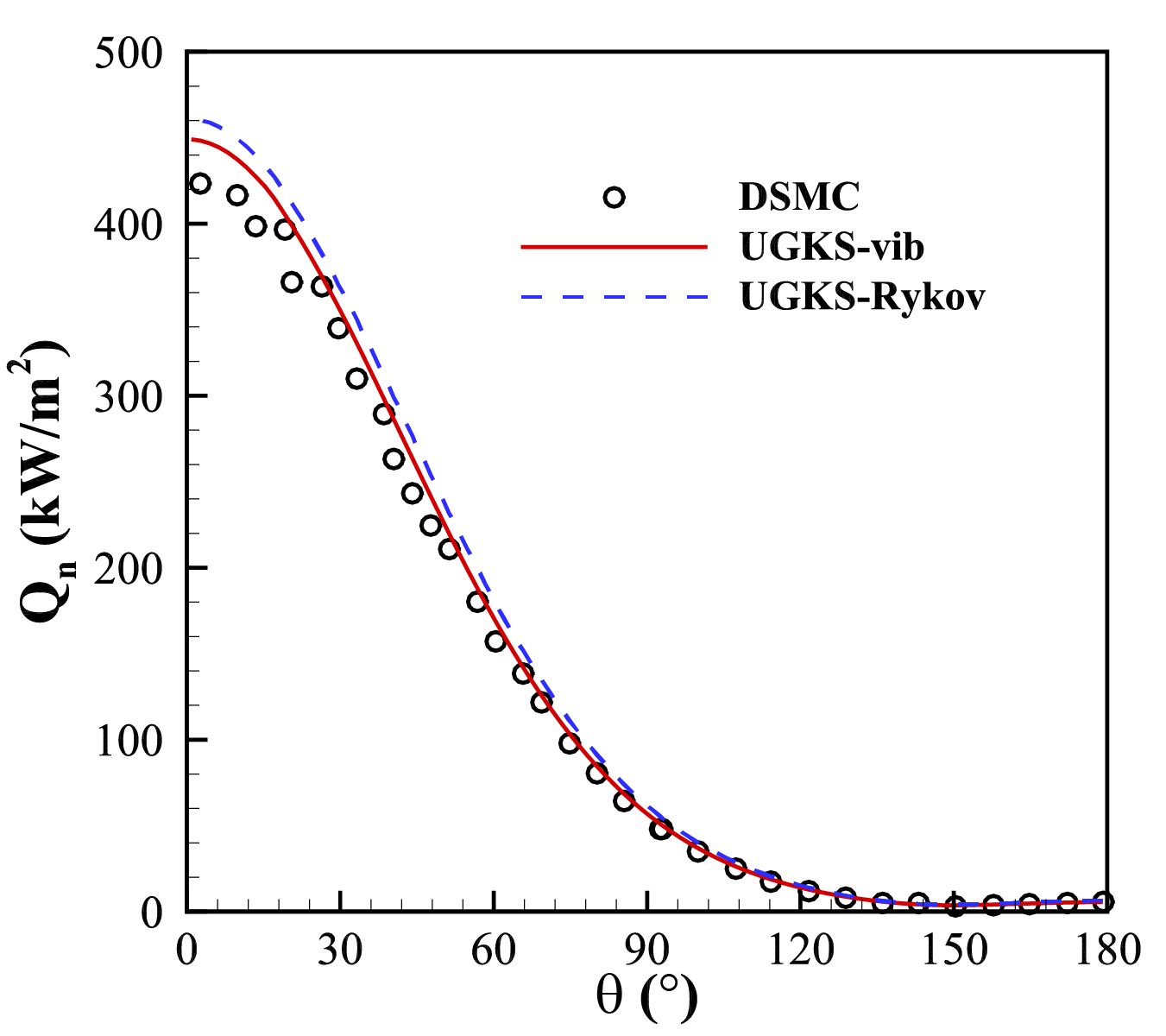}	    
		\end{minipage}
		\label{fig_cylinder_ma20_surfaceline_heatflux}
	}
	\subfigure[Translational temperature]{
		\begin{minipage}[!t]{0.45\textwidth}
			\centering
			\includegraphics[width=1.0\textwidth]{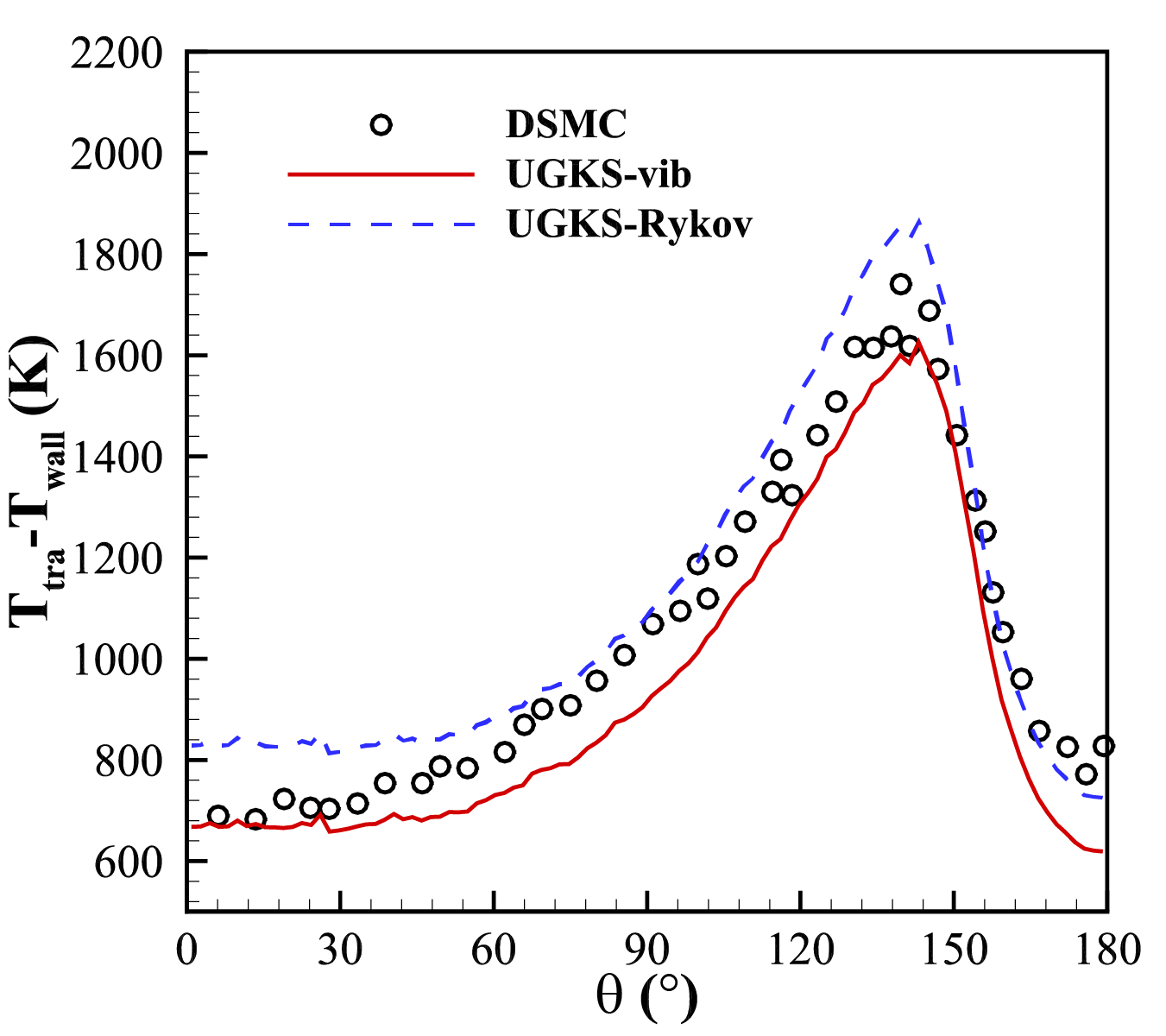}		
		\end{minipage}
		\label{fig_cylinder_ma20_surfaceline_temtra}
	}
	\subfigure[Rotational temperature]{
		\begin{minipage}[!t]{0.45\textwidth}
			\centering
			\includegraphics[width=1.0\textwidth]{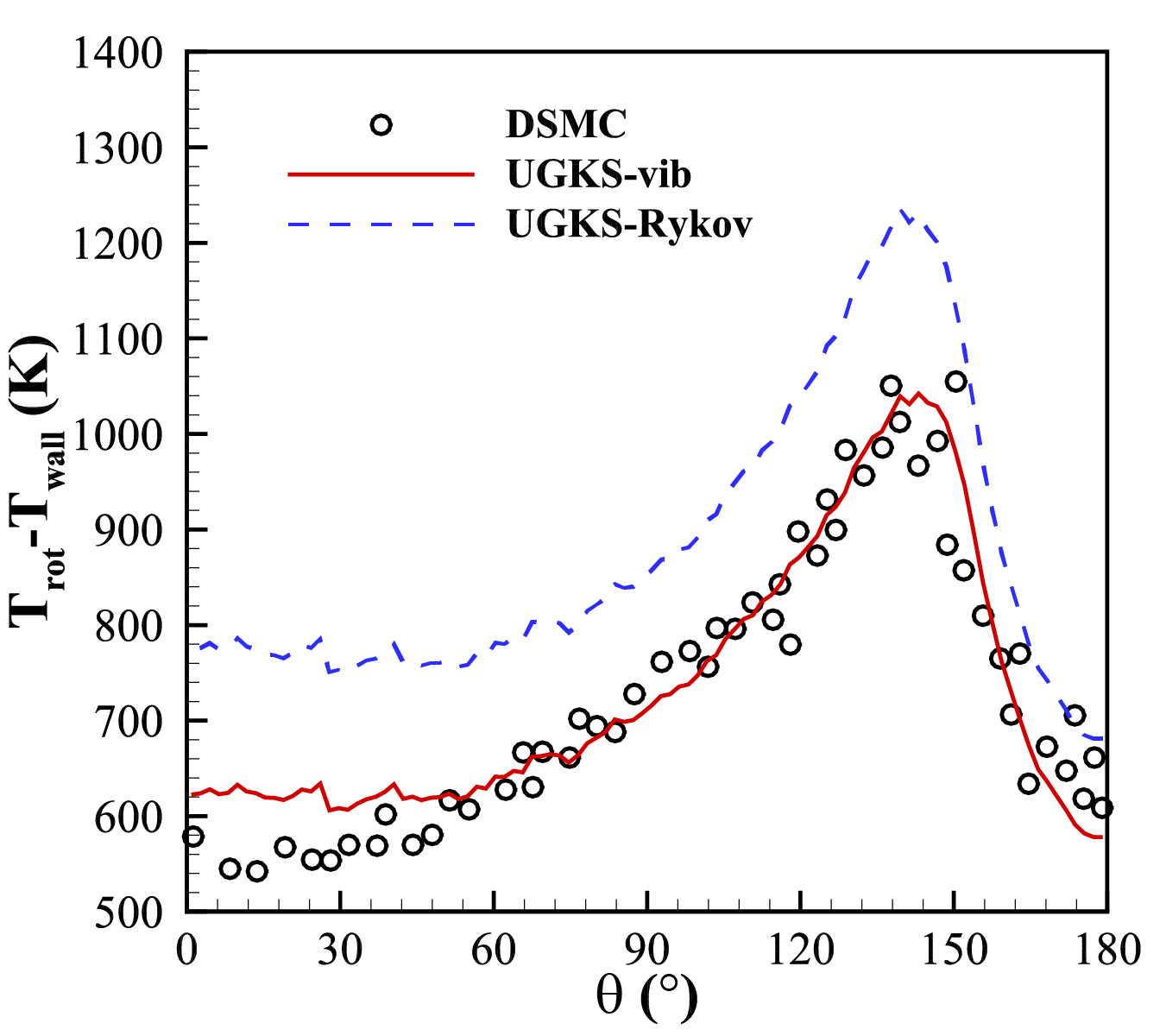}		
		\end{minipage}
		\label{fig_cylinder_ma20_surfaceline_temrot}
	}
	\caption{Distributions of the pressure, heat flux, translational and rotational temperatures
		on the wall surface for the cylinder at $\rm Ma=20$.}
	\label{fig_cylinder_ma20_surfaceline}
\end{figure}

Fig.~\ref{fig_cylinder_ma20_surfaceline} compares the pressure, heat flux, translational and rotational temperature 
jumps on the cylinder surface from the present method and the results of DSMC~\citep{JunLinWu_2021} and UGKS-Rykov 
method~\citep{ShaLiu_2014} in detail. The pressure solved by both the present method and UGKS-Rykov method are almost 
identical with the result of DSMC (Fig.~\ref{fig_cylinder_ma20_surfaceline_pressure}). The results of heat flux 
around the stagnation region solved by both the present and UGKS-Rykov method are large than the DSMC’s result, while 
the current method can improve the result of heat flux on the wall surface compared with UGKS-Rykov method  
(Fig.~\ref{fig_cylinder_ma20_surfaceline_heatflux}). The translational and rotational temperature jumps on the cylinder 
surface are also qualitatively consistent with the results of DSMC (Figs.~\ref{fig_cylinder_ma20_surfaceline_temtra} 
and~\ref{fig_cylinder_ma20_surfaceline_temrot}), but the rotational temperature solved by UGKS-Rykov method show apparent 
deviations compared with the present and DSMC’s results. According to the study of Wu et al.~\citep{JunLinWu_2021}, 
the discrete quantum effect of vibrational energy cause about $2\%$ deviation in temperature when free-stream Mach is 
about 20. However, it is very time-consuming for solving the kinetic model equation with quantum vibrational energy in 
discrete physical and velocity space. Therefore, when the free-stream Mach is less than 20, it is more practical and 
economical to establish a kinetic model equation with continuous distribution modes of rotational and vibrational 
energies for the simulation of hypersonic flows, especially the three-dimensional flows. Meanwhile, the construction 
of kinetic model equation and the determination of model relaxation parameters are more important and it is about 
whether the present multi-scale method can accurately evaluate the heat flux on the wall surface in all flow regimes. 
\par

\subsection{Flow passing a flat plate}\label{Plate}
\par  
When the hypersonic gas flow passes through a flat plate, the features of flow filed include
shock-boundary interactions that cause a strong thermodynamic non-equilibrium between 
translational, rotational and vibrational temperatures near the surface wall.
The hypersonic rarefied nitrogen flow passing a flat plate with a sharp leading edge is simulated 
by the present method, and the simulation results are compared with the experimental 
measurements~\citep{Nobuyuki_Tsuboi_2005}.
\par
The run34 case in Ref.~\citep{Nobuyuki_Tsuboi_2005} is studied. The free-stream Mach number $\rm Ma_{\infty}$, 
pressure $p_{\infty}$, temperature $T_{\infty}$ and viscosity ${\mu _{\infty}}$ are 4.89, $2.12 Pa$, $116 K$ and 
$8.7783 \times {10^{-6}} {pa \cdot s}$, respectively. The temperature of the plate surface is $T_w=290K$. 
The viscosity in the present simulation is calculated with the VHS model and 
the viscosity index is $\omega = 0.74$, thus the free-stream mean free path $\lambda_{mfp,\infty}$ is 
around $7.9408 \times {10^{-4}} m$. In the calculations, the dimensionless quantities normalized by 
the reference length $L_{ref}=0.001m$, density $\rho_{ref}=\rho_{\infty}$, temperature $T_{ref}=T_{\infty}$, and 
velocity $U_{ref}=\sqrt{2RT_{ref}}$~\citep{ShaLiu_2014} are used, and the flow Knudsen number is around 0.79.
The rotational collision number is set as $Z_{rot}=3.5$, while the vibrational collision number is set 
as $Z_{vib}=30$.
\par
\begin{figure}[!t]
	\centering
	\subfigure[Physical space mesh ($3634$ cells)]{
		\begin{minipage}[!t]{0.45\textwidth}
			\centering
			\includegraphics[width=1.0\textwidth]{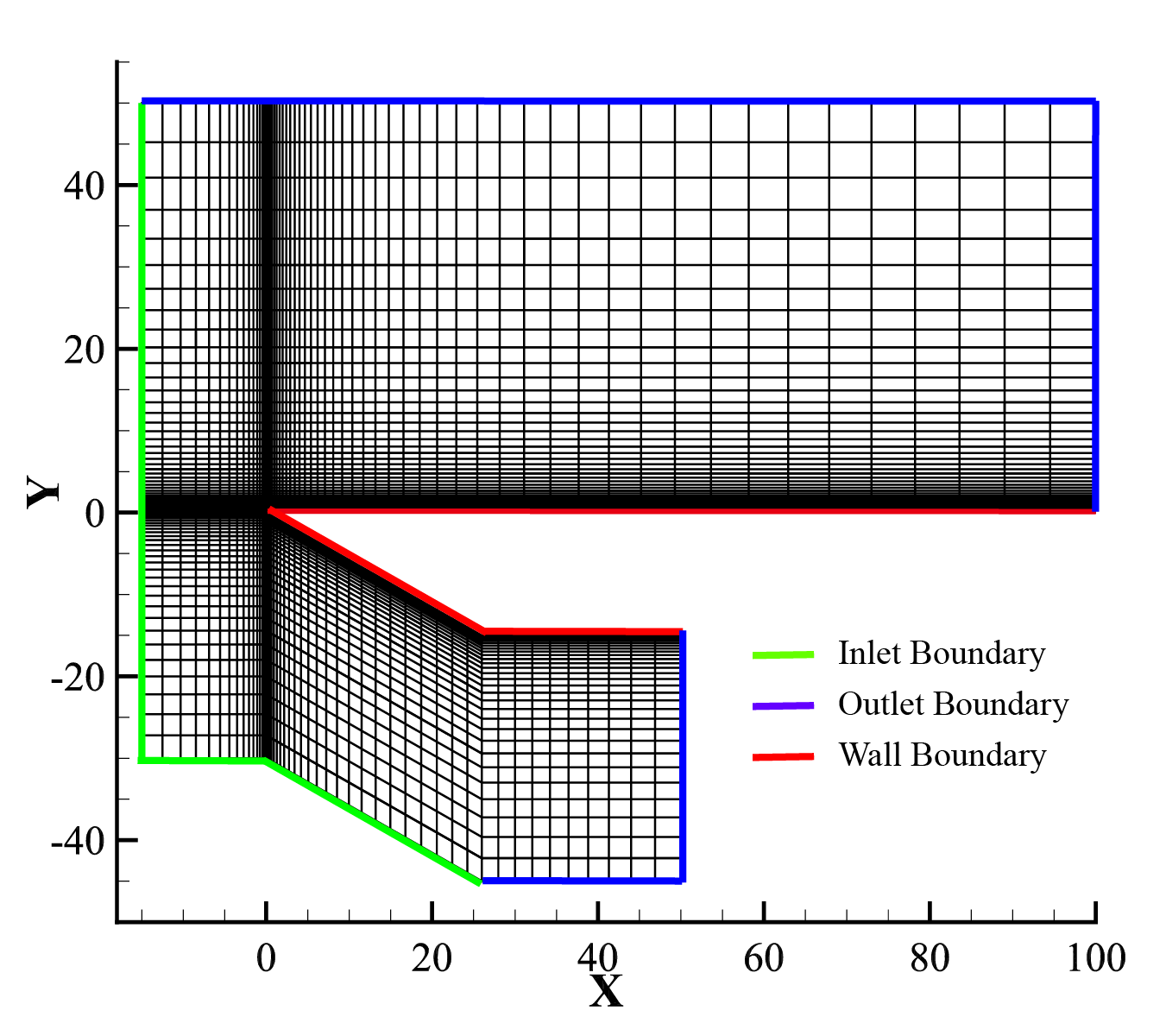}	    
		\end{minipage}
		\label{fig_flat-ma4.89-macmesh}
	}
	\subfigure[Unstructured discrete velocity space mesh ($2894$ cells)]{
		\begin{minipage}[!t]{0.45\textwidth}
			\centering
			\includegraphics[width=1.0\textwidth]{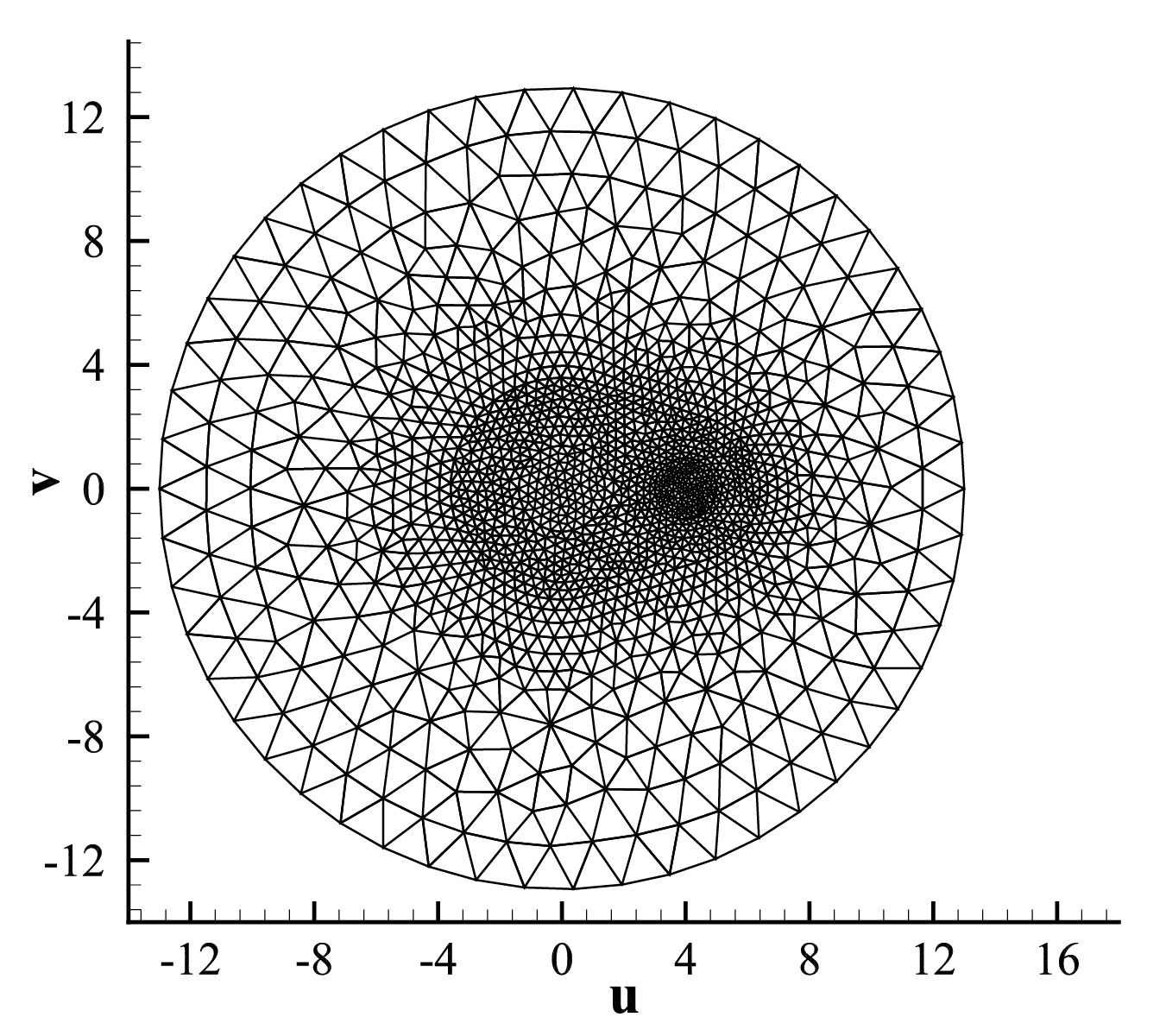}		
		\end{minipage}
		\label{fig_flat-ma4.89-micmesh}
	}
	\caption{The physical space mesh and unstructured discrete velocity space mesh for the flat plate.}
	\label{fig_flat-ma4.89-mesh}
\end{figure}
\begin{figure}[!t]
	\centering
	\subfigure[Pressure]{
		\begin{minipage}[!t]{0.45\textwidth}
			\centering
			\includegraphics[width=1.0\textwidth]{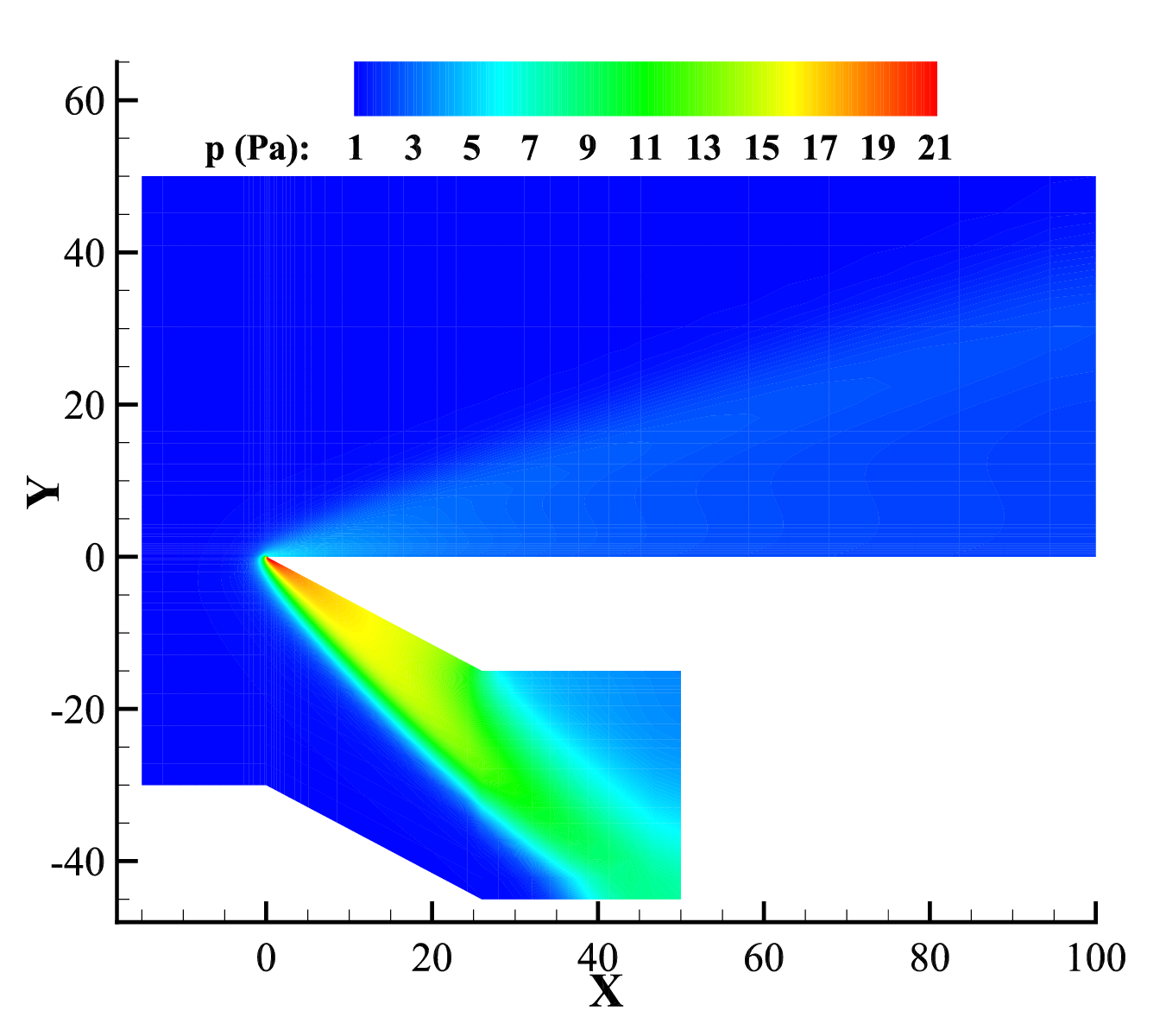}	    
		\end{minipage}
		\label{fig_flat-ma4.89-mac-pressure}
	}
	\subfigure[Mach number]{
		\begin{minipage}[!t]{0.45\textwidth}
			\centering
			\includegraphics[width=1.0\textwidth]{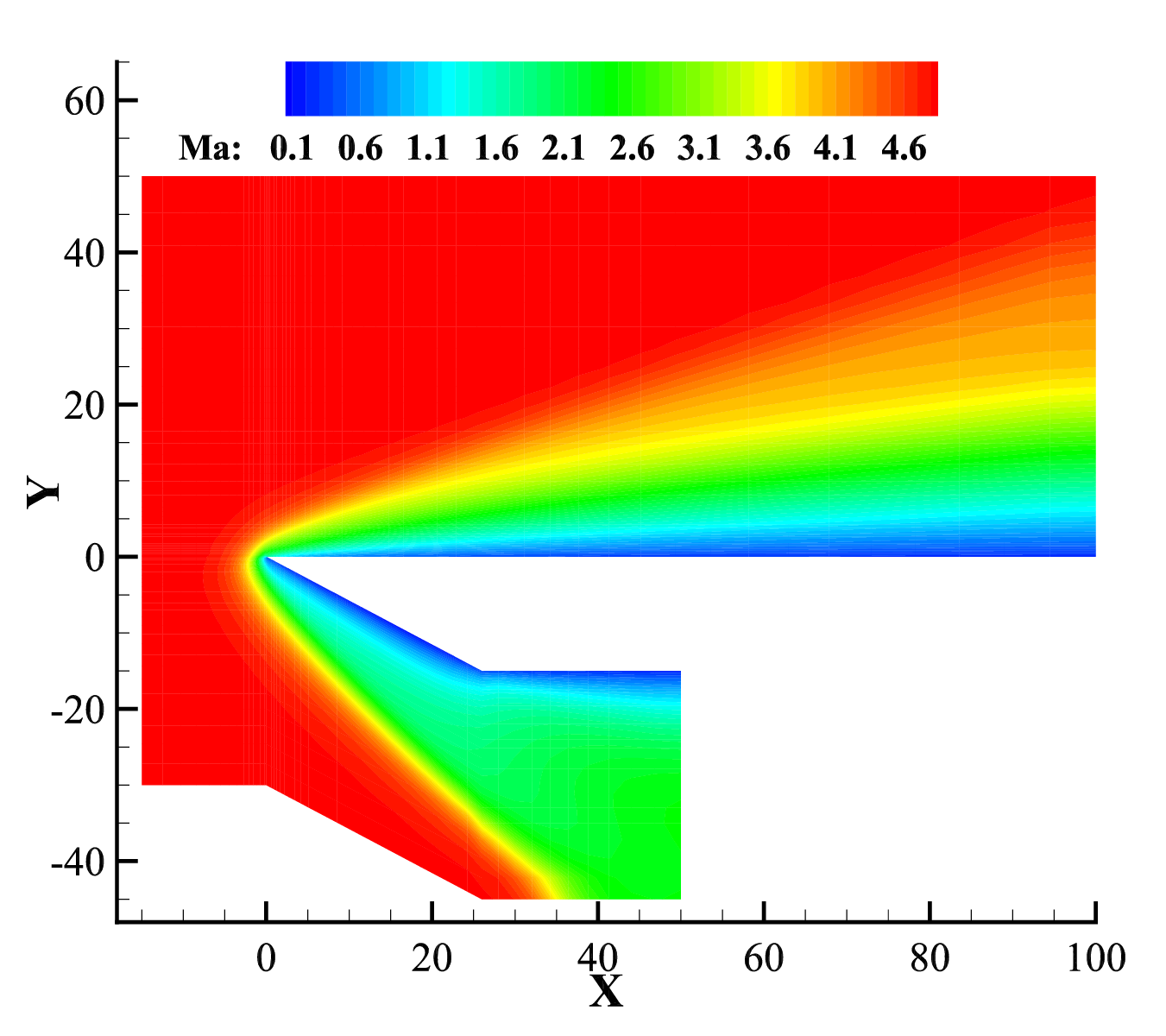}		
		\end{minipage}
		\label{fig_flat-ma4.89-mac-mach-number}
	}
	\subfigure[Equilibrium temperature]{
		\begin{minipage}[!t]{0.45\textwidth}
			\centering
			\includegraphics[width=1.0\textwidth]{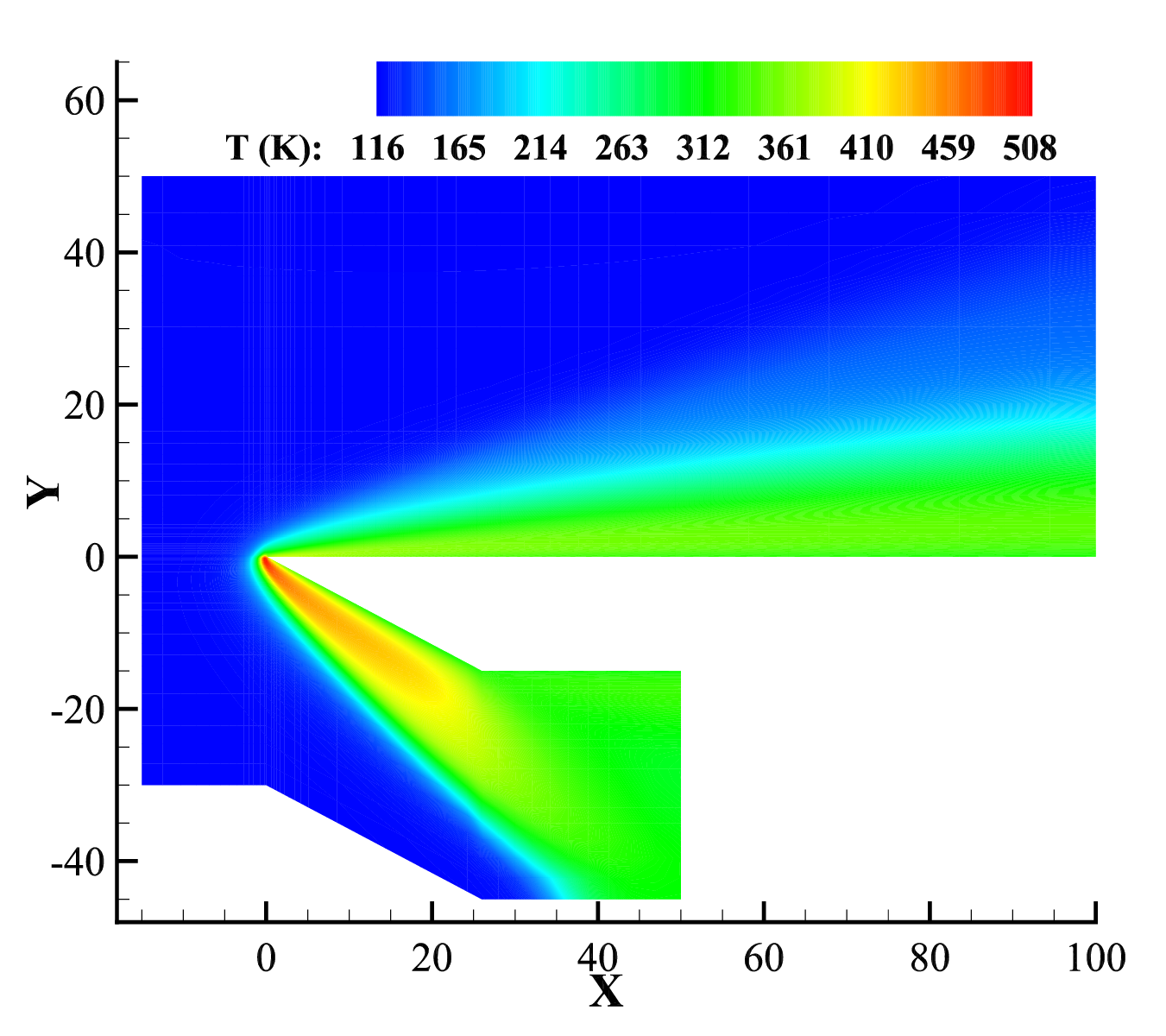}	    
		\end{minipage}
		\label{fig_flat-ma4.89-mac-temperature}
	}
	\subfigure[Translational temperature]{
		\begin{minipage}[!t]{0.45\textwidth}
			\centering
			\includegraphics[width=1.0\textwidth]{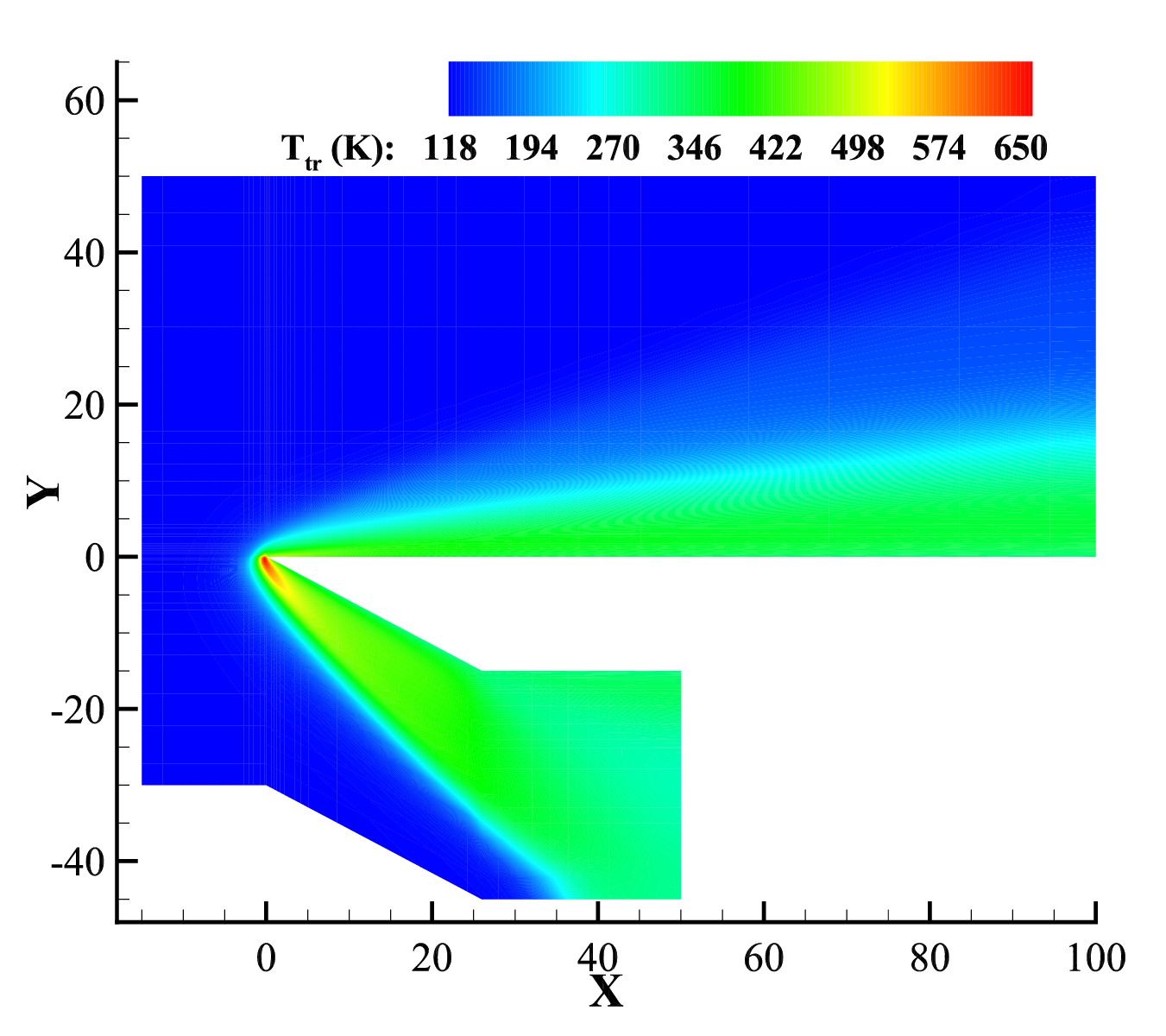}		
		\end{minipage}
		\label{fig_flat-ma4.89-mac-temperature-tra}
	}
	\subfigure[Rotational temperature]{
		\begin{minipage}[!t]{0.45\textwidth}
			\centering
			\includegraphics[width=1.0\textwidth]{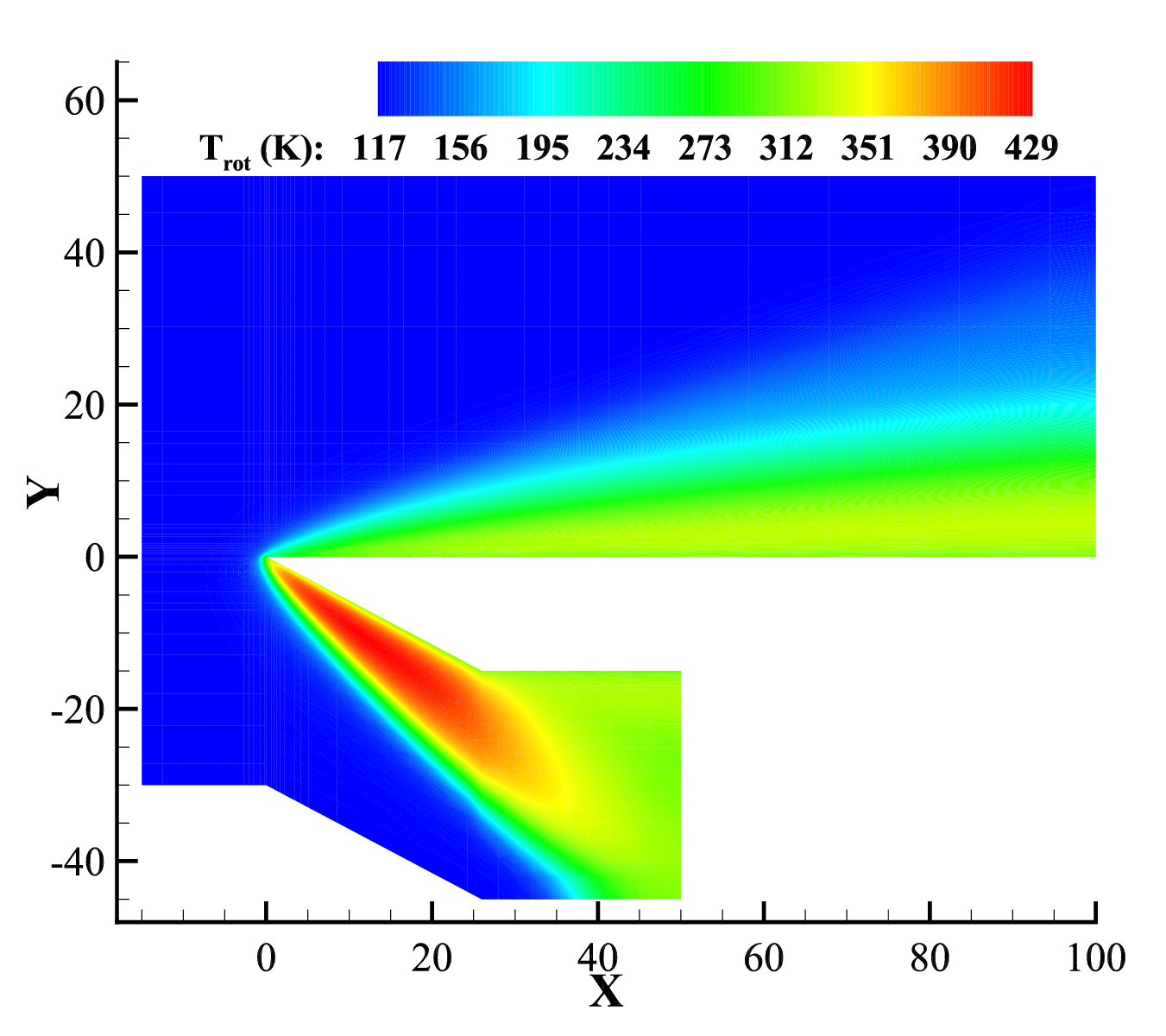}		
		\end{minipage}
		\label{fig_flat-ma4.89-mac-temperature-rot}
	}
	\subfigure[Vibrational temperature]{
		\begin{minipage}[!t]{0.45\textwidth}
			\centering
			\includegraphics[width=1.0\textwidth]{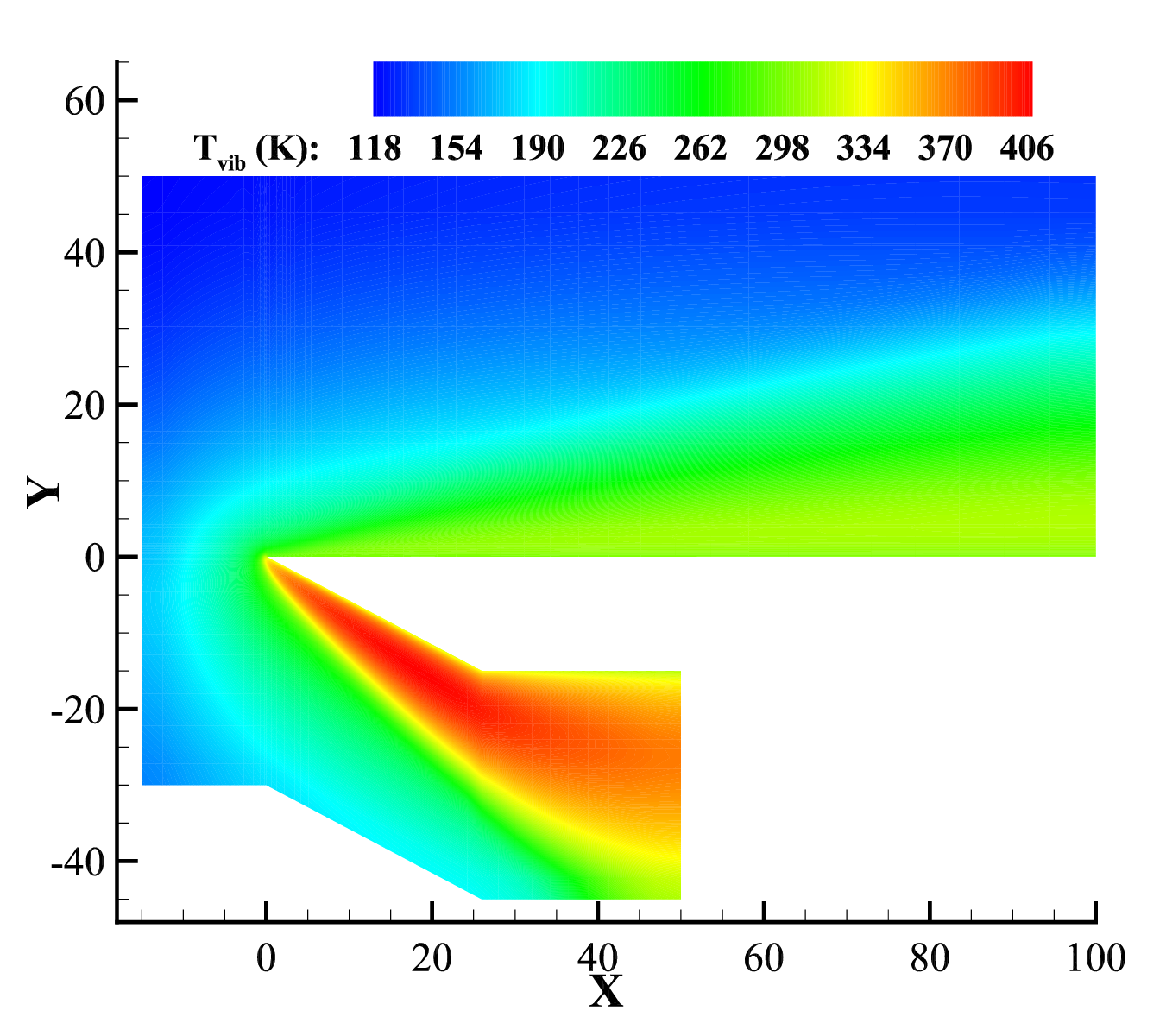}		
		\end{minipage}
		\label{fig_flat-ma4.89-mac-temperature-vib}
	}
	\caption{The contours of macroscopic flow variables for the hypersonic flow passing a flat plate.}
	\label{fig_flat-ma4.89-mac-field}
\end{figure}

Here 3634 elements are used in physical space with $58 \times 39$ cells above the plate and $49 \times 28$
below the plate, which is shown in Fig.~\ref{fig_flat-ma4.89-macmesh}. The unstructured DVS 
(2894 cells) is shown in Fig.~\ref{fig_flat-ma4.89-micmesh}. Fig.~\ref{fig_flat-ma4.89-mac-field} illustrates 
the contours of pressure, Mach number, equilibrium temperature, and translational, rotational and vibrational 
temperatures around the plate.
\par
\begin{figure}[!t]
	\centering
	\subfigure[$X=5mm$]{
		\begin{minipage}[!t]{0.45\textwidth}
			\centering
			\includegraphics[width=1.0\textwidth]{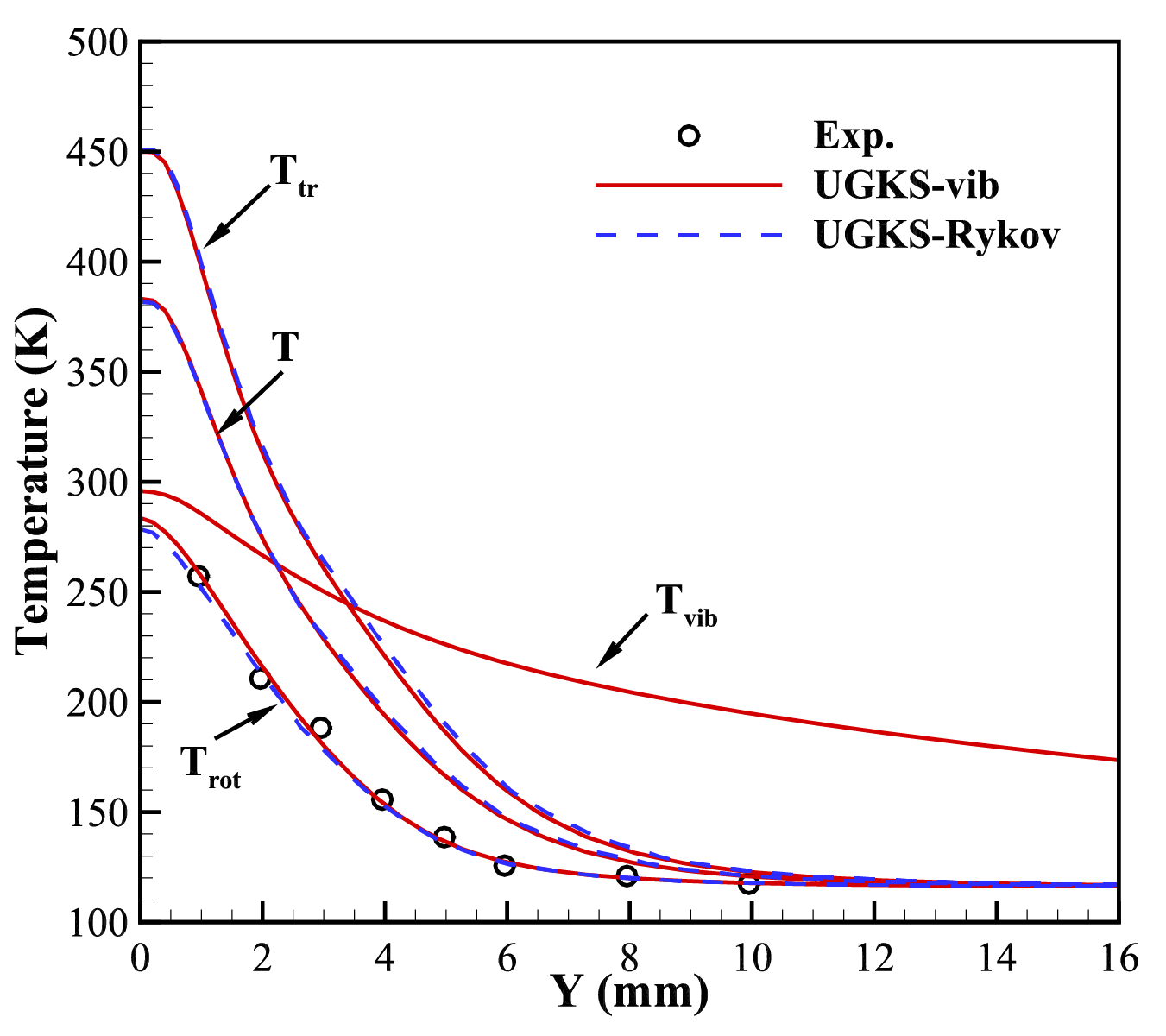}	    
		\end{minipage}
		\label{fig_flat-ma4.89-line-X-5}
	}	
	\subfigure[$X=10mm$]{
		\begin{minipage}[!t]{0.45\textwidth}
			\centering
			\includegraphics[width=1.0\textwidth]{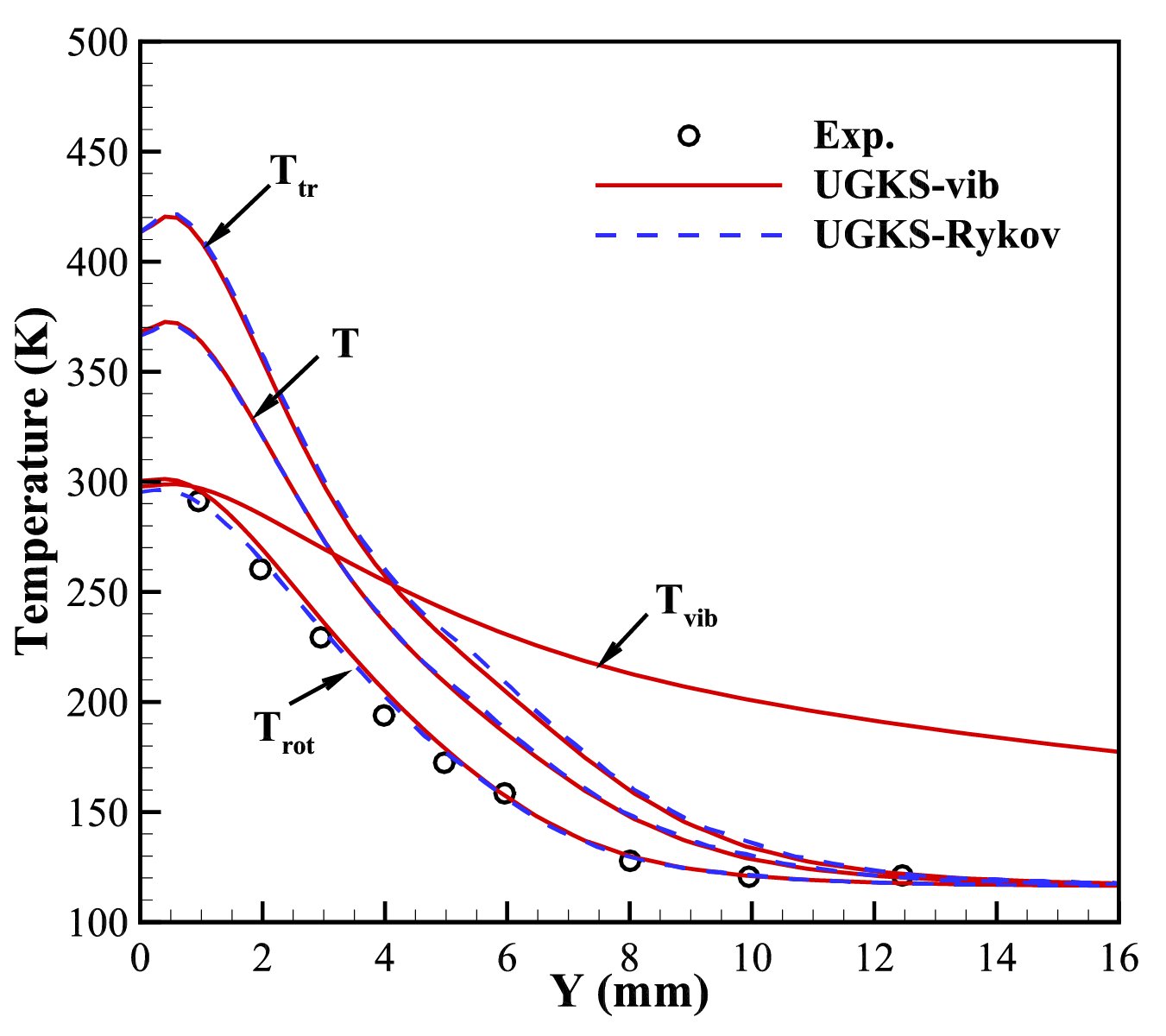}	    
		\end{minipage}
		\label{fig_flat-ma4.89-line-X-10}
	}
	\subfigure[$X=20mm$]{
		\begin{minipage}[!t]{0.45\textwidth}
			\centering
			\includegraphics[width=1.0\textwidth]{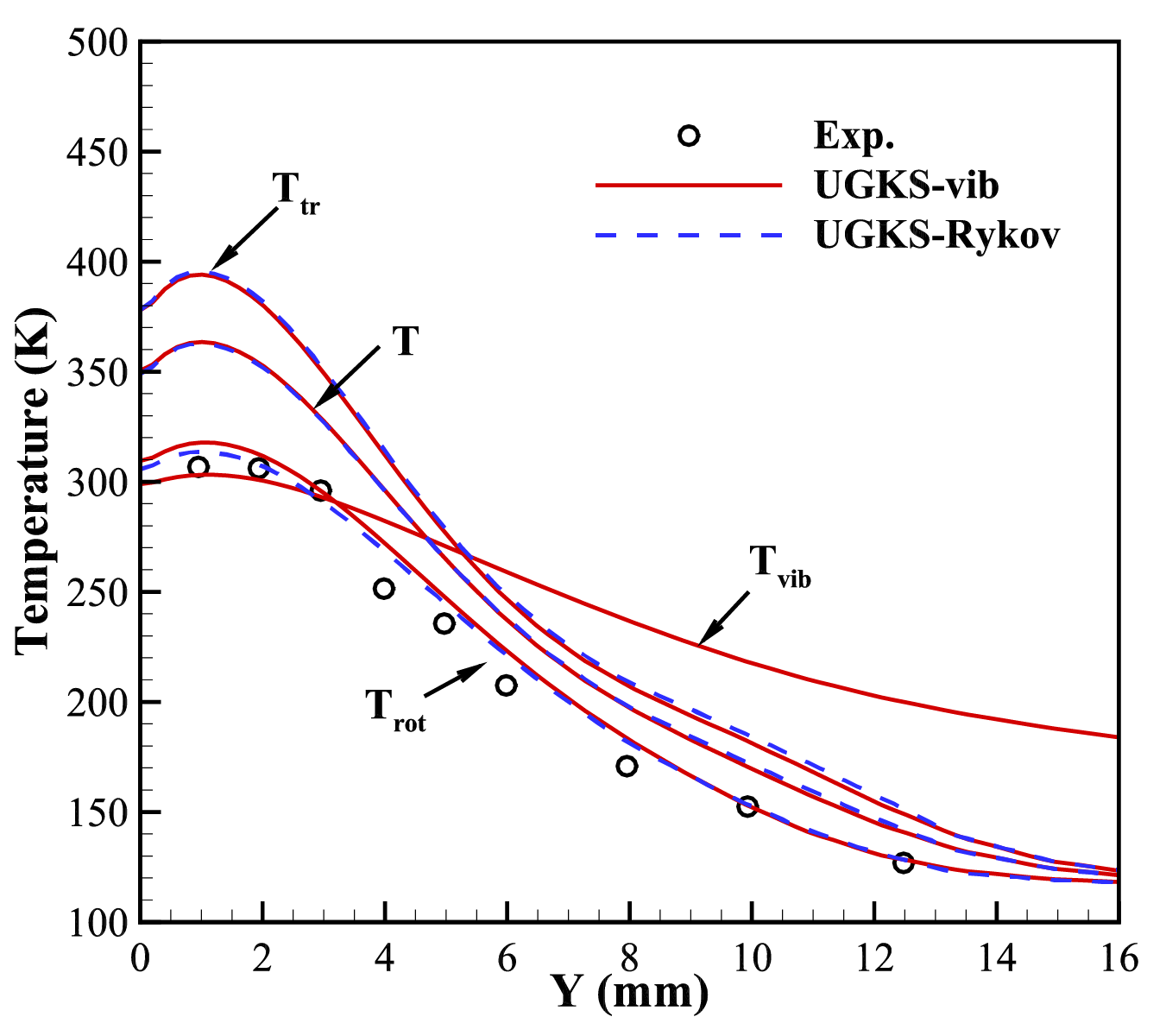}		
		\end{minipage}
		\label{fig_flat-ma4.89-line-X-20}
	}
	\subfigure[$Y=1mm$]{
		\begin{minipage}[!t]{0.45\textwidth}
			\centering
			\includegraphics[width=1.0\textwidth]{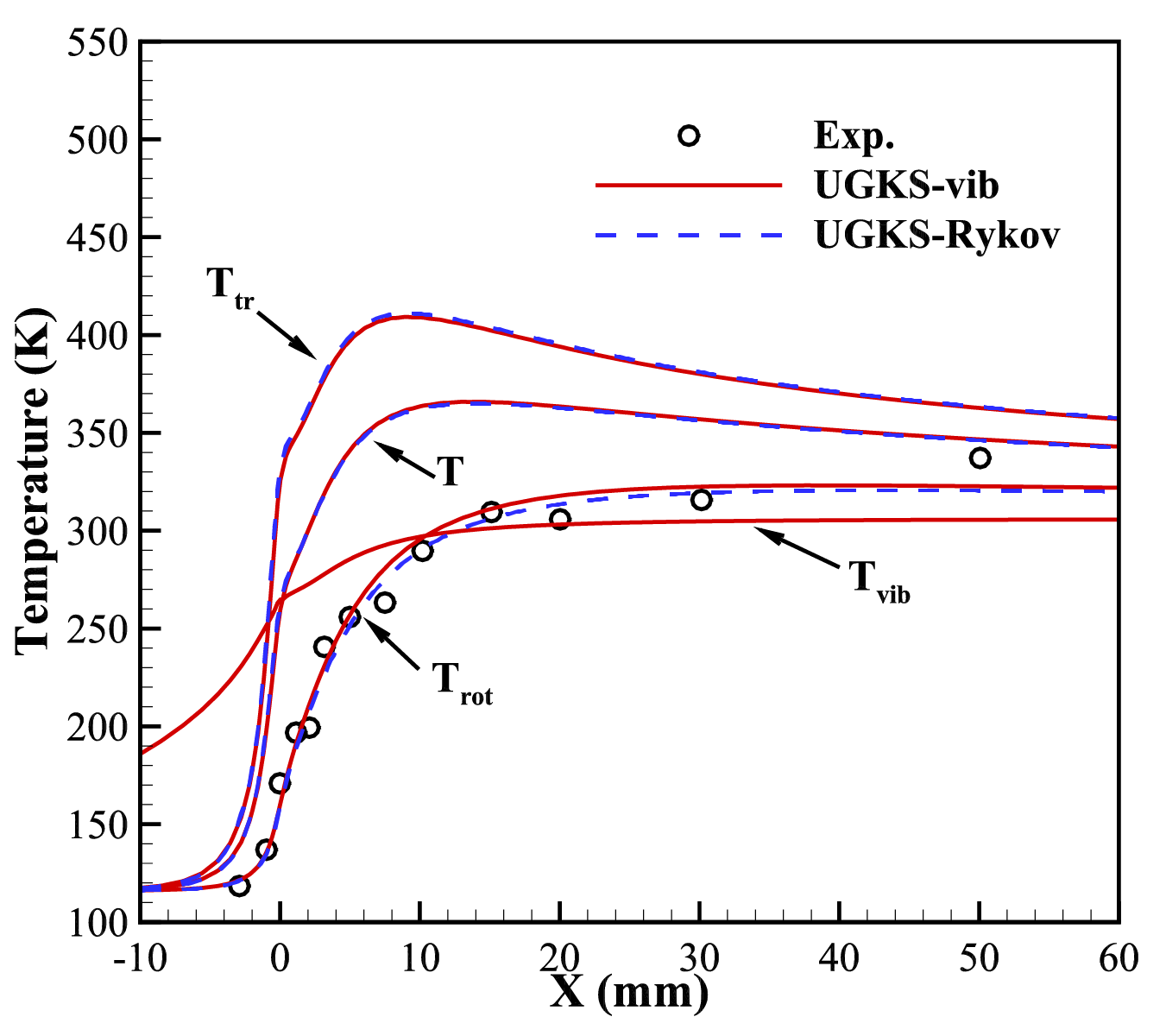}		
		\end{minipage}
		\label{fig_flat-ma4.89-line-Y}
	}
	\caption{Temperature profiles of the hypersonic flow passing a flat plate compared with the experimental rotational 
		temperature distributions over the flat plate along the vertical line $X=5mm$ (a), $X=10mm$(b), 
		$X=20mm$ (c) and along the horizontal line $Y=1mm$ (d).}
	\label{fig_flat-ma4.89-line-X}
\end{figure}

The temperature distributions above the upper surface of the plate along three vertical line $X=5mm$, $X=10mm$ 
and $X=20mm$ are shown in Figs.~\ref{fig_flat-ma4.89-line-X-5}, ~\ref{fig_flat-ma4.89-line-X-10} and 
~\ref{fig_flat-ma4.89-line-X-20}. 
The computed rotational temperatures match well with the 
experimental data measured by an electron beam fluorescence technique~\citep{Nobuyuki_Tsuboi_2005}. The thickness 
of the thermodynamic non-equilibrium layer without considering the vibrational temperature is around $11mm$ at $X=5mm$ 
while $17mm$ at $X=20mm$, which are very close to the values simulated by UGKS~\citep{ShaLiu_2014} with 
Rykov model~\citep{VA_Rykov_1975}. The reason 
is that the relatively low temperature result in very small molecular vibrational energy in this case. 
Therefore, the present results agree well with the UGKS-Rykov method without considering the excitation of 
vibrational degrees of freedom. The temperature profiles above the upper surface of the plate along the line $y=1mm$ 
is also illustrated in Fig.~\ref{fig_flat-ma4.89-line-Y}, and the rotational temperature is consistent with the
experimental data.
\par

\section{Conclusions}\label{Conclusions}
\par
In this paper, a new BGK-type kinetic model equation involving excited vibrational degrees of freedom for diatomic gases 
with thermodynamic non-equilibrium effect is proposed. Both the molecular rotational and vibrational energies are taken to 
be the continuous distribution in present model, which can substantially decrease the computational cost of simulations 
at high temperatures without a significant decrease in accuracy. Based on this proposed kinetic model equation, a efficient 
UGKS with simplified multi-scale numerical flux coupling the merits of UGKS and DUGKS is constructed to simulate the 
hypersonic thermodynamic non-equilibrium flows in all flow regimes. The strategy of updating both the macroscopic flow variables 
and microscopic gas distribution function in UGKS and the strategy of constructing multi-scale numerical fluxes in DUGKS are 
combined in present algorithm. Furthermore, the unstructured DVS with a quick and simple integration error correction are 
adopted to relieve the dimensional crisis caused by a large number of discrete velocity points in the hypersonic flow simulation. 
\par
In the numerical tests, the Sod’s shock tube problem from free-molecular regime to continuum one 
and the planar shock structures with high Mach number are computed, and the present results agree well with the 
analytical and validated DSMC solutions. In the simulation of hypersonic flow around a circular cylinder, the 
thermodynamic non-equilibrium phenomena in the shock wave zone are accurately computed. Besides, the pressure, 
heat flux, velocity slip and temperature jump near the solid wall are directly and accurately captured. 
Finally, the test cases of hypersonic rarefied flow passing a flat plate is performed and compared with the results 
of experiment, in which the present method shows good accuracy in capturing thermodynamic non-equilibrium between 
translational, rotational and vibrational temperatures near the surface wall. In conclusion, the numerical results 
show that the current UGKS with simplified multi-scale numerical flux involving the excitation of molecular 
vibrational degrees is accurate and efficient for computing the distribution of pressure on the wall surface and 
has a significant effect on improving the result of heat flux. 
\par

\section*{Acknowledgments}
The authors thank Prof. Kun Xu at Hong Kong University of Science and Technology and 
Prof. Zhaoli Guo at Huazhong University of Science and Technology for discussions of the 
direct modeling of multi-scale flows. Rui Zhang thanks Dr. Ruifeng Yuan at Southern University of Science 
and Technology for discussions in constructing multi-scale numerical methods. Sha Liu thanks Prof. Lei Wu 
at Southern University of Science and Technology for discussion about the non-equilibrium modeling. 
The present work was supported by the National Natural Science Foundation of China 
(Grants No. 12172301, No. 11902266, No. 12072283 and No. 11902264) and the 111 Project of China (No. B17037).


\bibliography{mybibfile}

\newpage
\begin{table*}[!t] 
	\centering			
	\caption{The parameters of pre-shock equilibrium state and post-shock equilibrium state for nitrogen shock structure.}
	\label{tbl:ss_parameters}
	\begin{tabular}{ccccccccccc}
		\toprule[0.5mm]
		& $\rm Ma_1$ & $\rho_1$ & $T_1$ & $u_1$ & $\gamma_1$ & $\rm Ma_2$ & $\rho_2$ & $T_2$ & $u_2$ & $\gamma_2$ \\
		\midrule
		$\rm Ma=10$ & 10 & 1.0000 & 1.0000 & 8.3666 & 1.4000 & 0.3532 & 7.0544 & 17.174 & 1.1860 & 1.3127 \\
		$\rm Ma=15$ & 15 & 1.0000 & 1.0000 & 12.550 & 1.4000 & 0.3426 & 7.5345 & 36.391 & 1.6656 & 1.2987 \\
		\bottomrule[0.5mm]
	\end{tabular}
\end{table*}

\begin{table*}[!t] 
	\centering			
	\caption{The dimensionless gas flow parameters at infinity and the wall surface temperatures
		for hypersonic flow around a circular cylinder.}
	\label{tb2:cylinder_parameters}
	\begin{tabular}{cccccc}
		\toprule[0.5mm]
		& $\rho_{\infty}$ & $p_{\infty}$ & $U_{\infty}$ & $T_{\infty}$ & $T_{w}$ \\
		\midrule
		$\rm Ma_{\infty}=5 $ & 1.0 & 0.5 & 4.1814  & 1.0 & 1.0 \\
		$\rm Ma_{\infty}=20$ & 1.0 & 0.5 & 16.7254 & 1.0 & 4.0 \\
		\bottomrule[0.5mm]
	\end{tabular}
\end{table*}

\end{document}